\pdfoutput=1
\documentclass[a4paper,12pt]{article}

\usepackage{amsfonts}
\usepackage{mathrsfs}
\usepackage{amsmath}
\usepackage{amssymb}
\usepackage{framed}

\usepackage[medium]{titlesec}
\usepackage{bm}
\usepackage{cite}
\usepackage{stmaryrd}

\usepackage[normalem]{ulem}
\usepackage{extarrows}
\usepackage{slashed}
\usepackage{isodateo}
\usepackage{graphicx}
\usepackage[dvipsnames]{xcolor}
\usepackage[bookmarksnumbered=true,bookmarksopen=true]{hyperref}
 \hypersetup{colorlinks,%
             linkcolor=NavyBlue, %
             citecolor=PineGreen, %
             urlcolor=PineGreen}
\usepackage[hmargin=.7in,vmargin=1.1in]{geometry}
\usepackage{indentfirst}
\usepackage{booktabs}
\usepackage{makecell}
\usepackage{subfig}

\usepackage{bbm}

\newcommand{\FR}[2]{\displaystyle\frac{\,{#1}\,}{#2}}
\newcommand{\fr}[2]{\mbox{$\frac{\,{#1}\,}{#2}$}}
\newcommand{\n}{\nonumber}

\makeatletter
\newcommand{\subalign}[1]{%
  \vcenter{%
    \Let@ \restore@math@cr \default@tag
    \baselineskip\fontdimen10 \scriptfont\tw@
    \advance\baselineskip\fontdimen12 \scriptfont\tw@
    \lineskip\thr@@\fontdimen8 \scriptfont\thr@@
    \lineskiplimit\lineskip
    \ialign{\hfil$\m@th\scriptstyle##$&$\m@th\scriptstyle{}##$\hfil\crcr
      #1\crcr
    }%
  }%
}
\makeatother

\graphicspath{{fig/}}

\def\bge{\begin{equation}}
\def\ede{\end{equation}}
\def\bga{\begin{aligned}}
\def\eda{\end{aligned}}
\def\bgb{\begin{bmatrix}}
\def\edb{\end{bmatrix}}
\def\bgp{\begin{pmatrix}}
\def\edp{\end{pmatrix}}
\def\bgm{\begin{matrix}}
\def\edm{\end{matrix}}
\def\bgs{\begin{subequations}}
\def\eds{\end{subequations}}

\def\di{{\mathrm{d}}}

\def\mb{\mathbf}

\def\pd{\partial}

\def\la{\langle}\def\ra{\rangle}
\def\sla{\slashed}

\def\to{\rightarrow}
\def\To{\Rightarrow}
\def\ii{\mathrm{i}}

\def\al{\alpha}
\def\be{\beta}

\def\de{\delta}
\def\ep{\epsilon}

\def\lam{\lambda}

\def\si{\sigma}

\def\vt{\vartheta}
\newcommand{\ft}[1]{\big[{#1}\big]}

\def\aa{\mathsf{a}}
\def\bb{\mathsf{b}}
\def\cc{\mathsf{c}}

\def\ss{\mathsf{s}}

\def\Re{\mathrm{Re}\,}

\def\G{{\wt{\mathcal{G}}}}
\def\G{{\mathcal{G}}}
\def\E{{\mathcal{E}}}

\def\ec{{{\mathsf{A}}}}

\usepackage{mdframed} 
\usepackage{tikz}

\newmdenv[skipabove=0pt,%
          skipbelow=5pt,%
          leftmargin=0pt,%
          rightmargin=0pt,%
          innertopmargin=-5pt,%
          innerbottommargin=7pt,%
          innerleftmargin=2pt,%
          innerrightmargin=2pt,%
          splittopskip=0pt,%
          splitbottomskip=0pt,%
          linewidth=0pt,%
          nobreak=true]%
          {keyeqn2}

\newmdenv[backgroundcolor=gray!15,%
          skipabove=0pt,%
          skipbelow=5pt,%
          leftmargin=0pt,%
          rightmargin=0pt,%
          innertopmargin=-5pt,%
          innerbottommargin=7pt,%
          innerleftmargin=2pt,%
          innerrightmargin=2pt,%
          splittopskip=0pt,%
          splitbottomskip=0pt,%
          linewidth=0pt,%
          nobreak=true]%
          {keyeqn}
           
\newmdenv[font=\small,
		  linecolor=black,
          skipabove=10pt,%
          skipbelow=10pt,%
          leftmargin=0pt,%
          rightmargin=0pt,%
          innertopmargin=14pt,%
          innerbottommargin=14pt,%
          innerleftmargin=12pt,%
          innerrightmargin=12pt,%
          splittopskip=15pt,%
          splitbottomskip=5pt,%
          linewidth=0.8pt]%
          {boxedtext}

\usepackage{titlesec}          
\titleformat{\section}
{\normalfont\fontsize{15}{20}\bfseries}{\thesection}{1em}{}

\newcommand{\wt}[1]{\mkern 2mu \widetilde{\mkern -2mu #1 \mkern -2mu}\mkern 2mu}
\newcommand{\wh}[1]{\mkern 2mu \widehat{\mkern-2mu#1\mkern-2mu}\mkern 2mu}

\newcommand{\mft}[1]{\big\llbracket{#1}\big\rrbracket}

\newcommand{\fnemail}[1]{\footnote{Email: \href{mailto:#1}{\nolinkurl{#1}}}}

\begin{document}

\title{\Large\textbf{Massive Inflationary Amplitudes:\\New Representations and Degenerate Limits\\[2mm]}}

\author{Zhong-Zhi Xianyu$^{\,a,b\,}$\fnemail{zxianyu@tsinghua.edu.cn}~~~~~and~~~~~Jiaju Zang$^{\,a\,}$\fnemail{zangjj24@mails.tsinghua.edu.cn}\\[5mm]
$^a\,$\normalsize{\emph{Department of Physics, Tsinghua University, Beijing 100084, China} }\\ 
$^b\,$\normalsize{\emph{Peng Huanwu Center for Fundamental Theory, Hefei, Anhui 230026, China }}
}

\date{}
\maketitle

\vspace{20mm}

\begin{abstract}
\vspace{10mm}
The particle model building of cosmological collider physics often involves boost-breaking bilinear mixing between a heavy particle and the nearly massless inflaton mode. In cosmological correlators, such a mixing is obtained by taking a folded limit of a generic tree graph, which is a special case of degenerate kinematics. In this work, we continue our exploration of massive inflationary amplitudes with a focus on degenerate kinematics. With a suitable change of variables, we derive new differential equations and full analytical solutions for generic tree graphs, making it trivial to take the folded limit and partial-energy limit at a vertex. Our result shows that folded tree graphs generally involve functions of smaller transcendental weights than their nondegenerate counterparts. In particular, the inflaton bispectrum with triple massive exchanges can be expressed in terms of a trivariate Kampé de Fériet function and simpler hypergeometric functions. 
\end{abstract}

\newpage
\tableofcontents

\newpage
\section{Introduction}\label{sec_intro}

This work is the second in a series of works devoted to the study of massive inflationary amplitudes. Following the previous work on general tree graphs \cite{Liu:2024str}, here, we focus on a special situation where a tree-level massive inflation correlator involves bilinear mixings between some massive fields and the effectively massless external inflaton fluctuations. These bilinear mixings are well motivated from  phenomenological studies, and also display their own interesting mathematical structure. So, we think that they deserve separate treatment. 

The study of massive inflationary amplitudes grows partly from the active explorations of cosmological collider (CC) physics in recent years \cite{Chen:2009we,Chen:2009zp,Arkani-Hamed:2015bza}\footnote{As in previous works, we use the term inflationary amplitudes or cosmological amplitudes to refer to both wavefunction coefficients and correlation functions. The two quantities are conceptually distinct but technically very similar. See \cite{Fan:2024iek} for their definitions and comparisons.}. On the observation side, the sensitivity to CC signals will be significantly increased by large-scale structure data in the near future \cite{Achucarro:2022qrl} and many studies appear recently trying to bridge CC models with data \cite{Cabass:2024wob,Goldstein:2024bky,Sohn:2024xzd,Philcox:2025bvj,Bao:2025onc,Goldstein:2025eyj,Anbajagane:2025uro}. On the theory frontier, many models and scenarios from particle physics have been found that naturally generate large CC signals \cite{Chen:2016nrs,Chen:2016uwp,Chen:2016hrz,Lee:2016vti,An:2017hlx,Iyer:2017qzw,Kumar:2017ecc,Tong:2018tqf,Chen:2018sce,Chen:2018xck,Wu:2018lmx,Saito:2018omt,Li:2019ves,Lu:2019tjj,Liu:2019fag,Hook:2019zxa,Hook:2019vcn,Kumar:2018jxz,Kumar:2019ebj,Wang:2019gbi,Wang:2020uic,Li:2020xwr,Wang:2020ioa,Aoki:2020zbj,Bodas:2020yho,Lu:2021gso,Sou:2021juh,Lu:2021wxu,Pinol:2021aun,Cui:2021iie,Tong:2022cdz,Reece:2022soh,Qin:2022lva,Chen:2022vzh,Niu:2022quw,Chen:2023txq,Chakraborty:2023qbp,Tong:2023krn,Jazayeri:2023xcj,Jazayeri:2023kji,Aoki:2023tjm,Craig:2024qgy,Pajer:2024ckd,McCulloch:2024hiz,Wu:2024wti,Quintin:2024boj,Bodas:2024hih,Aoki:2024jha,An:2024zfi,Wang:2025qww,Bodas:2025wuk,Kumar:2025anx}. The inflationary correlators are central observables in CC physics. It is thus crucial to develop tools for precise and efficient computations of inflationary correlators. At the same time, inflationary correlators have their own analytical structure, which may have a close relationship with fundamental properties of a quantum field theory (QFT) in dS spacetime such as locality, unitarity, and causality. All of these facts motivate us to pursue a systematic study of massive inflationary amplitudes. 

Inflationary correlators are equal-time expectation values of operator products in QFT in an inflationary spacetime. When perturbation theory holds, a convenient computation strategy is the diagrammatic expansion with the Schwinger-Keldysh path integral \cite{Schwinger:1960qe,Keldysh:1964ud,Feynman:1963fq,Weinberg:2005vy}; See \cite{Chen:2017ryl} for a review. Despite its inherent complications of massive exchanges, recent years have witnessed significant progress in this direction. At the tree level, early studies have successfully computed graphs with single exchanges \cite{Arkani-Hamed:2018kmz,Baumann:2019oyu,Sleight:2019mgd,Sleight:2019hfp}, which were generalized to boost-breaking dispersions in \cite{Pimentel:2022fsc,Jazayeri:2022kjy,Qin:2022fbv,Qin:2025xct} and double massive exchanges in \cite{Xianyu:2023ytd,Aoki:2024uyi}. A systematic method to compute tree graphs of arbitrary massive exchanges was first proposed in \cite{Xianyu:2023ytd}, and the full answer was finally obtained in \cite{Liu:2024str}. Progress has also been made for massive loop graphs \cite{Wang:2021qez,Xianyu:2022jwk,Qin:2023bjk,Qin:2023nhv,Liu:2024xyi,Qin:2024gtr,Zhang:2025nzd} and conformal-scalar amplitudes \cite{Arkani-Hamed:2017fdk,Arkani-Hamed:2023bsv,Arkani-Hamed:2023kig,Fan:2024iek,He:2024olr,Hillman:2019wgh,De:2023xue,Benincasa:2024leu,Benincasa:2024lxe,Baumann:2024mvm}. See also \cite{Stefanyszyn:2024msm,Ema:2024hkj,Werth:2024mjg,Chen:2024glu,Raman:2025tsg,Cheung:2025dmc,Jazayeri:2025vlv} for related works.  

In \cite{Liu:2024str}, a set of differential equations was identified and solved for general massive tree graphs. More importantly, a simple set of rules was derived for directly writing down the analytical solutions without doing computations. The full expression for a tree graph with $I$ massive internal lines can be written as a hypergeometric function of $2I$ variables and all of its cuts, which involves hypergeometric functions of fewer variables. The result suggests that we can define a transcendental weight for these multivariate hypergeometric functions. A function defined from a hypergeometric series of $n$-fold summations is said to be of (transcendental) weight-$n$. \cite{Liu:2024str} showed that a tree graph with $I$ internal lines is a hypergeometric function of weight-$2I$. With the results of \cite{Liu:2024str}, the analytical computation of tree-level massive inflationary correlators can be viewed as a solved problem. 

However, the situation is not fully satisfactory for tree graphs with bilinear mixings. In the terminology of \cite{Liu:2024str}, the momentum conservation enforced by the mixing implies that the corresponding vertex energy and line energy are set to equal.\footnote{A vertex energy is the magnitude sum of momenta of all external lines attached to a vertex, while a line energy is the magnitude of the momentum of an internal line. See App.\;\ref{app_masstree} for further explanations of the terminology.} Thus, the bilinear mixing belongs to degenerate kinematics where not all energies can be treated as independent variables. In principle, these degenerate kinematics can be reached by taking folded limits of generic graphs in \cite{Liu:2024str}. However, existing results for graphs with a single bilinear mixing suggest that these limits often lead to further simplifications that cannot be seen using the variables of \cite{Liu:2024str}. Given the phenomenological importance of the bilinear mixings, it seems to us useful to further elaborate on the simplifications of taking folded limits, a topic pursued in this work. In the rest of this introductory section, we provide phenomenological and theoretical motivations for studying graphs with bilinear mixings.

\paragraph{Phenomenology of bilinear mixings}
It is easy to see why bilinear mixing is important to phenomenology: The inflaton bispectrum (3-point correlator) is the lowest-point correlator that can carry scale-invariant CC signals, and bilinear mixing is the only option if we want a heavy particle to generate tree-level CC signals in the 3-point function. 

Indeed, the bilinear mixing between a massive field $\si$ and the massless inflaton fluctuation $\varphi$ is very common in CC models. The key observation is the following. During inflation, the inflaton field $\phi$ can be expanded around its slowly rolling background $\phi(t,\bm x)\simeq \phi_0+\dot\phi_0 t+\varphi(t,\bm x)$, where $\phi_0$ is an (unimportant) initial value of the inflaton background, $\dot\phi_0$ is the inflaton rolling speed and is almost constant, and $\varphi(t,\bm x)$ denotes the almost massless inflaton fluctuation. In the CC model building, it is typical to introduce couplings between $\phi$ to a massive field $\si$ in order to generate a CC signal of $\si$ in correlators of $\varphi$. The near scale invariance of the observed power spectrum at the CMB scale \cite{Planck:2018jri} suggests the existence of an approximate shift symmetry of $\phi$. For this reason, we typically want to couple $\phi$ to other fields through derivatives. (Direct couplings are certainly allowed but are slow-roll suppressed.) Then, a typical example involving $\phi$-$\si$ coupling is\footnote{Here we use the background metric $\di s^2=(-\di \tau^2+\di\bm x^2)/(H\tau)^2$, where $\tau\in(-\infty,0)$ is the conformal time and $H$ is the (constant) Hubble parameter. Also, $f'\equiv\di f/\di\tau$ and $\dot f\equiv \di f/\di t$ where $t$ is the comoving time. The conformal time $\tau$ is related to the comoving time $t$ via $a(t)=e^{Ht}=-1/(H\tau)$, where $a$ is the scale factor. Note that it is $\dot\phi_0$ rather than $\phi_0'$ that is nearly constant during slow-roll inflation.}:
\bge
\label{eq_dim5toMix}
  S\supset -\FR1{2\Lambda}\int\di^4x\,\sqrt{-g}(\pd_\mu\phi)^2\si\to -\FR{\dot\phi_0}{\Lambda}\int\di \tau\di^3\bm x\,a(\tau)^3 \varphi' \si.
\ede
Thus, at the fluctuation level, we have a bilinear mixing between $\varphi$ and $\si$, which respects the scale invariance but breaks the dS boosts, an important point to which we will come back soon. 

The phenomenological study of the above bilinear mixing has a long history. Well before the advent of the term ``cosmological collider,'' inflaton correlators with bilinear mixing were already studied in the context of quasi-single-field inflation (QSFI) \cite{Chen:2009we,Chen:2009zp,Chen:2012ge}. It was identified that the triple-exchange graph, i.e., the graph with 3 mixed propagators, makes the dominant contribution to the 3-point correlator. Later, it was realized that the mixing can be made strong without spoiling the loop expansion \cite{An:2017hlx,Iyer:2017qzw,Reece:2022soh}. The mixing is also essential in a scenario where the heavy scalar acquires a chemical-potential enhancement \cite{Bodas:2020yho}. More generally, there are interesting cases where the background inflaton has an oscillating component $\phi_0(t)=\dot\phi_0 t+\al(t/t_0)^{q} \cos(\omega t)$. This oscillation could generate scale-asymmetric correlators with enhanced CC signals via resonances \cite{Chen:2022vzh}. In such cases, there can be a bilinear mixing whose coefficient is oscillatory in time.

Phenomenologically interesting bilinear mixings typically involve a reduction of symmetry. If the mixing of two fields $\si$ and $\varphi$ respects the local Lorentz invariance, such as a kinetic mixing $(\pd\si)(\pd\varphi)$ or a mass mixing $\si\varphi$, it can be rotated away by a Lorentz-invariant field redefinition and thus is in a sense trivial.\footnote{Incidentally, for two scalars $\si_{1,2}$ of masses $m_{1,2}$ having a small mass mixing $\mu^2\si_1\si_2$, we can expand the mass rotation matrix to first order in $\mu^2$ and prove a curious relation for propagators $G_{1,2}$ of $\si_{1,2}$ in inflation: $\int_y G_1(x,y)G_2(y,z)=[G_1(x,z)-G_2(x,z)]/(m_1^2-m_2^2)$. This relation is not quite obvious in dS, but is easily seen with Euclidean signature. It's also trivially true in flat spacetime. See, e.g., \cite{Chen:2016hrz} for more discussions.} It is therefore important to develop techniques that do not rely on the full dS isometries. In this work, we will consider bilinear mixings of the following form:
\bge
\label{eq_mixing}
  S\supset \lam\int\di \tau\di^3x\,(-\tau)^{q}\varphi'\si,
\ede
This form covers a wide range of interesting mixings seen in phenomenological studies. The scale-invariant mixing in (\ref{eq_mixing}) corresponds to $q=3$, and the oscillatory coupling in the resonance scenario can be realized by choosing a $q$ with a nonzero imaginary part. 

\paragraph{Theoretical development} 
The reduction of symmetry in the bilinear mixing implies that the computation of graphs involving such mixings is not entirely trivial. Nevertheless, the numerical integration of graphs with mixings was successfully implemented in the early studies \cite{Chen:2009zp}. Another approach is to numerically solve the coupled equations of motion for the two mixed fields \cite{Chen:2015dga}, which also covers the case of strong mixing \cite{An:2017hlx,Iyer:2017qzw}. A similar approach was also adopted in a recent numerical package \cite{Werth:2023pfl,Pinol:2023oux,Werth:2024aui} which can handle tree graphs with mixings.

On the analytical frontier, an analytical expression for the mixed propagator was found in \cite{Chen:2017ryl}, which can speed up numerical integrations and was further explored in \cite{Pimentel:2022fsc}. The full analytical result for the graph with one mixing in (\ref{eq_mixing}) was obtained in \cite{Qin:2023ejc} using an improved bootstrap method. Curiously, \cite{Qin:2023ejc} showed that a change of variables can simplify the result, which was reproduced with a different method \cite{Liu:2024xyi}. Also, the change of variable was further explored in \cite{Aoki:2024uyi} to compute tree graphs with two mixed propagators. 

More generally, as mentioned above, we may want to compute tree graphs with any number of mixings by taking folded limits of the general results in \cite{Liu:2024str}. This is indeed possible, barring a minor issue of cancelling spurious folded divergences among different terms. However, less obvious is that the transcendental weight of the graph is reduced when we take a folded limit. This fact cannot be seen by directly taking the folded limits of results in \cite{Liu:2024str}. A change of variables is essential here, as noticed previously \cite{Qin:2022fbv,Qin:2023ejc,Aoki:2024uyi}. The reduction of transcendental weight not only helps to speed up numerical computation, but also points to potentially new mathematical relations linking hypergeometric functions of different numbers of variables. Thus, it would be interesting to see whether the weight reduction is a general phenomenon for graphs with an arbitrary number of mixings. It would also be interesting to use this fact to find an analytical expression for the good old triple-exchange graph. In this work, we will report our new progress along these directions.

\paragraph{Outline of this work} The rest of the work is structured as follows. First, to set the stage, we review in Sec.\;\ref{sec_single} the 3-point function with single massive exchange through a bilinear mixing. Here, we highlight a known fact that a suitable change of variable simplifies the final result, which serves as an initial motivation for this work. 

In Sec.\;\ref{sec_NewRep}, we introduce our kinematic variables for a general tree graph with arbitrary number of exchanges. The new representation begins with specifying a direction to each internal line, which can be arbitrarily chosen and has nothing to do with vertex energy orders or time orders. Then, we shift all vertex energies $E_i$ to dressed vertex energies $E_i\to \E_i=E_i+\sum \ss_\al K_\al$ where the sum goes over all internal lines attached to $E_i$ and $\ss_\al=+1$ (or $-1$) when the internal line $K_\al$ points against (or towards) $E_i$. Clearly, there are $2^I$ distinct ways to do so for a graph with $I$ internal lines. Then, we use a new dimensionless ratio $u_{\al i}\equiv 2K_\al/\E_i$ in place of the conventional choice $r_{\al i}=K_\al/E_i$. In Sec.\;\ref{subsec_PDE}, we derive the new PDE system with $u$-variables with the result given in (\ref{eq_DEinUvar}) and show that this PDE system features a first-order recursion relation. Then, we completely fix the solution to a generic PDE system (\ref{eq_DEinUvar}) with the result given in (\ref{eq_fullsolutionGprime}), which is a sum of a completely inhomogeneous solution (CIS) and its cuts. We derive the CIS in Sec.\;\ref{subsec_CIS} and its cuts in Sec.\;\ref{subsec_cut}.

Next, in Sec.\;\ref{sec_limits}, we use the new representations to study folded limits and partial-energy limits at a given vertex for an arbitrary tree graph. The folded limit, discussed in Sec.\;\ref{subsec_fold}, is a physical limit at any leaf site that can be reached by taking the colinear limit of all momenta; See Fig.\;\ref{fig_fold}. The folded limit leads to the bilinear mixing that is phenomenologically important. The new representation makes it very simple to reach folded limits with convergent hypergeometric series. Interestingly, the resulting hypergeometric function features a reduction of transcendental weights, as we have observed in the single-exchange case. Then, in Sec.\;\ref{subsec_partialenergy}, we very briefly discuss the partial-energy limit at a vertex. The partial-energy limit is unphysical and is reached by analytically continuing the energies. As is widely discussed, it is a useful way to probe the early-time/high-energy behavior of a subgraph. The new representation again makes it trivial to take the partial-energy limit, and the result is a sum of a factorized singular term and many regular terms, as summarized in (\ref{eq_PEL}).  

In Sec.\;\ref{sec_star}, we consider an important class of tree graphs, formed by $N$ mixed propagators joined at a single vertex, which we call a $N$-pointed degenerate star and denote by $\mathcal{S}_N$, as shown in Fig.\;\ref{sec_star}. Remarkably, the complete result of $\mathcal{S}_N$ has a rather compact form (\ref{eq_DegStarSol}), where the summation over $\ec_i\in\{0,\pm1\}$ encodes the three scaling dimensions of the $i$'th mixed propagator in the late-time limit, while the summation over $m_i\in\mathbb{N}$ keeps track of the corresponding descendent modes. Generally, these hypergeometric series sum to functions no more complicated than Lauricella functions and multivariate Kampé de Fériet (KdF) functions. In simple cases with $N\leq 3$, these expressions gain further simplifications, which we study in Sec.\;\ref{sec_DegStarExamples}. 

In Sec.\;\ref{sec_bispectrum}, we use the result of degenerate stars to study the inflaton bispectrum with single, double, and triple massive exchanges at the tree level, which are interesting processes predicted by many CC models. The single and double exchanges have been well studied in previous works, and here we only discuss their squeezed limits for completeness. On the other hand, we provide a full expression of the triple exchange using trivariate KdF functions and discuss numerical strategies to evaluate these functions. We also provide simple formulae describing the various modes in the squeezed limit. We further show that, due to the distinct mass dependences of these modes, a simultaneous measurement of both the oscillatory signal and the smooth background can be used to differentiate the single, double, and triple exchanges from each other. We conclude the work with further discussions in Sec.\;\ref{sec_conclusion}.

This work makes heavy use of results for generic massive tree graphs obtained in \cite{Liu:2024str}. We summarize a few key definitions and results in App.\;\ref{app_masstree}. Readers unfamiliar with \cite{Liu:2024str} are suggested to at least glance through App.\;\ref{app_masstree} before reading the rest of the paper. In the rest of the appendices, we give an inductive proof of the compact solution {(\ref{eq_DegStarSol}) for arbitrary degenerate star graph $\mathcal{S}_N$ in App.\;\ref{app_proof}, an exact series representation for an arbitrary tree graph in the small partial-energy region at any given site in App.\;\ref{app_PElimit}, a more explicit expression for the 2-pointed degenerate star $\mathcal{S}_2$ in App.\;\ref{app_S2}, and, finally, useful special functions in App.\;\ref{app_spcfun}.

\paragraph{Notations and conventions} Our notations and conventions largely follow previous works in the same direction, and are almost identical to \cite{Liu:2024str}. We reiterate a few special points for readers' convenience. We choose the natural unit throughout the paper and set the inflationary Hubble parameter $H=1$ except in Sec.\;\ref{sec_bispectrum} where $H$ is kept explicit. Summations are often abbreviated into stacked indices, e.g., $E_{ij}\equiv E_i+E_j$, $p_{1\cdots N}\equiv p_1+\cdots+p_N$. When no confusion occurs, we also use abbreviations such as $\bm r\equiv r_{1\cdots N}=r_1+\cdots+r_N$ and $\bm\vartheta\equiv \vartheta_{1\cdots N}=r_1\pd_{r_1}+\cdots+r_N\pd_{r_N}$. A tilted quantity such as $\wt q_j$ means the sum of $q$'s of Site $j$ and all her descendants, while $\wh q_j$ means the sum of $q$'s of descendants of Site $j$ (with Site $j$ excluded). However, \emph{there is a single exception to this rule}: $\wt\nu_j=\sqrt{m_j^2-9/4}$ stands for the mass parameter of Line $j$ and no summing over descendants is involved here.\footnote{This annoying clash of notations is due to a conventional definition of mass parameters $\nu\equiv\sqrt{9/4-m^2}$ and $\wt\nu\equiv\sqrt{m^2-9/4}$. It is convenient to use $\wt\nu$ for principal scalars for which $m>3/2$.} Other commonly used symbols, notations, and definitions are collected in App.\;\ref{app_masstree},

\section{Single Exchange Revisited}
\label{sec_single}

To better understand the folded limit, we review a known example of a single massive exchange, before considering more general situations in subsequent sections. 

The single massive exchange tree graph is the simplest example of massive inflationary correlators and has been extensively studied in recent years. Since the number of external lines associated with a vertex of a massive correlator is usually irrelevant \cite{Liu:2024str}, we can use the 4-point correlator as an example for the nondegenerate single exchange graph, and use the 3-point correlator to represent its degenerate limit, as shown in Fig.\;\ref{fig_single}. We will consider correlators with general twists. (See App.\;\ref{app_masstree} for the definition.) The full analytical results for 4-point and 3-point correlators with general twists were first obtained in \cite{Qin:2022fbv} and \cite{Qin:2023ejc}, respectively. Our treatment here mainly follows \cite{Liu:2024str} and \cite{Qin:2023ejc}. 

\begin{figure}[t]
\centering
\includegraphics[width=0.75\textwidth]{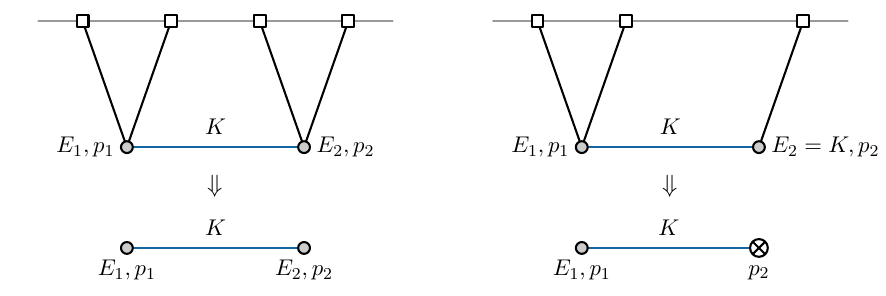} 
\caption{The 4-point (left) and 3-point (right) functions with a single massive exchange. The lower graphs show our convention for drawing inflation correlators: 1) All bulk-to-boundary propagators are removed; 2) A bilinear-mixing vertex is denoted by a large crossed circle.}
\label{fig_single}
\end{figure}

\paragraph{4-point correlator} With the bulk Schwinger-Keldysh diagrammatic rules \cite{Chen:2017ryl}, the 4-point correlator of a single massive exchange is given by the following integral:
\begin{align}
\label{eq_4pt}
  \G_{2}(E_1,E_2,K)=&-E_1^{1+p_1}E_2^{1+p_2}K^3\sum_{\aa,\bb=\pm}\aa\bb\int_{-\infty}^0\di\tau_1\di\tau_2\,(-\tau_1)^{p_1}(-\tau_2)^{p_2}e^{+\ii\aa E_1\tau_1+\ii\bb E_2\tau_2} D_{\aa\bb}(K;\tau_1,\tau_2)\n\\
  =&-\sum_{\aa,\bb=\pm}\aa\bb\int_{-\infty}^0\di z_1\di z_2(-z_1)^{p_1}(-z_2)^{p_2}\wt D_{\aa\bb}(r_1z_1;r_2z_2).
\end{align}
Here we have rescaled the graph as in (\ref{eq_Grescale}) so that $\G_2$ is dimensionless. The SK indices $\aa,\bb$ and the twists $p_{1,2}$ are introduced as usual. The massive scalar propagator $D_{\aa\bb}$ and the dimensionless version $\wt D_{\aa\bb}$ are the same as in \cite{Liu:2024str}. In the second line, we use the dimensionless time $z_i\equiv E_i\tau_i$ and dimensionless energy ratios $r_i\equiv K/E_i$. In \cite{Liu:2024str}, it was shown that $\G_2$ satisfies the following differential equations: 
\begin{align}
  &\Big[\big(\vartheta_{r_i}-\fr{3}{2}\big)^2+\wt\nu^2-r_{i}^2\big(\vartheta_{r_i}+p_i+2\big)\big(\vartheta_{r_i}+p_i+1\big)\Big]\G_2=\FR{r_1^{p_2+4}r_2^{p_1+4}}{r_{12}^{p_{12}+5}}2\cos(\pi p_{12}/2)\Gamma(p_{12}+5),
\end{align}
where $\wt\nu\equiv \sqrt{m^2-9/4}$ is the mass parameter. The solution to these equations is uniquely fixed by imposing the Bunch-Davies initial condition, and can be organized as a sum over the two-site CIS and all its cuts (See App.\;\ref{app_masstree} for a review of these concepts and notations):
\begin{align}
\label{eq_G2result} 
    \G_{2}=\mft{12}+\Big\{\mft{1^\sharp}\Big(\mft{2^\sharp}+\mft{2^\flat}\Big)+(\wt\nu\to-\wt\nu)\Big\}.
\end{align}
The two-site CIS is:
\begin{align} 
\label{eq_mft12}
    \mft{12}=\ &\sum_{\ell,m=0}^{\infty} 
      \FR{(-1)^{\ell}4\cos(\pi p_{12}/2)\Gamma(p_{12}+\ell+2m+5)}{\ell!\big(\fr{p_2+\ell}{2}+\fr{5}{4}\pm\fr{\ii\wt\nu}{2}\big)_{m+1}}\Big(\FR{K}{2E_1}\Big)^{2m+3}\Big(\FR{E_2}{E_1}\Big)^{p_2+\ell+1},
\end{align}
and the cut components are ($i=1,2$):
\begin{align} 
\label{eq_i_sharp}
  \mft{i^\sharp}=&- \FR{2^{3/2+p_i}\cos\big[\fr{\pi(p_i+\ii\wt\nu+3/2)}{2}\big]}{\sin(\pi\ii\wt\nu)}\Big(\FR{K}{E_i}\Big)^{3/2+\ii\wt\nu} {}_{2}\mathcal{F}_1\left[\bgm\fr{p_i}{2}+\fr{5}{4}+\fr{\ii\wt\nu}{2},\fr{p_i}{2}+\fr{7}{4}+\fr{\ii\wt\nu}{2}\\1+\ii\wt\nu\edm \middle| \FR{K^2}{E_i^2} \right], \\
\label{eq_2_flat}
  \mft{2^\flat}=&~\FR{2^{3/2+p_2}\cos\big[\fr{\pi(p_2+\ii\wt\nu+3/2)}{2}\big]}{\sin(\pi\ii\wt\nu)}\Big(\FR{K}{E_2}\Big)^{3/2-\ii\wt\nu} {}_{2}\mathcal{F}_1\left[\bgm\fr{p_2}{2}+\fr{5}{4}-\fr{\ii\wt\nu}{2},\fr{p_2}{2}+\fr{7}{4}-\fr{\ii\wt\nu}{2}\\1-\ii\wt\nu\edm \middle| \FR{K^2}{E_2^2} \right].
\end{align}

As noticed in \cite{Liu:2024str}, the result for this 4-point function has a special feature that all terms in (\ref{eq_G2result}) are hypergeometric functions of weight-2. However, there is a difference in the uncut and cut terms: The two summation variables in  the uncut term $\mft{12}$ are nested in the Gamma factors in the summand, resulting in a two-variable hypergeometric function after the summation, known as the Kampé de Fériet (KdF) function, henceforth KdF function for short. On the contrary, the two summation variables are factorized in the cut terms such as $\mft{1^\sharp}\mft{2^\sharp}$, and thus the results are a product of two single-variable hypergeometric functions ${}_2\mathcal{F}_1$. Interestingly, in the special case of the exchange particle being a conformal scalar with $m^2=2$ ($\wt\nu=\ii/2$), the two-variable hypergeometric function in $\mft{12}$ reduces to a dilogarithmic function $\text{Li}_2$ while the factorized terms reduce to $\log\times \log$ \cite{Fan:2024iek}. Thus, our counting of transcendental weight and the factorization properties remain valid in this limit. 

\paragraph{3-point correlator} Clearly, the 3-point correlator can be obtained from the 4-point correlator (\ref{eq_4pt}) by taking the limit $E_2\to K$, or equivalently, $r_2\to 1$. However, this limit is not completely trivial: The hypergeometric factors ${}_2\mathcal{F}_1$ in both (\ref{eq_i_sharp}) and (\ref{eq_2_flat}) become singular in this limit, but the singular terms cancel out in the combination $\mft{2^\sharp}+\mft{2^\flat}$: 
\begin{align}
  \lim_{K/E_2\to 1}\Big(\mft{2^\sharp}+\mft{2^\flat}\Big)=\FR{-\pi^2\sin\big[\fr\pi2(\fr12+p_2+\ii\wt\nu)\big]\big/[\cos(2\pi p_2)+\cosh(2\pi\wt\nu)]}{2^{1/2+p_2}\Gamma[3+p_2,-\fr32-p_2+\ii\wt\nu,-\fr32-p_2-\ii\wt\nu]}.
\end{align}
Interestingly, the result is not only finite but also contains no hypergeometric series. In other words, the transcendental weight of the combination $\mft{2^\sharp}+\mft{2^\flat}$ drops from 1 to 0 as we take $K/E_2\to 1$. As a result, the cut terms of $\G_2$ (terms in curly brackets in (\ref{eq_G2result})) give rise to a weight-1 function, with the only hypergeometric function coming from $\mft{1^\sharp}$. 

On the other hand, the limit $K/E_2\to 1$ is trivial for the uncut term $\mft{12}$. From (\ref{eq_mft12}), we have:
\begin{align}
\label{eq_mft12folded1}
  \mft{1\mathring{2}}\equiv\lim_{K/E_2\to 1}\mft{12}=\ &\sum_{\ell,m=0}^{\infty} 
      \FR{(-1)^{\ell}\cos(\pi p_{12}/2)\Gamma(p_{12}+\ell+2m+5)}{2^{2m+1}\ell!\big(\fr{p_2+\ell}{2}+\fr{5}{4}\pm\fr{\ii\wt\nu}{2}\big)_{m+1}}\Big(\FR{K}{E_1}\Big)^{2m+p_2+\ell+4} .
\end{align}
Somewhat surprisingly, we have one fewer kinematic variable here ($r_2=K/E_2=1$ fixed), but still two summation layers. At face value, it suggests that the weight of the uncut term is still 2 after sending the limit $K/E_2\to 1$. Were this true, $\G_2$ would no longer have a universal transcendental weight after taking the folded limit. 

However, it was realized in \cite{Qin:2023ejc} that the weight of the uncut term $\mft{12}$ also reduces if we make a change of variable. Using $u_1\equiv 2r_1/(1+r_1)$ instead of $r_1$, it was found that the uncut term can be written as:
\begin{align}
\label{eq_mft12folded2}
  \mft{1\mathring{2}}=\FR{\cos(\pi p_{12}/2)\Gamma(p_{12}+5)u_1^{p_{12}+5}}{2^{4+p_{12}}\big[(\fr52+p_2)^2+\wt\nu^2\big]}\,{}_3\text{F}_2\left[\bgm 1,p_2+3,p_{12}+5\\\fr72+p_2+\ii\wt\nu,\fr72+p_2-\ii\wt\nu\edm \middle|u_1\right],
\end{align}
which is weight-1. Certainly, (\ref{eq_mft12folded1}) and (\ref{eq_mft12folded2}) are identical. Still, the reduction of summation layers by changing variables appears a bit mysterious, at least to us who are not super familiar with various functional identities linking multivariate hypergeometric functions of different weights. Thus, it would be interesting to see if the reduction of transcendental weight still holds for more complicated tree graphs with more folded limits taken. Also, it remains unclear how to make the change of variables for more complicated graphs. We will explore these questions in the following sections. 

\section{New Representations for General Trees}
\label{sec_NewRep}

In this section, we return to the most general tree-level graphs reviewed in App.\;\ref{app_masstree}. With the insights from the previous section, we introduce a new set of variables to represent these graphs. In accordance with these new variables, we derive a new set of differential equations and solve them in a complete form, very similar to the procedure in \cite{Liu:2024str}. These new results will make it trivial to take folded limits at arbitrary ``leaf'' vertices, which is all we need to compute graphs with bilinear mixings. 

\subsection{New variables and differential equations}
\label{subsec_PDE}

For a general tree graph $\wh{\mathcal{G}}(\bm k_1,\cdots,\bm k_N)$ of $V$ vertices and $I=V-1$ internal lines, it proves convenient to rescale it into a dimensionless graph, of which the kinematics can be fully specified by $V$ vertex energies $E_i$ $(i=1,\cdots,V)$  and $I$ line energies $K_\al$ ($\al=1,\cdots,I$). Furthermore, these $2I+1$ energies must enter the dimensionless graph in ratios, and the overall scale must drop out. As a result, the dimensionless graph is specified by $2I$ independent energy ratios.

Conventionally, the $2I$ energy ratios are chosen to be $r_{\al i}\equiv K_\al/E_i$ and $r_{\al i}\equiv K_\al/E_j$, where $\al=1,\cdots I$, and $(i,j)$ denotes the two vertices connected by Line $\al$. It turns out convenient to derive and solve differential equations for the dimensionless graph with these $r$ variables, as reviewed in App.\;\ref{app_masstree}. 

However, other choices of energy ratios are possible, and we now consider some of these choices, which further simplify the PDE system. It turns out that we have plenty of freedom in making this choice. To parameterize this freedom, we give every internal line a direction, which can be arbitrarily chosen. Then, we define a \emph{sign label} $\mathsf{s}_{\al i}=\pm 1$ for every internal line $\al$ and one of its endpoint $i$: 
\begin{align}
\label{eq_lineDirection}
  \mathsf{s}_{\al i}=\begin{cases}
  +1, &\text{if Line $\al$ flows out of Vertex $i$;}\\
  -1, &\text{if Line $\al$ flows into Vertex $i$.}
  \end{cases}
\end{align}
Then, for every vertex, say Vertex $i$, we define a dressed vertex energy $\E_i$ as:
\bge
\label{eq_Ehat}
  \E_i\equiv E_i+\sum_{\al\in\mathcal{N}(i)}\mathsf{s}_{\al i}K_\al,
\ede 
where $\mathcal{N}(i)$ denotes the set of neighbor lines of Vertex $i$. That is, the sum goes over all lines connected to Vertex $i$. 

We emphasize that the direction of the line for introducing the sign label in (\ref{eq_lineDirection}) is completely arbitrary and has nothing to do with either the time ordering or energy ordering. To make this point clearer, we illustrate it with an example of 5-site MFT $\mft{123(4)(5)}'$,\footnote{Here we add a prime $\mft{\cdots}'$ in anticipation of a change of overall normalization of the MFT adapted to our new variables. See (\ref{eq_MFTrescaling}).} which can be represented by the following graph:
\bge
  \mft{123(4)(5)}'=\parbox{65mm}{ 
\includegraphics[width=65mm]{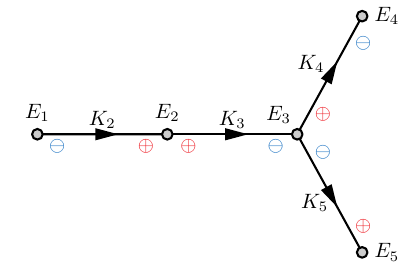}}.
\ede
Here, we use the arrows on the lines to show the family tree structure of the graph (as reviewed in App.\;\ref{app_masstree}), which is induced by asking Site 1 to be the maximal energy site. On the other hand, we use the label {\color{Red}$\oplus$} to denote $\ss_{\al i}=+1$ and {\color{RoyalBlue}$\ominus$} to denote $\ss_{\al i}=-1$, which is completely the same notation as in \cite{Fan:2024iek}. Explicitly, this assignment of sign labels gives:
\begin{align}
\label{eq_dressedE5site}
  &\E_1=E_1-K_2,
  &&\E_2=E_2+K_2+K_3,
  &&\E_3=E_3-K_3+K_4-K_5,\n\\
  &\E_4=E_4-K_4,
  &&\E_5=E_5+K_5.
\end{align}
Clearly, the sign labels have nothing to do with the family tree structure.\footnote{For readers familiar with conformal scalar amplitudes \cite{Arkani-Hamed:2023kig,Arkani-Hamed:2023bsv,Fan:2024iek}, the dressed energies in (\ref{eq_dressedE5site}) may look similar to the variables appearing in the conformal scalar amplitudes, which are also known as ``letters'' in that context. Despite the same appearance, our sign labels differ structurally from letters. Specifically, the ``letters'' for conformal amplitudes are determined by the structure of the wavefunction or correlator itself, and the expression for a single correlator is a sum of many family trees with many different assignments of letters. However, in the current case, a single assignment of sign labels is sufficient to express the whole correlator, as will be shown below. }

Now, with the dressed vertex energies $\E_i$ defined, we can introduce the $2I$ independent energy ratios as:
\bge
  u_{\al i}\equiv\FR{2K_\al}{\E_i}.
\ede
For convenience, we also use:
\begin{align}
  &\bm u_i\equiv \sum_{\al\in\mathcal{N}(i)} \mathsf{s}_{\al i} u_{\al i},
  &&\bm r_i\equiv \sum_{\al\in\mathcal{N}(i)} \mathsf{s}_{\al i} r_{\al i}
\end{align}
We define the corresponding Euler operators to be:
\begin{align}
  &\eta_{\al i}\equiv u_{\al i}\FR{\pd}{\pd u_{\al i}},
  &&\bm \eta_{i}\equiv \sum_{\al\in\mathcal{N}(i)} \eta_{\al i}.
\end{align}
The relation between $r$ variables and $u$ variables is simple:
\begin{align}
\label{eq_ur}
  &u_{\al i}=\FR{2r_{\al i}}{1+\bm r_i},
  &&r_{\al i}=\FR{u_{\al i}}{2-\bm u_i},
  &&\bm u_i=\FR{2\bm r_i}{1+\bm r_i},
  &&\bm r_i=\FR{\bm u_i}{2-\bm u_i}.
\end{align}
We emphasize that the definition of $u$ variables depends on the line directions. Since every internal line can have either direction, there are $2^I$ distinct definitions of $u$ variables. 

With $u$ variables in use, it is natural to work with a new dimensionless graph $\G'$, defined from the original graph $\wh\G$ via (compare with (\ref{eq_Grescale}))  
\bge
\label{eq_Grescaleu}
  \wh{\mathcal{G}}(\bm k_1,\cdots,\bm k_N)=\bigg(\prod_{n=1}^N\FR{-\tau_f}{2k_n}\bigg)\bigg(\prod_{i=1}^V\FR{1}{\E_{i}^{1+p_i}}\bigg)\bigg(\prod_{\al=1}^I \FR{1}{(2K_\al)^3}\bigg) \mathcal{G}'(\{ u \}).
\ede
Also, we define dimensionless time variables $y_i\equiv \E_i\tau_i$ for each vertex. Then, the Schwinger-Keldysh time integral for $\G'$ can be written as:
\begin{align}
\label{eq_Gprime}
  \G'(\{u\})=&\sum_{\aa_1,\cdots,\aa_V=\pm}\int_{-\infty}^0\prod_{i=1}^V\Big[\di y_i\, \ii\aa_i (-y_i)^{p_i}e^{\ii \aa_i (1-\bm u_i/2)y_i}\Big]\prod_{\al=1}^I \wt{D}_{\aa_{i}\aa_{j}}^{(\wt\nu_\al)}(\fr12 u_{\al i}y_{i},\fr12 u_{\al j}y_{j}).
\end{align}
The main result of this subsection is that the rescaled graph $\G'$ in (\ref{eq_Gprime}) satisfies the following set of $2I$ differential equations:
\begin{keyeqn}
\begin{align}
\label{eq_DEinUvar}
  \bigg[\Big(\eta_{\al i}-\FR32\Big)^2+\wt\nu_\al^2-\mathsf{s}_{\al i}u_{\al i}(\eta_{\al i}-1)(\bm\eta_i+p_i+1)\bigg]\G'=\FR{u_{\al i}^{p_j+4}u_{\al j}^{p_i+4}}{(u_{\al i}+u_{\al j})^{p_{ij}+5}} \mathsf{C}_\al\big[\G'\big].
\end{align}
\end{keyeqn}
On the right-hand side, we use the \emph{contraction} of a graph, introduced in \cite{Liu:2024str}. It means to remove Line $\al$ in $\G_\al'$, to pinch its two endpoints $i$ and $j$, and to assign the pinched site a new (dressed) vertex energy $\E_{ij}$ and a twist $p_{ij}+4$. As shown in Fig.\;\ref{fig_PDE}, this equation has exactly the same structure as (\ref{eq_generalDE}) derived in \cite{Liu:2024str}, the only difference being the change of variables. 
\begin{figure}[t]
\centering
\includegraphics[width=0.7\textwidth]{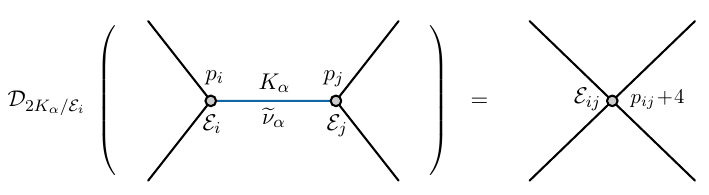} 
\caption{An illustration of the differential equation (\ref{eq_DEinUvar}).}
\label{fig_PDE}
\end{figure}

\paragraph{Deriving the differential equations} The derivation of (\ref{eq_DEinUvar}) closely parallels with the PDEs in $r$-variable in (\ref{eq_generalDE}). Thus, we will be brief. 

In the integral expression (\ref{eq_Gprime}), we pick up an arbitrary internal line, say Line $\al$, which connects Vertex $i$ and Vertex $j$. We then insert the Klein-Gordon operator in front of the propagator for Line $\al$ from the side of $i$, and get
\begin{align}
\label{eq_eomInU}
   \Big(\vartheta_{y_i}^2-3\vartheta_{y_i}+\FR{u_{\al i}^2y_i^2}4+m_\al^2\Big)\wt D_{\aa_i\aa_j}^{(\wt\nu_\al)}\Big(\FR{u_{\al i}y_i}2,\FR{u_{\al j}y_j}2\Big) 
   = -\ii\aa_i(u_{\al i}y_i u_{\al j}y_j)^2\de(u_{\al i}y_i-u_{\al j}y_j)\de_{\aa_i\aa_j},
\end{align}
where $\vartheta_{y_i}\equiv y_i\pd/\pd y_i$. To turn this equation into an equation for the graph, on the left-hand side, we want to pull the differential operator out of the integral, and on the right-hand side, we can perform a layer of integral with the help of $\de$ functions. It can be proved that pulling the operator out of the integral is equivalent to the following substitutions:
\begin{align}
\label{eq_thetaToeta}
  &\vartheta_{y_i}\to \eta_{\al i}- \FR{1}{2}\ss_{\al i}u_{\al i}(p_i+1+\bm\eta_i), 
  &&\vartheta_{y_i}^2+\FR14u_{\al i}^2y_i^2\to \eta_{\al i}^2 - \ss_{\al i}u_{\al i}\Big(\eta_{\al i}+\FR12\Big) (p_i+1+\bm\eta_i).
\end{align}
So, using $m_\al^2=\wt\nu_\al^2+9/4$, the left hand side of the equation becomes
\begin{align}
  \bigg[\Big(\eta_{\al i}-\FR32\Big)^2+\wt\nu_\al^2-\ss_{\al i}u_{\al i}(\eta_{\al i}-1)(\bm\eta_i+p_i+1)\bigg]\G'. 
\end{align}
\begin{boxedtext}
\noindent\textbf{Proof of (\ref{eq_thetaToeta})~~} For convenience let us write $X\equiv (-y_i)^{p_i}e^{\ii\aa_i(1-\bm u_i/2)y_i}$ and $\int_{y_i}\cdots \equiv\int_{-\infty}^0\di y_i\cdots $. Then, a direct computation shows  
\bge
\label{eq_intthetaD}
  \int_{y_i} X (\vartheta_{y_i}\wt D_\al)\cdots = \int_{y_i}\big(\eta_{\al i}+\FR12 \ii\aa_i\ss_{\al i}u_{\al i}y_i\big)X  \wt D_\al\cdots,
\ede
where $\cdots$ denotes other propagators ending at Vertex $i$. 
Next, consider the following integral:
\begin{align}
  0=\int_{y_i}\pd_{y_i}\bigg[y_i X\prod_{\be\in\mathcal{N}(i)}\wt D_\be\bigg].
\end{align}
By acting $\partial_{y_i}$ on each $y_i$-dependent factor, we get:
\begin{align}
\label{eq_intyXProdD}
  &\int_{y_i} y_iX\prod \wt D_\be 
  = -\FR{1}{\ii\aa_i(1-\bm u_i/2)}\int_{y_i}X\big(p_i+1+\bm\eta_i\big)\prod \wt D_\be\n\\
  =&-\FR{1}{\ii\aa_i(1-\bm u_i/2)}\int_{y_i}\big(p_i+1+\bm\eta_i+\FR12\ii\aa_i\bm u_i y_i\big)X\prod \wt D_\be,
\end{align}
where we have used $\vartheta_{y_i}\prod\wt D_\beta=\bm\eta_i\prod \wt D_\beta$ in the first equality. Then, comparing the first and last expressions of (\ref{eq_intyXProdD}), we have
\begin{align}
  \int_{y_i} y_i X\prod \wt D_\al  
  = -\FR{1}{\ii\aa_i }\big(p_i+1+\bm\eta_i \big)\int_{y_i}X\prod \wt D_\al.
\end{align}
Using this equality to replace the last term on the right hand side of (\ref{eq_intthetaD}), we get: 
\begin{align}
   \int_{y_i}X (\vartheta_{y_i}\wt D_\al)\cdots  
  = \Big[\eta_{\al i}-\FR12 \ss_{\al i}u_{\al i}(p_i+1+\bm\eta_i)\Big]\int_{y_i} X \wt D_\al\cdots
\end{align}
This proves the first substitution in (\ref{eq_thetaToeta}).

Similarly, by distributing the Euler operator $\eta_{\al i}^2$ over all factors in the integral $\int_{y_i} \eta_{\al i}^2\big[X\wt D_\al\cdots \big]$, we can prove the second part of (\ref{eq_thetaToeta}):
\begin{align}
   &\int_{y_i}X\Big[\big(\vartheta_{y_i}^2+\FR14 u_{\al i}^2y_i^2\big)\wt D_\al\Big]\cdots 
  = \int_{y_i}\Big[\eta_{\al i}^2+ \ii\aa_i\ss_{\al i} u_{\al i}\Big(\eta_{\al i}+\FR12\Big)y_i\Big]X\wt D_\al\cdots\n\\
  =&~\Big[\eta_{\al i}^2-\ss_{\al i}u_{\al i}\Big(\eta_{\al i}+\FR12\Big) (p_i+1+\bm\eta_i)\Big]\int_{y_i}X \wt D_\al\cdots.
\end{align}

\end{boxedtext}
Next, we consider the right hand side of (\ref{eq_eomInU}) and restore the $y_i$ and $y_i$ integrals: 
\label{eq_RHSint}
\begin{align}
  &\sum_{\aa_i,\aa_j=\pm}(-\aa_i\aa_j)\int\di y_i\di y_j\,(-y_i)^{p_i}(-y_j)^{p_j}e^{\ii\aa_i(1-\bm u_i/2)y_i+\ii\aa_j(1-\bm u_j/2)y_j}\n\\
  &\times(-\ii\aa_i)(u_{\al i}y_i u_{\al j}y_j)^2\de(u_{\al i}y_i-u_{\al j}y_j)\de_{\aa_i\aa_j}\cdots\n\\
  =&\sum_{\aa_i=\pm} \ii\aa_i \FR{u_{\al i}^{p_j+4}}{u_{\al j}^{p_j+1}} \int\di y_i\,(-y_i)^{p_{ij}+4}e^{\ii\aa_i(1-\bm u_i/2)y_i+\ii\aa_i(1-\bm u_j/2)(u_{\al i}/u_{\al j})y_i}\cdots
\end{align}
The exponent in the exponential factor above can be massaged into the following form.
\begin{align}
\label{eq_pinchedexponent}
  &1-\fr12\bm u_i+(1-\fr12\bm u_j)\FR{u_{\al i}}{u_{\al j}}=\FR{u_{\al i}+u_{\al j}}{u_{\al j}}(1-\fr12\bm u_{ij}),
  &&\bm u_{ij}\equiv\FR{u_{\al i}\bm u_j+u_{\al j}\bm u_i}{u_{\al i}+u_{\al j}}.
\end{align}
The meaning of these factors is more transparent in original energy variables.  For instance,
\bge
\label{eq_UinE}
  \bm u_{ij}=2\bigg(\sum_{\be\in\mathcal{N}(i)}\ss_{\be i}K_\be+\sum_{\be\in\mathcal{N}(j)}\ss_{\be j}K_\be \bigg)\bigg/ \E_{ij}
\ede
is the $u$ variable defined for a new vertex obtained by pinching Vertices $i$ and $j$ together, with their dressed vertex energies summed. Importantly, the line energy $K_\al$ of the pinched line drops out in the above expression since $s_{\al i}+s_{\al j}=0$.  

Back to the exponent in (\ref{eq_pinchedexponent}), we see that it suggests the change of variable
\bge
y_i\to \FR{u_{\al j}}{u_{\al i}+u_{\al j}} y_i = \FR{\E_i}{\E_{ij}}y_i. 
\ede
With the rescaled $y_i$, the integral (\ref{eq_RHSint}) is further reduced to:
\begin{align}
  &\FR{u_{\al i}^{p_j+4}u_{\al j}^{p_i+4}}{(u_{\al i}+u_{\al j})^{p_{ij}+5}}\sum_{\aa_i}\ii\aa_i\int\di y_i\,(-y_i)^{p_{ij}+4}e^{\ii\aa_i(2-\bm u_{ij})y_i}\cdots .
\end{align}
The upshot is that all $y_i$-dependence in the original integral is replaced by $(\E_i/\E_{ij}) y_i$, and all $y_j$ replaced by $(\E_j/\E_{ij})y_j$. So, we get a new graph with Vertices $i$ and $j$ pinched together, with a new dressed vertex energy $\E_{ij}$ and a new twist $p_{ij}+4$, which is nothing but the contraction $\mathsf{C}_\al[\G']$. This completes the derivation of (\ref{eq_DEinUvar}). 

To complete this subsection, we comment on our definition of dressed vertex energy (\ref{eq_Ehat}). One may wonder if $\ss_{\al i}$ can be arbitrary real numbers instead of just being $\pm 1$. This question can be easily answered by examining the steps leading to the differential equations (\ref{eq_DEinUvar}). First, the disappearance of line energy $K_\al$ in the contracted graph $\mathsf{C}_\al[\G']$ requires $s_{\al i}+s_{\al j}=0$; See (\ref{eq_UinE}). Second, when we go from the equation for the propagator (\ref{eq_eomInU}) to the equation for the correlator (\ref{eq_DEinUvar}), there is a subtle simplification. To explain this point, consider the Klein-Gordon operator in the original equation (\ref{eq_eomInU}). If we count the dimension of each term in the differential operator under the rescaling $y_{i}\to \lam y_i$, the Euler operator $\vartheta_{y_i}$ and the mass term $m_\al^2$ are dimensionless, while the term $u_{\al i}^2y_i^2/4$ is dimension 2. Thus, if we use a series ansatz $\sim \sum c_n y^n$ to solve this equation, the Klein-Gordon operator will lead to a second-order recursion relation for $c_n$. In contrast, in the final correlator equation (\ref{eq_DEinUvar}), all terms in the differential operator have $u$-dimension either 0 or 1, implying that the corresponding recursion is of first order. This reduction of recursion order is originated from the second substitution in  (\ref{eq_thetaToeta}). The above derivations show that the resulting operator is of first order only when $\ss_{\al i}^2=1$. It is for this reason that we choose to define dressed energies as in (\ref{eq_Ehat}). Other definitions with more general $\ss_{\al i}$ may also lead to valid differential equations for correlators, although we expect them to be more complicated.

\subsection{Completely inhomogeneous solution as massive family trees}
\label{subsec_CIS}

In the current and next sections, we solve the equation (\ref{eq_DEinUvar}) to get the full expression for the graph $\G_V'$. Again, the procedure is very similar to \cite{Liu:2024str}. The solution is the sum of a completely inhomogeneous solution (CIS) and appropriate homogeneous solutions. In this subsection, we derive the CIS in terms of MFT functions; In the next subsection, we derive the homogeneous solutions as cuts of the CIS. 

\begin{figure}[t]
\centering
\includegraphics[width=0.33\textwidth]{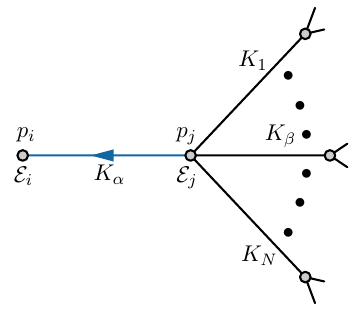} 
\caption{An illustration of the differential equation (\ref{eq_DEinUvar}).}
\label{fig_recursion}
\end{figure}
\paragraph{Recursion formulae}
The CIS is constructed by recursively adding new lines to an existing graph. This is achieved by a recursive formula linking the CISs of two graphs before and after the new line is added. As shown in Fig.\;\ref{fig_recursion}, we consider an arbitrary tree graph $\G_V'$. We want to add a new line to Vertex $j$, which we call Line $\al$, whose other endpoint is called $i$. Then, the new graph $\G_{V+1}'$ satisfies the following equation:
\begin{align}
  \label{eq_ujequation_1}
  \Big[(\eta_{{\alpha j}}-\fr32)^2+\wt\nu_\alpha^2-\ss_{\alpha j}u_{\alpha j}(\eta_{\alpha j}-1)(\bm{\eta}_{j}+p_j+1)\Big]{\mathcal{G}}_{V+1}' =\FR{u_{\alpha i}^{p_j+4}u_{\alpha j}^{p_i+4}}{ (u_{\al i}+u_{\al j})^{p_{ij}+5}}{\mathcal{G}}_{V}'\Big|_{\subalign{&E_j\to E_{ij}\\&p_j\to p_{ij}+4}}.
\end{align}
Suppose the source term has the following series solution:
\begin{align}
  \label{eq_sourceterm_0}
  {\mathcal{G}}_{V}'\Big|_{p_j\to p_{ij}+4}
   = \sum_{\{n\}} c_{\{n\}} \prod_{\be\in\mathcal{N}(j)} u_{\be j}^{q_\be} \times \cdots ,
\end{align}
Here and below, we use the abbreviation $\sum\limits_{\{n\}}=\sum\limits_{n_1=0}^\infty \sum\limits_{n_2=0}^\infty\cdots$ to collect all summation variables. Also, we only show $u$ variables containing Vertex $j$ since they are the only relevant variables when doing the recursion. The exponents $q_\be$ can depend on summation variables $\{n\}$ and other parameters, so is the coefficient $c_{\{n\}}$.

The energy shift $E_j\to E_{ij}$ required in (\ref{eq_ujequation_1}) is effected by the change of variable $u_{\be j}\to u_{\be j}u_{\al i}/(u_{\al i}+u_{\al j})$. Then, we work with the region of $u_{\al j}<u_{\al i}$ and express the source as a series of small $u_{\al j}$. 
\begin{align}
  \label{eq_sourceterm}
  \FR{u_{\alpha i}^{p_j+4}u_{\alpha j}^{p_i+4}}{ (u_{\al i}+u_{\al j})^{p_{ij}+5}}{\mathcal{G}}_{V}'\Big|_{\subalign{&E_j\to E_{ij}\\&p_j\to p_{ij}+4}} 
  = &\sum_{\ell,\{n\}} \frac{(-1)^\ell (\ell+1)_{q_{1\cdots N}+p_{ij}+4}}{ \Gamma(q_{1\cdots N}+p_{ij}+5)}c_{\{n\}}\n\\
  &\times u_{\alpha j}^3 \Big(\frac{u_{\alpha j}}{u_{\alpha i}}\Big)^{\ell+p_i+1}  \prod_{\be\in\mathcal{N}(j)} u_{\be j}^{q_\be} \times \cdots,
\end{align}
The form of this source motivates us to consider the following ansatz for the inhomogeneous solution of $\G_{V+1}'$:
\begin{align}
\label{eq_ougoingAnsatz}
  \mathop{\text{Inh}}_{K_\alpha} \big[{\mathcal{G}}_{V+1}'\big] = \sum_{\ell,m,\{n\}} d_{\ell m\{n\}} u_{\alpha j}^{m+3}\Big(\FR{u_{\alpha j}}{u_{\alpha i}}\Big)^{\ell+p_i+1}\prod_{\substack{\be\in\mathcal{N}(j)\\ \be\neq \al}} u_{\be j}^{q_\be} \times \cdots,
\end{align}
Putting (\ref{eq_ougoingAnsatz}) and (\ref{eq_sourceterm}) into both sides of the equation (\ref{eq_ujequation_1}), and requiring all powers of $u$ to match, we get the coefficients of the series solution for $\G_{V+1}'$ as simple functions of the coefficients of the series solution for the contracted graph:
\begin{align}
\label{eq_recursion}
  d_{\ell m\{n\}} = \FR{(-1)^{\ell}\ss_{\alpha j}^m (\ell+p_i+3)_m (q_{1\cdots N}+p_{ij}+5)_{\ell+m}}{\ell!(\ell+p_i+\fr52\pm \ii \wt\nu_\alpha)_{m+1}}c_{\{n\}}.
\end{align}
Since all expressions here are expanded around small $u_{\al j}$, the recursion relation (\ref{eq_recursion}) works when $\E_j>\E_i$. As in \cite{Liu:2024str}, we use the concept of ``energy flow'' defined for every internal line, which flows from the site of larger energy to the smaller. In this terminology, we can say that (\ref{eq_recursion}) allows us to add an outgoing line to any existing tree graph. We stress that, \emph{a priori}, the energy flow has nothing to do with the sign labels $\ss_{\al i}$ that were introduced to define $u$ variables.

Likewise, we can build an ``ingoing'' recursion relation which works when $\E_j<\E_i$. We can use it to add an ingoing new line to any existing tree graph. However, as proved in \cite{Liu:2024str}, adding an ingoing line yields the same result for the CIS as adding an outgoing line. Therefore, we will neglect the discussion of ingoing recursion, and consider only outgoing cases. The resulting solution will automatically apply to graphs with ingoing directions.  

\paragraph{Massive family tree} As indicated above, we can find the CIS of $\G_V'$ in the following way. We first choose a site of maximal (dressed) vertex energy, say $\E_1$. Then, the choice of the maximal energy site automatically generates a unique energy flow for the entire tree graph. Then, starting from the maximal energy site $\E_1$, we can use the outgoing recursion relation (\ref{eq_recursion}) to recursively add new lines, which are always outgoing, until we reach the full graph $\G_V'$. The starting point of this recursive construction is a one-site graph:
\bge
\label{eq_G1prime}
  \G_1'=\sum_{\aa_1=\pm}\ii\aa_1\int_{-\infty}^0\di y_1\,(-y_1)^{p_1}e^{\ii\aa_1 y_1}=2\cos(\pi p_1/2)\Gamma(p_1+1).
\ede
At this point, it is useful to use the notation for MFT $\mft{\cdots}$ introduced in \cite{Liu:2024str}. See App.\;\ref{app_masstree} for an introduction. Due to the different overall factors involved in defining dimensionless graphs $\G_V$  (for $r$ variables) and $\G_V'$ (for $u$ variables), we rescale the MFT accordingly, and also add a prime like $\mft{\cdots}'$ for distinction. Explicitly, the relation between the two versions is:
\begin{align}
\label{eq_MFTrescaling}
  \mft{\mathscr{P}(1\cdots V)}' = 2^{3I}\prod_{i=1}^V \Big( \FR{\E_i}{E_i} \Big)^{1+p_i} \mft{\mathscr{P}(1\cdots V)}.
\end{align}
Here $\mathscr{P}$ denotes a particular partial order of the family tree. 
For the 1-site graph, the two versions happen to be the same, $\mft{1}'=\mft{1}$. Then, we can repeat the exercise in \cite{Liu:2024str} and construct CIS for tree graphs. This time, it is of no surprise that we also get a closed-form CIS for any $V$-site tree graphs $\G_V'$ in $u$ variables. To write down this expression, we choose the partial order induced by the maximal energy site $\E_1$, denoted as $\mathscr{P}(\wh 1\cdots V)$. With this partial order, we can identify the label $\al$ of an internal line with the label $i$ of the vertex to which the line points (i.e., the daughter site). In particular, every sign label can all be written as $\ss_i=\ss_{ii}$, which means the sign label of Line $i$ with respect to its \emph{daughter site}, and it is automatically understood that the sign label of Site $i$ with respect to its mother site is $-\ss_i$. Then, for each but the maximal-energy site, we assign two summation labels $\ell_i,m_i\in\mathbb{N}$ ($i=2,\cdots, V$), together with a \emph{family parameter} $q_i$ which encodes the family-tree structure $\mathscr{P}$:
\bge
   q_i\equiv \wh{\ell}_i+\wh{m}_i+\wh{p}_i+4N_i,
\ede
where a hatted parameter, say $\wh\ell_i$, means the sum of corresponding parameters over all descendant sites of $i$. With this notational preparation, we can now give the expression for the CIS of $\G_V'$ as:
\begin{keyeqn}
\begin{align}
\label{eq_dressedCIS}
    \text{CIS}\,\big[\G_V'\big] = &~\mft{\mathscr{P}(\wh 1\cdots V)}'= 2\cos(\pi p_{1\cdots V}/2)  \sum_{\{\ell,m\}}\Gamma(p_1+q_1+1) \n\\ &\times \prod_{i=2}^V \frac{\ss_i^{m_i} (-1)^{\ell_i}(\ell_i+q_i+p_i+3)_{m_i}}{\ell_i!(\ell_i+q_i+p_i+\fr52\pm\ii \wt{\nu}_i)_{m_i+1}}\Big(\frac{2K_i}{\E_1}\Big)^{m_i+3}\Big(\frac{\E_i}{\E_1}\Big)^{\ell_i+p_i+1}.
\end{align}
\end{keyeqn}
As in previous work \cite{Liu:2024str}, this series can be viewed as the definition of a $2(V-1)$-variate hypergeometric function in the sense of Horn. The series is well convergent at least when the maximal energy $\E_1$ is much larger than all other energies, while the relative sizes of other energies are not important. Beyond the region of convergence, the function is defined by analytic continuation. Importantly, the choice of the maximal energy site is arbitrary. In a region where some other $\E_i$ instead of $\E_1$ is maximal, we can write down the corresponding CIS expanded in large $\E_i$. The two expressions are analytical continuations of each other up to homogeneous solutions to be determined below. Thus, the many ways ($V$ ways for $\G_V'$) of choosing the maximal energy site provide many transformation-of-variable formulae which partially achieve the analytical continuation of the original series (\ref{eq_dressedCIS}) beyond its convergence region.

Curiously, (\ref{eq_dressedCIS}) has a similar yet distinct form with the previously obtained MFT using $r$ variables in (\ref{eq_MFT}). In both cases, a family tree is written as $2I$-variate hypergeometric series, but the variables are different, and are related by (the square of) linear fractional transformation as in (\ref{eq_ur}). Thus, we immediately get yet another set of infinitely many transformation-of-variable formulae for these hypergeometric functions through (\ref{eq_MFTrescaling}), which are essentially multivariate generalizations of known quadratic transformations of Gauss hypergeometric function $_{2}\text{F}_1$ \cite{nist:dlmf}. 

\subsection{Homogeneous solutions as cuts of family trees }
\label{subsec_cut}
Next, we consider homogeneous solutions. As in \cite{Liu:2024str}, we mean by a homogeneous solution the piece in the full graph that is annihilated by at least one of $2I$ differential operators in (\ref{eq_DEinUvar}). Given Line $\al$ connecting Vertices $i$ and $j$, it is always true that, if a homogeneous solution solves the sourceless differential equation for $u_{\al i}$, it also does for $u_{\al j}$. In this case, we call Line $\al$ is \emph{cut}, because the corresponding time integrals over $y_i$ and $y_j$ are fully factorized. In the \emph{signal region} where all vertex energies are larger than all line energies, we have a one-to-one relation between the solution being homogeneous/inhomogeneous and the line being cut/uncut. So the homogeneous solutions can be specified by the cut lines, and thus there are a total of $2^I-1$ distinct homogeneous solutions. 

\paragraph{Single cut}
Now we briefly describe how to get a single-cut solution from the CIS, i.e., a piece annihilated by nothing but $u_{\al i}$ and $u_{\al j}$ operators. As always, the homogeneous solutions are fixed by appropriate boundary conditions, which can be conveniently imposed from various squeezed limits. At these limits, the time integrals become analytically computable, so let us turn to these time integrals. 

The cut over Line $\al$ produces two subgraphs, one containing Vertex $i$ and the other containing Vertex $j$. With this bipartition, the time integrals over $y_i$ and $y_j$ can be written as:
\bge 
\label{eq_yiyjint}
  \G'=\sum_{\aa_i,\aa_j=\pm}\int_{-\infty}^0\prod_{n=i,j}\Big[\di y_n\,\ii\aa_n(-y_n)^{p_n}e^{\ii\aa_n(1-\bm u_n/2) y_n}\Big]\wt D_{\aa_i\aa_j}^{(\wt\nu_\al)}\big(\fr12 u_{\al i}y_i,\fr12 u_{\al j}y_j\big) \mathcal{I}_{\aa_i}^\text{(L)}\mathcal{I}_{\aa_j}^\text{(R)},
\ede
where $\mathcal{I}^{(L)}_{\aa_i}$ and $\mathcal{I}^{(R)}_{\aa_j}$ collect remaining time integrals of left and right subgraphs, respectively. Without loss of generality, we assume that the maximal energy site of the whole graph (Site 1 with energy $E_1$) belongs to the left subgraph after the cut, so that the (locally) maximal energy of the right subgraph is smaller than that of the left. Anticipating that the integral will be factorized by the cut, we rearrange the exponential factors in the integrand as:
\begin{align}
\label{eq_useparation}
  & \ii(1-\fr12\bm u_n)= \ii(1-\fr12\wh{\bm u}_n )-\fr12 \ii\,\ss_{\al n}u_{\al n},
  &&\wh{\bm u}_n\equiv \bm u_n-\ss_{\al n}u_{\al n}.~~~(n=i,j)
\end{align}
Since the locally maximal energy is larger on the left subgraph, we apply the cutting rule $\wt{D}_{\pm\pm}^{(\wt\nu_\al)}\To \wt{D}_{\mp\pm}^{(\wt\nu_\al)}$ to get the single cut of the graph \cite{Liu:2024str}:
\begin{align}
    \label{eq_singlecut}
    \mathop{\text{Cut}}_{K_\al}\big[\G'\big] = &- \int_{-\infty}^0
   \di y_i\di y_j\,(-y_i)^{p_i}(-y_j)^{p_j} \Big[e^{\ii(1-\wh{\bm{u}}_i/2)y_i} \mathcal{I}^{\text{(L)}}_+   e^{-\ii \ss_{ i}u_{i}y_i/2}-e^{-\ii(1-\wh{\bm{u}}_i/2)y_i} \mathcal{I}^{\text{(L)}}_-  e^{\ii \ss_{ i}u_{i}y_i/2}\Big]  \n\\
    &\times\mathcal{I}^{\text{(R)}}_+ e^{\ii(1-\wh{\bm{u}}_j/2)y_j}  e^{-\ii \ss_{ j}u_{\alpha j}y_j/2}\wt{D}^{(\wt{\nu}_\al)}_{-+}\big(\fr12 u_{ i}y_i,\fr12 u_{j}y_j\big)+\text{c.c..}
\end{align}
Notably, this asymmetric cutting rule endows the cut with a direction that flows from the left subgraph to the right, i.e., from the larger locally maximal energy to the smaller. Then, the above integrals factorize into two separated integrals for $y_i$ and $y_j$, since the opposite-sign propagators $\wt D_{-+}^{(\wt\nu_\al)}$ are factorized:
\begin{align}
  \wt{D}^{(\wt{\nu})}_{-+}\big(\fr12 u_{i}y_i,\fr12 u_{j}y_j\big) = \sigma\big(\fr12 u_{i}y_i\big)\sigma^*\big(\fr12 u_{j}y_j\big),
\end{align} 
where $\si$ is the mode function for the massive scalar in the Line $\al$ \cite{Chen:2017ryl}. As a result, the integral (\ref{eq_singlecut}) is computable by expanding the mode functions in the small-$y$ limit. The new feature here, compared to \cite{Liu:2024str}, is that we need to expand the mode function together with additional exponential factors arising due to (\ref{eq_useparation}). Explicitly, we need: 
\begin{align}
    \label{eq_latetimeexpansion}
    &e^{-\ii\bb u y/2} \sigma\big(\fr12 u y\big) = \sum_{\cc=\pm}\sum_{m=0}^\infty A_{m,\cc}^{\bb}(u)(-y)^{m+\ii \cc \wt\nu+3/2}, 
\end{align}
where the coefficients $A_{m,\cc}^{\bb}(u)$ are:
\begin{align}
    \label{eq_latetimeexpansionA}
    &A_{m,\cc}^{\bb}(u) \equiv \frac{\ii}{2} e^{\pi \cc \wt\nu/2} \,\text{csch}\,(\pi \cc \wt\nu) \Gamma\Bigg[ \bgm \fr12 +m+\ii \cc \wt\nu\\ 1+m+2 \ii \cc \wt\nu \edm \Bigg]\frac{(\ii\bb)^m}{m!}u^{m+\ii \cc \wt\nu+3/2}.
\end{align}
The coefficients satisfy a useful relation $A_{m,\cc}^{\bb *}(u) =e^{\pi \cc \wt\nu}A_{m,-\cc}^{-\bb}(u)$ which follows directly from its explicit form.

Then, we can substitute the expansion \eqref{eq_latetimeexpansion} into the single-cut integral \eqref{eq_singlecut} and get:
\begin{align}
    \label{eq_cutG}
    \mathop{\text{Cut}}_{K_\alpha}\,\big[\mathcal{G}'\big] = \sum_{\{\cc,m\}}\Big[ A^{+\ss_i}_{m_1,\cc_1}(u_i)A^{-\ss_j *}_{m_2,-\cc_2}(u_j)\mathcal{L}^{\cc_1}_+ \mathcal{R}^{\cc_2}_+ + A^{-\ss_i}_{m_1,\cc_1}(u_i)A^{-\ss_j *}_{m_2,-\cc_2}(u_j)\mathcal{L}^{\cc_1}_- \mathcal{R}^{\cc_2}_+\Big]+\text{c.c.}.
\end{align}
Here, the two integrals $\mathcal{L}^{\cc_1}_\pm$ and $\mathcal{R}^{\cc_2}_\pm$ are defined by:
\begin{align}
    \mathcal{L}^{\cc_1}_\pm \equiv& \pm \ii \int_{-\infty}^0 \mathrm{d} y_i(-y_i)^{p_i+m_1+\ii\cc_1\wt\nu+3/2}e^{\pm\ii (1-\bm{u}^{\text{(L)}}_i/2)y_i}\mathcal{I}^{\text{(L)}}_\pm,\\
    \mathcal{R}^{\cc_2}_\pm \equiv& \pm \ii \int_{-\infty}^0 \mathrm{d} y_j(-y_j)^{p_j+m_2+\ii\cc_2\wt\nu+3/2}e^{\pm\ii (1-\bm{u}^{\text{(R)}}_j/2)y_j}\mathcal{I}^{\text{(R)}}_\pm.
\end{align}
They are the time integrals of two subgraphs after the cut. Here, $\cc_{1,2}=\pm1$ are parameters labeling the sign in front of $\pm\ii \wt\nu$, to which we refer as the \textit{shadow pair}. On the other hand, the subscripts in $\mathcal{L}_\pm^{\cc_1}$ and $\mathcal{R}_\pm^{\cc_2}$ are SK indices. As shown in \cite{Liu:2024str}, these integrals can be rewritten into an SK-branch-independent form using the identity in (\ref{eq_branchRelation}). As a result, we can rewrite the single cut of the graph as the sum of products of \textit{tuned} CISs of two subgraphs as:
\begin{align}
    \label{eq_cuttingRule}
    \mathop{\text{Cut}}_{K_\alpha}\, \big[\mathcal{G}'\big] = \mft{\wh{1}\cdots i^{\sharp} \cdots}'\Big\{ \mft{\cdots j^{\sharp} \cdots}' +\mft{\cdots j^{\flat} \cdots}'\Big\}+(\wt\nu_\alpha \to -\wt\nu_\alpha).
\end{align}
Here, the term ($\wt\nu_\al\to-\wt\nu_\al$) is the shadow conjugate of the displayed terms, and we have used the \emph{augmentation} ($\sharp$) and \emph{flattening} ($\flat$) of a rescaled MFT, defined by:
\begin{align}
    \label{eq_augment}
    \mft{\cdots i^{\sharp}\cdots}' \equiv &~ \FR{-1}{2\cos(\pi\Delta_\al^+)}\sum_{m=0}^{\infty} \Gamma\bgb \Delta_\alpha^+ +m-1\\ 2\Delta_\alpha^+ +m- 2\edb\FR{\ss_i^m}{m!}\Big(\FR{2K_\al}{\E_i}\Big)^{m+\Delta_\al^+} \n\\ &\times\bigg\{\FR{\cos\big[\fr{\pi(p_\text{tot}-m)}{2}\big]}{\cos\big(\fr{\pi p_\text{tot}}{2}\big)}\mft{\cdots i\cdots}'\bigg\}_{\subalign{&p_i\to p_i+m+\Delta_\al^+\\&E_i\to E_i+\ss_i K_\alpha}},\\
    \label{eq_flat}
    \mft{\cdots i^{\flat}\cdots}' \equiv &~\FR{-1}{2\cos(\pi\Delta_\al^-)} \sum_{m=0}^{\infty}  \Gamma\bgb \Delta_\alpha^- +m-1\\ 2\Delta_\alpha^- +m- 2\edb\FR{\ss_i^m}{m!}\Big(\FR{2K_\al}{\E_i}\Big)^{m+\Delta_\al^-} \n\\ &\times\bigg\{\FR{\cos\big[\fr{\pi(p_\text{tot}+2\ii\wt\nu_\al-m)}{2}\big]}{\cos\big(\fr{\pi p_\text{tot}}2\big)}\mft{\cdots i\cdots}'\bigg\}_{\subalign{&p_i\to p_i+m+\Delta_\al^-\\&E_i\to E_i+\ss_i K_\alpha}}.
\end{align}
Here, $p_\text{tot}$ means the sum of all twists of the MFT $\mft{\cdots i\cdots}'$ (\emph{before} the shift $p_i\to p_i+m+\Delta_\al^\pm$ indicated in the subscript). Also, we have used the scaling dimension $\Delta_\al^\pm\equiv \fr32\pm\ii\wt\nu_\al$ for notational simplicity. Note that these definitions are different from \cite{Liu:2024str} due to the change of variables and the different scaling factors in the definition of $\mft{\cdots}'$ and $\mft{\cdots}$. 

The augmentation and flattening are sort of dressing an MFT by ``one half'' of an extra massive line. In the above definition, we are dressing the site $i$ with a massive mode with mass $\wt\nu_\al$ and line energy $K_\al$. Thus, $\ss_i$ is the sign label of the endpoint $i$. Notably, in this version of tuning an MFT, we shift the vertex energy $E_i$ to $E_i+\ss_i K_\alpha$, so that the dressed energy $\E_i$ remains the same as that of the original graph. In other words, taking a cut does not change the variables. 

We emphasize that the single cut
\eqref{eq_cuttingRule} is manifestly asymmetric with respect to $i\leftrightarrow j$. As explained in \cite{Liu:2024str}, this asymmetry is ultimately from the asymmetric assignment of (dressed) vertex energies. Specifically, whenever we execute a cut over Line $\al$ which separates the whole graph into two subgraphs, we single out the largest (dressed) vertex energy from each of the two subgraphs (called the locally maximal dressed vertex energy) and compare the two.\footnote{It was explained in \cite{Liu:2024str} why we should compare the two locally maximal vertex energies of the two subgraphs, rather than compare the two vertex energies $E_i$ and $E_j$ as one may naively expect.} The direction of the cut is always from the subgraph with the larger locally maximal dressed vertex energy ($\mft{\wh 1\cdots i\cdots}'$ in (\ref{eq_cuttingRule})) to the other subgraph ($\mft{\cdots j\cdots}'$ in (\ref{eq_cuttingRule})). The former subgraph is augmented, while the latter is both augmented and flattened in the single cut.

\paragraph{Multiple cuts}
The above prescription for taking the single cut generalizes directly to multiple cuts, as detailed in \cite{Liu:2024str}. Briefly, when taking multiple cuts over $C$ lines, we are breaking the original graph into $C+1$ disjoint subgraphs. Then, we compare the $C+1$ locally maximal dressed vertex energies of the $C+1$ subgraphs to determine the directions of all cuts. Once again, for any specific cut line, the direction is from the subgraph with the larger locally maximal dressed vertex energy to the other. Then, the result of the $C$ cuts is the sum and products of the tuned (augmented or flattened) CISs of $C+1$ subgraphs. Explicitly, every cut generates a pair of terms together with their shadow conjugates, in which the “outgoing” vertex is augmented in both terms, and the “ingoing”  vertex is respectively augmented and flattened in the two terms. Note that a vertex can be shifted multiple times with respect to different cut lines. 

It is also direct to find the expression for a multiply tuned MFT. Suppose that an arbitrary MFT $\mft{\cdots}$ is augmented $A$ times at vertices $E_1,\cdots,E_A$ (with twists $p_1,\cdots,p_A$) by cut lines $K_1,\cdots,K_A$ (with mass $\wt\nu_1,\cdots,\wt\nu_A$), and flattened $F$ times at vertices $E'_1,\cdots,E'_F$ (with twists $p'_1,\cdots,p'_F$) by cut lines $K'_1,\cdots,K'_F$ (with mass $\wt\nu'_1,\cdots,\wt\nu'_F$). Then the tuned MFT is:
\begin{align}
\label{eq_multituning}
    \mft{(\cdots)^{\sharp_1\cdots\sharp_A \flat_1\cdots\flat_F}}' = &\sum_{\{m,m'\}}  \bigg\{\FR{\cos\big[\fr{\pi(p_\text{tot}-m_{1\cdots A}-m'_{1\cdots F}+2\ii\wt\nu_{1\cdots F}')}2 \big]}{\cos(\pi p_\text{tot}/2)}\mft{\cdots}'\bigg\}_{\subalign{p_{\al}&\to p_{\al}+m_\al+\Delta_\al^+,E_\al \to E_\al+\ss_\al K_\alpha\\ p_{\be}'&\to p_{\be}'+m'_\be+\Delta_\be^-{}',E'_\be \to E'_\be+\ss'_\be K'_\be}}\n\\
    &\times\prod_{\al=1}^A \frac{-1}{2\cos(\pi\Delta_\alpha^+)} \Gamma\bgb \Delta_\al^++m_\al-1 \\ 2\Delta_\al^++m_\al-2\edb\FR{\ss_\al^{m_\al}}{m_\al!}\Big(\FR{2K_\al}{\E_\al}\Big)^{m_\al+\Delta_\al^+}\n\\
    &\times \prod_{\be=1}^F \frac{-1}{2\cos(\pi\Delta_\be^-)}  \Gamma\bgb \Delta_\be^-{}' +m'_\be-1\\ 2\Delta_\be^-{}'+m'_\be-2\edb \FR{\ss'^{m'_\be}_\be}{m'_\be!} \Big(\FR{2K_\be}{\E_\be}\Big)^{m'_\be+\Delta_\be^-{}'}.
\end{align}
With all the above results known, it is straightforward to write down the full solution to our differential equations with appropriate BD boundary conditions as the sum of the CIS (i.e., the MFT) and all of its cuts, as given in (\ref{eq_fullsolutionG}) with trivial replacement $\G\to \G'$:
\begin{keyeqn}
\begin{align}
\label{eq_fullsolutionGprime}
  \G'=\text{CIS}\,\big[\G'\big]+\sum_\al \mathop\text{Cut}_{K_\al} \big[\G'\big]+\sum_{\al\neq\be} \mathop\text{Cut}_{K_\al,K_\be} \big[\G'\big]+\cdots+\mathop\text{Cut}_{\text{all~}K} \big[\G'\big].
\end{align}
\end{keyeqn}

\section{Folded and Partial-Energy Limits}
\label{sec_limits}

As advocated before, a main advantage of our new representation of correlators is that taking certain limits becomes a straightforward task. In this section, we consider two kinematic limits. One is the folded limit, or colinear limit, which can be taken at any leaf site and is relevant to our discussion of degenerate correlators with bilinear mixings. The other is the partial-energy limit, which is an unphysical limit (unreachable in physical regions) but is physically interesting as it probes the short-distance behavior of a subgraph. In the following, we will be more detailed in folded limits and only briefly touch upon partial-energy limits.

\subsection{Folded limit}
\label{subsec_fold}

As mentioned, a folded limit is defined with respect to any leaf site. A leaf site is a site to which only one internal line is attached. It is possible to use our new variables to consider similar folded limits but we will leave them to interested readers. Let Site $i$ be an arbitrary leaf site with vertex energy $E_i$, and it is attached to a line with line energy $K_i$. (Since the line is unique, we can identify the subscript of $E_i$ and $K_i$ without ambiguities.) Then, the folded limit refers to $E_i-K_i\to 0$; See Fig.\;\ref{fig_fold}. 
\begin{figure}[t]
\centering
\includegraphics[width=0.95\textwidth]{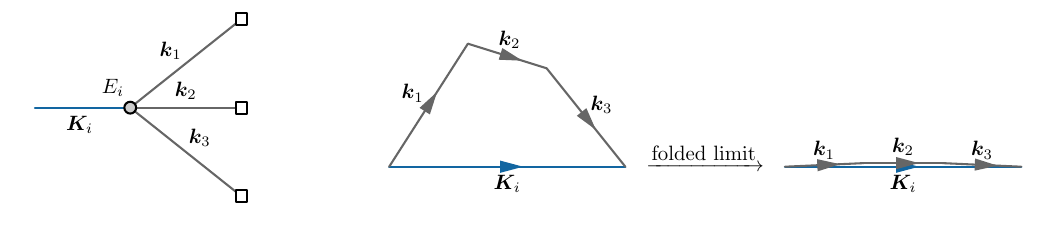} 
\caption{The momentum configuration of a folded limit (colinear limit) at a folded leaf (Site $i$). Here, we are considering an example with 3 external lines attached to Site $i$ with momenta $\bm k_i$ ($i=1,2,3$). Thus we have $E_i=|\bm k_1|+|\bm k_2|+|\bm k_3|$, $K_i=|\bm k_1+\bm k_2+\bm k_3| $, and $E_i\to K_i$ in the folded limit.}
\label{fig_fold}
\end{figure}

The folded limit suggests that we take the sign label at Site $i$ to be $-1$, i.e., we define $\E_i\equiv E_i-K_i$ and the folded limit corresponds to $\E_i\to 0$. An immediate but inessential problem is that the originally defined graph $\G'$ in (\ref{eq_Grescaleu}) becomes divergent, and we need to rescale it to make it well-defined in the folded limit. Thus, we define:
\begin{align}
    \mathop{\text{Fold}}_{i}\,\big[\mathcal{G'}\big] = \lim_{E_i\to K_i} \Big(\frac{2K_i}{E_i-K_i}\Big)^{p_i+1} \mathcal{G'},
\end{align}
where we introduce the factor $\big(\frac{2K_i}{E_i-K_i}\big)^{p_i+1}$ to compensate for the singular behavior of $\G'$. One can certainly take multiple folded limits at many leaf sites; The generalization is trivial, and we will do them in the next section.

Given our full solution to the graph in (\ref{eq_fullsolutionGprime}), taking folded limits can be separately done for the CIS and its cuts, on which we elaborate below.

\paragraph{CIS}
Using our new variables, the folded limit for a CIS is trivial. Recall that, once we have chosen a maximal dressed vertex energy, say $\E_1$, the CIS of a graph $\G'$ is a massive family tree with Site 1 being the root. Using our notation already introduced in Sec.\;\ref{sec_single}, we use an over-circle to denote the leaf site being taken the folded limit:
\begin{align}
    \mft{\wh{1}\cdots \mathring{i}\cdots}' \equiv \mathop{\text{Fold}}_{i}\,\mft{\wh{1}\cdots i\cdots}' .
\end{align}
Now, let us consider an arbitrary MFT with an arbitrary nonzero number of leaves taking folded limits, called \emph{folded leaves}. We denote the set of folded leaves by $\mathcal{F}$. Then, the folded limit of the CIS follows trivially from (\ref{eq_dressedCIS}):
\begin{align}
\label{eq_foldedCIS}
    \mft{\wh{1}\cdots\mathring{\mathcal{F}}}'
    = ~&2\cos(\pi \wt p_1/2)  \sum_{\{\ell,m\}}\Gamma(p_1+q_1+1) \prod_{i\in\mathcal{F}} \frac{(p_i+3)_{m_i}}{(p_i+\fr52\pm\ii \wt{\nu}_i)_{m_i+1}}\Big(\frac{2K_i}{\E_1}\Big)^{m_i+p_i+4} \n\\ &\times \prod_{j\notin\mathcal{F}} \frac{\ss_j^{m_j} (-1)^{\ell_j}(\ell_j+q_j+p_j+3)_{m_j}}{\ell_j!(\ell_j+q_j+p_j+\fr52\pm\ii \wt{\nu}_j)_{m_j+1}}\Big(\frac{2K_j}{\E_1}\Big)^{m_j+3}\Big(\frac{\E_j}{\E_1}\Big)^{\ell_j+p_j+1}.
\end{align}
That is, the summation over $\ell_j$ disappears when $j\in\mathcal{F}$, which confirms in general cases the earlier observation that the transcendental weight of a CIS decreases by $F$ after taking $F$ folded limits.

\paragraph{Cut} Next, we consider the folded limits for the cut part of the graph. Again, we identify the subscript of the line energy $K_i$ with the associated leaf-site vertex energy $E_i$ when $i\in\mathcal{F}$. Then, nothing special happens for the cut over Line $\al$ for $\al\notin\mathcal{F}$ when taking the folded limit: In such cases, we can take folded limits of subgraphs after the cuts in exactly the same way as we do for CISs. So, the only interesting possibility is when the cut line is associated with a folded leaf, which we consider below.

Without loss of generality, we can consider a single cut over Line $K_i$ which is connected to a mother site $E_j$ and a folded leaf $E_i$. Since the dressed vertex energy $\E_i\to 0$ in the folded limit, the direction of the cut is naturally from the rest of the graph to the folded leaf. Thus, according to (\ref{eq_cuttingRule}), we have:
\begin{align}
  \mathop{\text{Fold}}_{E_i}\Big\{\mathop{\text{Cut}}\limits_{K_i}\big[\G'\big]\Big\}=\mft{\wh 1\cdots j^\sharp\cdots}'\lim_{E_i\to K_i}\Big(\FR{2K_i}{E_i-K_i}\Big)^{p_i+1}\Big\{\mft{i^\sharp}'+\mft{i^\flat}'\Big\}+(\wt\nu_i\to-\wt\nu_i).
\end{align}
From now on, the discussion parallels Sec.\;\ref{sec_single}: each of the two tuned single-site MFTs is divergent by itself in the folded limit, but the sum is finite. More explicitly, using (\ref{eq_augment}), we have:
\begin{align}
    \Big(\frac{2K_i}{E_i-K_i}\Big)^{p_i+1}\mft{i^\sharp}' = &~ \frac{4\sin\big[\frac{\pi}{2}\big(p_i+\ii \wt\nu_i+\frac12\big)\big]}{\sin(\ii\pi \wt\nu_i)} {}_2 \mathcal{F}_1\Bigg[ \bgm \frac{\fr72 +p_i+\ii\wt\nu_i}{2},\frac{\fr52 +p_i+\ii\wt\nu_i}{2}\\ 1\pm\ii \wt\nu_i \edm \Bigg| \FR{K_i^2}{E_i^2} \Bigg]\Big(\frac{K_i}{E_i}\Big)^{3/2+\ii\wt\nu_i} 
    \label{eq_foldedright},  
\end{align}
and a similar equation with $i^\sharp\to i^\flat$ on the left-hand side and all $\ii\wt\nu_i\to-\ii\wt\nu_i$ on the right-hand side. Then, taking the sum, we get a generally finite result:
\begin{align}
    \lim_{E_i\to K_i} \Big(\frac{2K_i}{E_i-K_i}\Big)^{p_i+1}\Big\{  \mft{ i^\sharp}' +\mft{i^\flat}'\Big\} = 2\cos\big[\fr{\pi}{2}\big(p_i+\ii \wt\nu_i+\fr32\big)\big]\Gamma\Bigg[ \bgm \fr52 +p_i\pm\ii\wt\nu_i\\ 3+p_i\edm \Bigg] .
\end{align}
Then, we have the following folded limit for cut leaves:
\begin{align}
    \label{eq_foldedCut}
    \mathop{\text{Fold}}_{E_i}\Big\{\mathop{\text{Cut}}_{K_i}\big[\mathcal{G'}\big]\Big\} =2\cos\big[\fr{\pi}{2}\big(p_i+\ii \wt\nu_i+\fr32\big)\big]\Gamma\bgb \fr52 +p_i\pm\ii\wt\nu_i\\ 3+p_i\edb \mft{\wh{1}\cdots j^\sharp \cdots}' +(\wt\nu_i \to -\wt\nu_i).
\end{align}
This result can be easily generalized to the multiple folded limits for arbitrary cuts, as:
\begin{align}
    \label{eq_foldedMultipleCuts}
    \mathop{\text{Fold}}_{\mathcal{F}}\Big\{\mathop{\text{Cut}}_{\mathcal{F}}\big[\mathcal{G'}\big]\Big\} =\mft{\cdots \mathcal{N}(\mathcal{F})^\sharp}' \prod_{i\in\mathcal{F}}2\cos\big[\fr{\pi}{2}\big(p_i+\ii\wt\nu_i+\fr32\big)\big]\Gamma\bgb \fr52 +p_i\pm\ii\wt\nu_i\\ 3+p_i\edb+\text{shadows}.
\end{align}
Here we use the notation $\mathcal{N}(\mathcal{F})$ to denote the set of all neighbor sites of folded leaves, and $\mathcal{N}(\mathcal{F})^\sharp$ means to augment all these sites accordingly.

Again, we see that the transcendental weight of the single cut drops from 1 to 0 as we take the folded limit. So the whole graph still has a uniform but reduced transcendental weight after taking a folded limit.

\subsection{Partial-energy limit}
\label{subsec_partialenergy}

As a second example, we consider very briefly the partial-energy limit at any given site of the graph. In general, for an arbitrary tree graph $\mathcal{G}'$, a partial-energy limit is discussed with respect to a subgraph $\mathcal{H}'\subset \mathcal{G}'$ and defined to be the limit of the magnitude sum of all momenta flowing into $\mathcal{G}'$ going to zero while each momentum is held finite. Clearly, this is an unphysical limit, but it probes the early-time limit (and thus the high-energy/short-distance/flat-space limit) of an inflationary correlator, as has been discussed a lot in the literature. 

Here, we only comment that our new variables make it trivial to take the partial-energy limit of a graph at any given \emph{site}. That is, we only consider subgraphs $\mathcal{H}'$ containing only one site, say Site $i$. Then, we can choose the sign labels $\ss_{\al i}=+1$ for all lines attached to this site and define:
\bge
  \E_i=E_i+\sum_{\al\in\mathcal{N}(i)}K_\al.
\ede  
Then, the partial-energy limit at Site $i$ is simply $\E_i\to 0$. 

It is by now standard knowledge that a correlator may develop singularities in the partial-energy limit. Here, we present a complete (not only the leading order) expression for the singular part of the correlator at $\E_i\to 0$. To properly present this singularity, it is useful to go back to the unprimed graph $\G$ defined in (\ref{eq_Grescale}). This is because the primed graph $\G'$ is related to the correlator by, among other factors, a factor of $1/\E_i^{1+p_i}$ which is by itself singular when $\E_i\to 0$; See (\ref{eq_Grescaleu}). 

As is well known, the partial-energy singularity at $\E_i\to 0$ can be understood as a failure of the convergence of the time integral (\ref{eq_dimlessG}): There are terms in the integrand of (\ref{eq_dimlessG}) whose early-time behavior is controlled by the factor $e^{\ii\E_i\tau_i}$. When $\E_i\to 0$, the $\ii\ep$-prescription fails to work, and we get a divergent result. Now, our method of decomposing a correlator into an MFT and its cuts says that, whenever Site $i$ is connected to another site $j$, the integral over $\tau_i$ is always bounded by $\tau_i>\tau_j$ and there are never divergences in this case, given that all other energy combinations remain finite. Thus, the only terms that can possibly contribute to partial-energy singularities at $\E_i\to 0$ must have all lines attached to Site $i$ cut.\footnote{See \cite{PartialEnergy_ongoing} for a more complete discussion of partial-energy limits of arbitrary massive inflationary correlators at both tree and loop levels, and \cite{Fan:2025scu} for related discussions on partial-energy singularities of (normal) family trees.}

We can make use of the above observation to write down a complete expression for the singular part of $\G$ at $\E_i\to 0$. In general, after cutting all $K_\al$ with $\al\in\mathcal{N}(i)$, $\G'\backslash\{i\}$ breaks into $M\geq 1$ (properly augmented) disjoint subgraphs $\mathcal{H}_1'^{\,\sharp_1},\cdots , \mathcal{H}_M'^{\,\sharp_M}$. Then, the partial-energy limit $\E_i\to 0$ of $\G'$ can be written as:
\begin{align}
\label{eq_PEL}
  \lim_{\E_i\to 0}\G=\FR{1}{2^{3(V-1)}}\prod_{j=1}^V\Big(\FR{E_j}{\E_j}\Big)^{1+p_j}\times \mathcal{H}_1'^{\,\sharp_1} \cdots \mathcal{H}_M'^{\,\sharp_M}\sum_{T_1,\cdots,T_M=\sharp,\flat}\mft{i^{T_1\cdots T_M}}'+\text{regular terms}.
\end{align}
Here, the prefactors $1/2^{3(V-1)}\prod_j(E_j/\E_j)^{1+p_j}$ account for the difference between $\G$ and $\G'$. The $T$-fold tuned (augmented or flattened) single-site tree $\mft{i^{T_1\cdots T_M}}'$ can be found from the general expression (\ref{eq_multituning}) of multi-fold tuning. However, the resulting series is divergent since the partial energy $\E_i\to 0$ sits in the denominator. Nevertheless, for this particular example, it is easy to analytically continue the divergent series and find a healthy series expression as $\E_i\to 0$. The healthy expression is actually a sum of several hypergeometric series, among which one is singular at $\E_i\to 0$ and all others are regular. Since we are only interested in the partial-energy singularity, we only write down the singular series: 
\begin{align}
\label{eq_PESof1site}
     & \FR{1}{\E_i^{p_i+1}} \sum_{T_1,\cdots,T_M=\sharp,\flat} \mft{i^{T_1\cdots T_M}}'  =  \FR{2\cos\big[\fr\pi2(p_i+\Delta^+_{1\cdots M})\big]}{\E_i^{p_i+1}}  \n\\
    &\times \sum_{\{m\}}\Gamma[p_i-m_{1\cdots M}+M+1]\prod_{\alpha=1}^M \FR{\Gamma[m_\al-1+\Delta^\pm_\al]}{m_\al!\text{sech}(\pi\wt\nu_\al)} \Big(\frac{-\E_i}{2 K_\alpha}\Big)^{m_\alpha-1}.
\end{align}
The regular terms, as well as the derivation of (\ref{eq_PESof1site}), are collected in App.\;\ref{app_PElimit}.

\section{Degenerate Stars}
\label{sec_star}

Given the importance of bilinear mixings to CC phenomenology, we use this section to discuss a simple set of tree graphs with bilinear mixings, which we call \emph{degenerate stars}. In a degenerate star, all internal massive particles are linearly mixed with external massless or conformal scalars with arbitrary twists, as shown in Fig.\;\ref{fig_star}. In other words, all leaves are folded in a degenerate star. Thus, we can directly use our new representation to work out its full solution. In this section, we change the convention slightly and call the root site $E_0$ (instead of $E_1$ previously). 
\begin{figure}[t]
\centering
\includegraphics[width=0.8\textwidth]{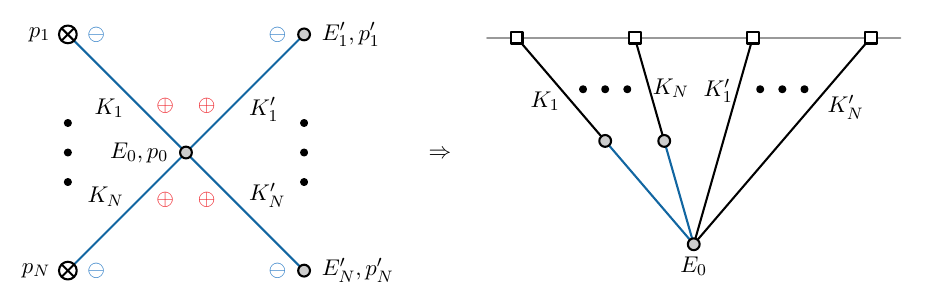} 
\caption{The degenerate star graph with $N$ mixed propagators.}
\label{fig_star}
\end{figure}

We denote the correlator of an $N$-pointed degenerate star graph by $\mathcal{S}_N$, and we adopt yet another overall factor to make $\mathcal{S}_N$ dimensionless. Explicitly, the SK time integral for $\mathcal{S}_N$ is given by:
\begin{align}
    \mathcal{S}_N (E_0;K_1,\cdots,K_N) \equiv& ~\E_0^{p_0+1}\prod_{j=1}^N \Big[(2K_j)^{p_j+4}\Big] \sum_{\aa_0,\cdots,\aa_N = \pm}\int^0_{-\infty} \di\tau_0\,\ii\aa_0(-\tau_0)^{p_0}e^{\ii \aa_0 E_0 \tau_0}\n\\ 
  &\times\int^0_{-\infty}\prod_{i=0}^{N} \Big[\di \tau_i\,(\ii \aa_i) (-\tau_i)^{p_i} e^{\ii \aa_i K_i \tau_i} D_{\aa_0\aa_i}(K_i; \tau_0,\tau_i)\Big],
\end{align}
where $\E_0=E_0+K_{1\cdots N}$. 
We don't have to do this integral again, as it is simply the folded limit of an $(N+1)$-site star graph $\G_{N+1}'$ whose CIS is $\mft{\wh 0(1)\cdots (N)}'$:
\begin{align}
\label{eq_GtoS}
    \mathcal{S}_N =  \mathop{\text{Fold}}_{E_1,\cdots,E_N}\big[ \mathcal{G}'_{N+1}\big].
\end{align}
Under this limit, a complete set of $u$-variables to specify the kinematics of $\mathcal{S}_N$ are given by $u_i \equiv 2K_i/\E_0, ~i =1,\cdots,N$. Incidentally, the degenerate star $\mathcal{S}_N$ satisfies a simple set of $N$ coupled second-order PDEs, which follows directly from the general PDEs in \eqref{eq_DEinUvar}:
\begin{align}
\label{eq_DEforStarinUvar}
  \bigg[\Big(\eta_i-\FR32\Big)^2+\wt\nu_i^2-u_i(\eta_i-1)(\bm\eta+p_0+1)\bigg]\mathcal{S}_N=\mathsf{C}_i\big[\mathcal{S}_N\big],
\end{align}
where $\eta_i=u_i\pd_{u_i}$ and $\bm\eta=\eta_{1\cdots N}$. 
It should be noted that this limit does not change the analytical property with respect to $u_\alpha$. Therefore, after taking the limit, the CIS and cuts of $\mathcal{G}'_{N+1}$ become those of $\mathcal{S}'_N$ respectively. 

\paragraph{Complete solution} Once again, the solution to the equation (\ref{eq_DEforStarinUvar}) with BD boundary conditions can be expressed as the sum of the CIS and all of the cuts:
\begin{align}
\label{eq_SNCutSum}
  \mathcal{S}_N=\text{CIS}\,\big[\mathcal{S}_N\big]+\sum_{j=1}^N\mathop{\text{Cut}}_{K_j} \big[\mathcal{S}_N\big]+\sum_{i\neq j}\mathop{\text{Cut}}_{K_i,K_j} \big[\mathcal{S}_N\big]+\cdots+ \mathop{\text{Cut}}_{\text{all~}K} \big[\mathcal{S}_N\big].
\end{align}
All these components can be gathered from the last section. First, the CIS is an MFT with all leaves folded, and can be obtained from \eqref{eq_foldedCIS}:
\begin{align}
\label{eq_CISStar}
    \text{CIS}\,\big[\mathcal{S}_N\big] \equiv &~\mft{0 ({\mathring{1}})\cdots  ({\mathring{N}})}'\n\\
    =&~ 2\cos(\fr{\pi}2\wt\Delta_0^0) \sum_{\{m\}}\Gamma(\wh m_{0}+\wt\Delta_0^0-3) \prod_{i=1}^N\FR{(\Delta_i^0-1)_{m_i}u_i^{m_i+\Delta_i^0}}{(\Delta_i^0+\Delta_i^\pm-3)_{m_i+1}}.
\end{align} 
For a reason that will be clear soon, we introduced a parameter $\Delta_i^0\equiv p_i+4$ ($i=0,1,\cdots,N$), and the scaling dimensions $\Delta_i^\pm=\fr32\pm \ii\wt\nu_i$ are defined as before. To remind the reader of our notation, we have $\wt\Delta_0^0\equiv\Delta_{01\cdots N}^0=p_{01\cdots N}+4N+4$, $\wh\Delta_0^0\equiv\Delta_{1\cdots N}^0=\Delta_{1\cdots N}^0=p_{1\cdots N}+4N$, and $\wh m_0=m_{1\cdots N}$. Also, we use the shorthand $(\Delta_i^0+\Delta_i^\pm-3)_{m_i+1}\equiv (\Delta_i^0+\Delta_i^+-3)_{m_i+1}(\Delta_i^0+\Delta_i^--3)_{m_i+1}$ which is derived from a similar shorthand used before. 

Next, let us look at the single cut over $K_j$ for any $j=1,\cdots,N$, whose result can be easily found by combining \eqref{eq_augment}, \eqref{eq_foldedCut}, \eqref{eq_GtoS} and \eqref{eq_CISStar}: 
\begin{align}
    \label{eq_starSingleCut}
  &\mathop{\text{Cut}}_{K_j} \big[\mathcal{S}_N\big] = \mathop{\text{Fold}}_{E_j}\Big\{\mathop{\text{Cut}}_{K_j}\,\big[\G_{N+1}'\big]\Big\}\n\\
=&~2\cos\big[\fr\pi2(\wt\Delta_0^0+\Delta_j^+ - \Delta_j^0)\big] \sum_{\{m\}}^\infty\Gamma(\wh m_0+\wt\Delta_0^0-\Delta_j^0+\Delta_j^+-3)  \prod_{i\neq j}\Bigg[\FR{(\Delta_i^0-1)_{m_i}u_i^{m_i+\Delta_i^0}}{(\Delta_i^0+\Delta_i^\pm-3)_{m_i+1}}\Bigg]  \n\\
    &\times\FR{\cos\big[\fr\pi2(\Delta_j^0+\Delta_j^+)\big]}{\cos\big[\pi(\Delta_j^+ +1)\big]}\Gamma\bgb \Delta_j^0+\Delta_j^\pm-3 \\ \Delta_j^0-1\edb\Gamma\bgb \Delta_j^++m_j-1\\ 2\Delta_j^++m_j-2\edb\FR{u_j^{m_j+\Delta_j^+}}{m_j!}  +(\wt\nu_j\to-\wt\nu_j).
\end{align}
Here, we have put the result into a particular form, so that a comparison between (\ref{eq_CISStar}) and (\ref{eq_starSingleCut}) suggests that we can obtain the single cut $\mathop{\text{Cut}}_{K_j}\big[\mathcal{S}_N\big]$ from $\text{CIS}\big[\mathcal{S}_N\big]$ by simple replacement of $\Delta_j^0\to \Delta_j^\pm$ in a few places. Then, this ``cut by replacement'' trick will also work for any number of cuts, suggesting that we can put the whole solution into a rather simple and compact form. This is the reason for our introducing the notation $\Delta_i^0$ above. Now, let us make this replacement more explicit and systematic by introducing the following symbol:
\bge
  \Delta_i^\ec=
  \begin{cases}
    p_i+4, &(\ec=0\text{~and~}i=0,1,\cdots,N)\\[2mm]
    \FR32\pm\ii\wt\nu_i. &(\ec=\pm\text{~and~}i=1,\cdots,N)
  \end{cases}
\ede
We can view $\Delta_i^\ec$ as a scaling dimension for each site. In particular, every folded site (i.e., Site $i$ with $i=1,\cdots,N$) is associated with an internal propagator of mass $\wt\nu_i$ and a twist $p_i$. So, it has three possible values of $\Delta_i^\ec$: $\ec=\pm$ corresponds to the scaling behavior of two late-time modes of the massive line, while $\ec=0$ is controlled by the twist $p_i$. On the other hand, the root site (Site 0) has only one scaling $\Delta_0^\ec$ with $\ec=0$, which is determined by the root twist $p_0$. Note that the tri label $\ec_i$ is defined with respect to each site, and it can take different values at different sites. So, when we write $\Delta_i^\ec$, we actually mean $\Delta_i^{\ec_i}$. Similarly, we also write $\wt\Delta_0^\ec=\Delta_0^0+\sum\limits_{i=1}^N\Delta_i^{\ec_i}$.

With all the above preparations, we can now write down the complete solution of $\mathcal{S}_N$ in a very simple form:
\begin{keyeqn}
\begin{align} 
\label{eq_DegStarSol}
    \mathcal{S}_N = &\sum_{\ec_1,\cdots,\ec_N=0,\pm1}\sum_{m_1,\cdots,m_N=0}^\infty {2\cos(\fr{\pi}2\wt \Delta_0^{\ec})} 
   \Gamma(\wh m_0+\wt\Delta_0^{\ec}-3)
    \n\\
    &\times\prod_{i=1}^N \FR{\cos[\fr\pi2\big(\Delta_i^0+\Delta_i^{\ec})]}{\cos[\pi(\Delta_i^{\ec}+\ec_i)]}\FR{(\Delta_i^0-1)_{m_i+\Delta_i^\ec-\Delta_i^0}u_i^{m_i+\Delta^{\ec}_i}}{(\Delta_i^0+\Delta_i^\pm-3)_{m_i+\Delta_i^\ec-\Delta_i^0+1}} ,
\end{align}
\end{keyeqn}
One can easily check that this expression includes all terms on the right-hand side of (\ref{eq_SNCutSum}). In particular, taking all $\ec_i=0$ reproduces the CIS in (\ref{eq_CISStar}) and taking one $\ec_j=\pm$ and all other $\ec_i=0$ reproduces the single cut in (\ref{eq_starSingleCut}). So, the pattern is clear: cutting Line $i$ is simply achieved by setting $\ec_i=\pm$. We give an inductive proof of (\ref{eq_DegStarSol}) in App. \ref{app_proof}.

The physical meaning of the scaling dimension $\Delta_i^\ec$ is now quite clear: Whenever we consider a soft limit (usually called the squeezed limit) where $u_i$ ($i=1,\cdots,N$) goes to 0, the degenerate star $\mathcal{S}_N$ gives rise to three leading fall-offs $\propto (K_i/\E_0)^{\Delta_i^\ec}$, where, in the terminology of CC physics, $\Delta_i^\pm=\fr32\pm\ii\wt\nu_i$ corresponds to oscillatory signal and $\Delta_i^0=p_0+4$ to the smooth and monotonic background. So, in a squeezed limit $K_i\to 0$, the relative importance between the signal and the background is controlled by the twist at the mixing vertex. As an example, the familiar bilinear mixing (\ref{eq_dim5toMix}) in the original QSFI corresponds to $p_i=-2$. So, we have the signal scaling like $K_i^{3/2\pm\ii\wt\nu_i}$ while the background like $K_i^2$. That is, the background is more suppressed than the signal by a factor of $K_i^{1/2}$, which is a well-known fact in CC physics. 

Finally, we briefly comment on the hypergeometric functions required to express degenerate stars (\ref{eq_DegStarSol}). According to our classification based on transcendental weights, an $N$-pointed degenerate star has a uniform transcendental weight-$N$. Here, being uniform means that all series in (\ref{eq_DegStarSol}) for all choices of $\ec_i$ $(i=1,\cdots,N)$ have the same weight, as is true in the generic (nondegenerate) case. However, there is still a milder distinction of complexity of these functions among different series in (\ref{eq_DegStarSol}), due to the downstairs Pochhammer factors in (\ref{eq_DegStarSol}):
\bge
  \FR{1}{(\Delta_i^0+\Delta_i^\pm-3)_{m_i+\Delta_i^\ec-\Delta_i^0+1}}=
  \begin{cases}
    \Gamma\bgb \Delta_i^0+\Delta_i^+-3,\Delta_i^0+\Delta_i^--3\\m_i+1,m_i+2\Delta_i^+-2\edb, &(\ec=+1) \\[5mm]
    \Gamma\bgb \Delta_i^0+\Delta_i^+-3, \Delta_i^0+\Delta_i^--3\\m_i+1,m_i+2\Delta_i^--2\edb, &(\ec=-1) \\[5mm]
    \Gamma\bgb \Delta_i^0+\Delta_i^+-3, \Delta_i^0+\Delta_i^--3\\m_i+\Delta_i^++\Delta_i^0-2,m_i+\Delta_i^-+\Delta_i^0-2\edb, &(\ec=0)
  \end{cases}
\ede
where we have used $\Delta_i^++\Delta_i^-=3$. The important point is that this pair of downstairs Pochhammer factors provides a function of summation variable $m_i$ in the form of $(m_i+a_i)^{-1}(m_i+b_i)^{-1}$ where $a_i$ and $b_i$ eventually become lower parameters in the hypergeometric functions. When $\ec=\pm 1$, we see that one of the two parameters becomes 1, signaling a reduction of the number of parameters compared to $\ec=0$. This implies that the functions required to express the cut part of the degenerate stars are generally simpler than the non-cut part. From Table\;\ref{tab_hypergeo}, we see that we can use Lauricella functions to express the all-cut part of the graph, as noticed before, but have to use multivariate KdF functions when any line is uncut in $\mathcal{S}_N$. Also, for 1-pointed and 2-pointed degenerate stars, we can express the entire results with more familiar hypergeometric functions. (We will show them below.) This explains, for instance, why previous works such as \cite{Qin:2023ejc} can obtain closed form expressions for 3-point correlators with a single massive exchange.

\begin{table}[htp]
\centering
\caption{Type of hypergeometric functions for degenerate stars}
\begin{tabular}{c|lll}
\toprule
  \makecell{No.\ of lower parameters \\ per variable} & univariate & bivariate & multivariate \\
\hline
 1 & Gauss ${}_2\text{F}_1$ & Appell & Lauricella \\
 2 & ${}_3\text{F}_2$ & KdF & multivariate KdF \\
 \bottomrule
\end{tabular} 
\label{tab_hypergeo}
\end{table}%

\subsection{Simple examples}
\label{sec_DegStarExamples}

The expression (\ref{eq_DegStarSol}) for general degenerate stars is rather condensed. It is useful to unpack it for a few simple examples and obtain more explicit results. Below we spell out these simple examples. They will reproduce the known results in the literature for 1-pointed and 2-pointed degenerate stars, and generate a new explicit expression for the 3-pointed degenerate star. In these examples, we finish all the summations in terms of (dressed) hypergeometric functions listed in Table \ref{tab_hypergeo}. For this purpose, it is useful to rewrite (\ref{eq_DegStarSol}) as:
\begin{align} 
    \label{eq_DegStarSolv2}
    \mathcal{S}_N = &\sum_{\{\ec,m\}} {2\cos(\fr{\pi}2\wt \Delta_0^{\ec})} 
   \Gamma(\wh m_0+\wt\Delta_0^{\ec}-3) \prod_{i=1}^N \mathcal{C}^\ec_i \Gamma \bgb m_i+\Delta_i^\ec-1 \\ m_i+\Delta_i^\ec+\Delta_i^\pm-2\edb u_i^{m_i+\Delta^{\ec}_i} ,
\end{align}
where we have defined the factor $\mathcal{C}_i^\ec$ as:
\begin{align}
    \mathcal{C}_i^{\ec}= \FR{\cos[\fr\pi2(\Delta_i^0+\Delta_i^{\ec})]}{\cos[\pi(\Delta_i^{\ec}+\ec_i)]}\Gamma \bgb \Delta_i^0+\Delta_i^{\pm}-3 \\ \Delta_i^0-1\edb.
\end{align}
More explicitly,
\begin{align}
  \mathcal{C}_i^\aa=&~\FR{\cos\big[\fr\pi2(p_i+\fr32+\ii\aa\wt\nu_i)\big]}{\cos\big[\pi(\fr12+\ii\aa\wt\nu_i)\big]}\Gamma\bgb p_i+\fr52+\ii\wt\nu_i,p_i+\fr52-\ii\wt\nu_i \\ p_i+3\edb,~~~(\aa=\pm1)\\
  \mathcal{C}_i^0=&~\Gamma \bgb p_i+\fr52+\ii\wt\nu_i,p_i+\fr52-\ii\wt\nu_i \\ p_i+3\edb.
\end{align}

\paragraph{Single exchange}
The 1-pointed star $\mathcal{S}_1$ is, up to an overall factor, nothing but the single-exchange graph considered in Sec.\;\ref{sec_single}. Taking $N=1$ in (\ref{eq_DegStarSol}), we get three hypergeometric series corresponding to $\ec_1=0,\pm1$:
\bge
\label{eq_S1}
  \mathcal{S}_1(E_0;K_1)=\underbrace{\text{CIS}\big[\mathcal{S}_1\big]}_{\ec_1=0}+\underbrace{\mathop{\text{Cut}}_{K_1}\big[\mathcal{S}_1\big]}_{\ec_1=\pm1}.
\ede
The series with $\ec_1=0$ is the CIS, or the background in the terminology of CC physics, and can be summed into a dressed ${}_3\text{F}_2$ function:
\begin{align}
\label{eq_CISofS1}
  \text{CIS}\big[\mathcal{S}_1\big]   = 2\mathcal{C}^0_1\cos(\fr\pi2 p_{01})\times {}_3\mathcal{F}_2\Bigg[\bgm 1,p_1+3,p_{01}+5\\ p_1+\fr72+\ii\wt\nu_1 ,p_1+\fr72-\ii\wt\nu_1\edm \Bigg| u_1 \Bigg]u_1^{p_1+4},
\end{align}
where $u_1=2K_1/(E_0+K_1)$. Next, we compute the homogeneous part by specifying $\ec_1= \pm1$:
\begin{align}
\label{eq_CutofS1}
  \mathop{\text{Cut}}_{K_1}\big[\mathcal{S}_1\big]   =&~ 2\mathcal{C}^{1}_1\cos\big[\fr{\pi}2(p_0+\fr32+\ii\wt\nu_1)\big]{}_2\mathcal{F}_1\Bigg[\bgm p_0+\fr52+\ii\wt\nu_1,\fr12+\ii\wt\nu_1 \\ 1+2\ii\wt\nu_1\edm\Bigg|u_1\Bigg]u_1^{3/2+\ii\wt\nu_1}
  +(\wt\nu_1\to-\wt\nu_1).
\end{align}
It is straightforward to check that our result agrees with known results in the literature \cite{Qin:2023ejc,Liu:2024xyi}. In practice, the one-pointed star graph $\mathcal{S}_1$ is most useful for computing a three-point function $\la\varphi_{\bm k_1}\varphi_{\bm k_2}\varphi_{\bm k_3}\ra'$ of inflaton fluctuation $\varphi$ with single massive exchange. In this case, we have $K_1=k_1$ and $E_0=k_2+k_3$ where $k_i=|\bm k_i|$ ($i=1,2,3$). In the interior of the physical region where all $|\bm k_i|$ are greater than zero and satisfy the triangle inequalities, we have $u_1=2k_1/k_{123}\in(0,1)$. Within this interval, the numerical implementation of (\ref{eq_S1}) is straightforward. A further folded limit $u_1\to 1$ is not quite trivial since it involves the cancellation of fake folded poles from both (\ref{eq_CISofS1}) and (\ref{eq_CutofS1}). This limit has been carefully analyzed in \cite{Qin:2023ejc}, and we refer readers to this work for further details. 

\paragraph{Double exchange} Next we consider the double-exchange graph $\mathcal{S}_2$. Taking $N=2$ in (\ref{eq_DegStarSol}), the whole graph can be written as the sum of 9 hypergeometric series, which can all be expressed in terms of (dressed) KdF functions ${}^{1+2}\mathcal{F}_{0+2}$ whose definition is collected in \eqref{eq_KdFfun2}:
\begin{align}
\label{eq_S2}
  \mathcal{S}_2(E_0;K_1,K_2) 
  =&\sum_{\ec_1,\ec_2= 0,\pm 1}{2\cos\big(\fr{\pi}2\Delta_{012}^{\ec}\big)}\mathcal{C}^{\ec}_1 \mathcal{C}^{\ec}_2 u_1^{\Delta^{\ec}_1}u_2^{\Delta^{\ec}_2}\n\\
  &~\times {}^{1+2}\mathcal{F}_{0+2}\Bigg[\bgm \Delta_{012}^{\ec}-3\\-\edm\Bigg|\bgm \Delta_1^\ec-1,1;\Delta_2^\ec-1 ,1\\ \Lambda_{1+}^\ec,\Lambda_{1-}^\ec; \Lambda_{2+}^\ec,\Lambda_{2-}^\ec  \edm\Bigg|u_1,u_2\Bigg],
\end{align} 
where $u_1=2K_1/(E_0+K_{12})$, $u_2=2K_2/(E_0+K_{12})$, and we define $\Lambda_{i\pm}^\ec\equiv\Delta^{\ec}_i+\Delta_i^\pm-2$ ($i=1,2$). The numerical evaluation of KdF functions is not straightforward unless $|u_{1,2}|\ll 1$, where the series representation (\ref{eq_DegStarSol}) itself provides a good numerical strategy. 

However, whether this strategy is applicable to the kinematic space of interest (say, the physical region) depends on specific processes under consideration. For instance, if we consider a four-point function $\la\varphi_{\bm k_1}\varphi_{\bm k_2}\varphi_{\bm k_3}\varphi_{\bm k_4}\ra'$ with two massive internal lines respectively mixing with $\varphi_{\bm k_1}$ and $\varphi_{\bm k_2}$, we have $K_1=k_1$, $K_2=k_2$, and $E_0=k_{34}$. Then, $|u_{1,2}|\ll 1$ can be easily satisfied in the physical region where $k_{1,2}\ll k_{3,4}$.

On the other hand, for a three-point function $\la\varphi_{\bm k_1}\varphi_{\bm k_2}\varphi_{\bm k_3}\ra'$ with two massive propagators mixing respectively with $\varphi_{\bm k_1}$ and $\varphi_{\bm k_2}$, we have $K_1=k_1$, $K_2=k_2$, and $E_0=k_3$. In this case, the physical region is constrained by the triangle inequalities of $k_{1,2,3}$, and has no overlap with the region of $|u_{1,2}|\ll 1$. Therefore, the numerical evaluation of this three-point function in its physical region requires a numerical strategy other than the series (\ref{eq_DegStarSol}). Fortunately, for this particular example, it is possible to recast (\ref{eq_S2}) into a partially resummed series which is numerically computable in a subspace of the physical region with $k_1\ll k_2\simeq k_3$ (or, equivalently, $|u_1|\ll 1$ and $u_2\lesssim 1$), thanks to the existing knowledge about the generalized hypergeometric function ${}_p\mathcal{F}_q$ \cite{Aoki:2024uyi}. We collect this partially resummed form in App.\;\ref{app_S2} for completeness.

\paragraph{Triple exchange} As a final example, we consider the three-pointed degenerate star $\mathcal{S}_3$. Taking $N=3$ in (\ref{eq_DegStarSol}), we can write all series there in terms of trivariate KdF functions \cite{srivastava1985multiple}, whose definition can be found in \eqref{eq_KdFfun3}:
\begin{align}
\label{eq_S3inKdF}
    &\mathcal{S}_3[E_0;K_1,K_2,K_3] 
    =   \sum_{\ec_1,\ec_2,\ec_3= 0,\pm 1} ~2\cos(\fr{\pi}2 \Delta_{0123}^{\ec})\mathcal{C}^{\ec}_1\mathcal{C}^{\ec}_2 \mathcal{C}^{\ec}_3 u_1^{\Delta^{\ec}_1}u_2^{\Delta^{\ec}_2}u_3^{\Delta^{\ec}_3}\n\\
  &\times {}^{1+2}\mathcal{F}_{0+2}\Bigg[\bgm \Delta_{0123}^{\ec}-3\\-\edm\Bigg|\bgm \Delta^{\ec}_1-1,1;\Delta^{\ec}_2-1,1; \Delta^{\ec}_3-1,1 \\ \Lambda_{1+}^{\ec},\Lambda_{1-}^{\ec}; \Lambda_{2+}^{\ec},\Lambda_{2-}^{\ec}; \Lambda_{3+}^{\ec},\Lambda_{3-}^{\ec}\edm\Bigg|u_1,u_2,u_3\Bigg],
\end{align} 
where $\Lambda_{i\pm}^\ec\equiv\Delta^{\ec}_i+\Delta_i^\pm-2$ ($i=1,2,3$) are as defined before. 
However, from the viewpoint of numerical evaluations, this expression is no more useful than the original series (\ref{eq_DegStarSol}) unless we understand the trivariate KdF function well enough, but unfortunately, we do not. Thus, numerically speaking, (\ref{eq_S3inKdF}) is physically useful only when $|u_{1,2,3}|\ll 1$, which is the case in, for instance, a five-point correlator $\la\varphi_{\bm k_1}\cdots\varphi_{\bm k_5}\ra'$ where the three internal massive propagators mix respectively with $\varphi_{\bm k_1}$, $\varphi_{\bm k_2}$, and $\varphi_{\bm k_3}$. For the case of the three-point function with triple massive exchanges, which is phenomenologically more interesting, we have $u_{123}=2$. We could set $u_1\ll 1$, but this then implies  $u_2\sim u_3\sim 1$. To our knowledge, there is no completely satisfactory solution to this problem. In the next section, we will look into the details of this problem and come up with a partial solution.

\section{Inflaton Bispectrum with Massive Exchanges}
\label{sec_bispectrum}

An interesting application of our result is the inflaton bispectrum (three-point function) with massive exchanges, especially the triple massive exchanges, as shown in Fig.\;\ref{fig_triple}. In this section, we apply the above results of degenerate stars and provide new expressions, both in complete hypergeometric form and simpler approximations in the squeezed limit, for inflaton bispectra with single, double, and triple massive exchanges. For simplicity, we assume all massive internal modes have the same mass. A generalization to unequal masses is trivial. 

The triple-exchange bispectrum is a well-motivated process from the original quasi-single-field inflation \cite{Chen:2009zp}. There have been extensive studies of this process from various perspectives. The model realization of this process is not unique. A possible locally Lorentz invariant model is provided by the following Lagrangian \cite{Wang:2019gbi}:\footnote{See also \cite{Kumar:2017ecc,Wu:2018lmx,Aoki:2024jha} for other related models.}
\bge
\label{eq_SQSFI}
  S=-\int\di^4x\,\sqrt{-g}\bigg\{\FR12\Big[(\pd_\mu\phi)^2+V(\phi)+(\pd_\mu\Sigma)^2+ m_0^2\Sigma^2\Big]+\FR{1}{4\Lambda^2}(\pd\phi)^2\Sigma^2+\FR{\lam}{24}\Sigma^4\bigg\}.
\ede
Here $V(\phi)$ is a slow-roll potential of the inflaton $\phi$ whose precise form is irrelevant, and $\Sigma$ is a real massive scalar spectator. Important is that the inflaton has a slow-roll background $\phi_0(t)\simeq \dot\phi_0 t+\text{const.}$, so that the dim-6 operator $(\pd\phi)^2\Sigma^2/\Lambda^2$ becomes a wrong-sign mass term for $\Sigma$. It is easy to get models where this wrong-sign mass squared $-\dot\phi_0^2/(2\Lambda^2)\gg m_0^2$ so that we can ignore the $m_0^2$-term from now on. Then, the wrong-sign mass term and $\Sigma^4$ term together generate a nonzero vev $\Sigma_0$ for the heavy field $\Sigma$. Thus the fluctuations $\si\equiv\Sigma-\Sigma_0$ and $\varphi\equiv\phi-\phi_0$ have the following interactions:
\begin{align}
\label{eq_QSFIcouplings}
  S\supset \int\di\tau\di^3\bm x\,\Big[a^3(\tau)\mu\varphi'\si-\FR12 \lam_1 a^2(\tau)\varphi'^2\si-\FR{1}{2}\lam_2a^3(\tau)\varphi'\si^2-\FR{1}{6}a^4(\tau)\lam_3\si^3\Big],
\end{align}
where the coupling strengths ($\mu$, $\lam_1$, $\lam_2$, $\lam_3$) are derived from (\ref{eq_SQSFI}) as $\mu= \dot\phi_0\Sigma_0/\Lambda^2$, $\lam_1=-\Sigma_0/\Lambda^2$, $\lam_2=-\dot\phi_0/\Lambda^2$, and $\lam_3=\lam\Sigma_0$, but we treat them as free parameters. Of course, there are other interactions, but the ones listed here already generate inflaton bispectrum with single, double, and triple exchanges, as shown in Fig.\;\ref{fig_triple}. It is possible to choose model parameters in (\ref{eq_SQSFI}) such that single-, double-, and triple-exchanges contribute similarly; Alternatively, one can also consider the parameter region where the triple-exchange graph dominates the CC signal as in the original work \cite{Chen:2009zp}. 

It is customary to express the inflaton three-point function in terms of a dimensionless \emph{shape function} $\mb{S}(k_1,k_2,k_3)$ which is invariant under a global scaling of all three momenta $k_i\to \lam k_i$ ($i=1,2,3$). The shape function is related to the inflaton correlator at the leading order in the perturbation theory via:
\bge
  \mb{S}(k_1,k_2,k_3)=-\FR{1}{2\pi P_\zeta^{1/2}}\FR{(k_1k_2k_3)^2}{H^3}\la\varphi_{\bm k_1}\varphi_{\bm k_2}\varphi_{\bm k_3}\ra',
\ede
where we have restored the Hubble parameter in this equation for clarity; $P_\zeta$ is the magnitude of the curvature power spectrum and is observed to be $P_\zeta \simeq 2\times 10^{-9}$ at the CMB scales \cite{Planck:2018jri}. 

\begin{figure}[t]
\centering
\includegraphics[width=0.85\textwidth]{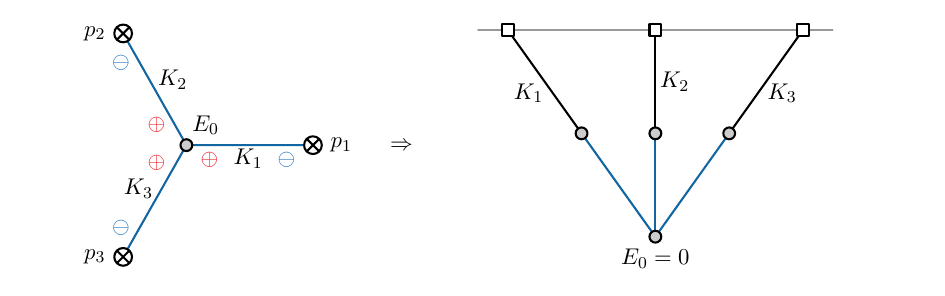} 
\caption{The degenerate star graph with 3 mixed propagators (left) and the inflaton correlator with triple massive exchanges (right). The right correlator is obtained from the left tree graph by setting $E_0=0$, $p_0=-4$, and $p_1=p_2=p_3=-2$. }
\label{fig_triple}
\end{figure}

The single- and double-exchange graphs are essentially solved problems both analytically and numerically; See Sec.\;\ref{sec_DegStarExamples}. We have little to add about these processes except a small comment: In the final inflaton bispectrum, we need to include all permutations of momenta. For instance, the bispectrum from the single-exchange graph has the following form:
\begin{align}
  \mb{S}_{\si^1} \propto (k_1k_2k_3)^2\bigg[\FR{\mathcal{S}_1(k_{123};k_1)}{k_1^3k_2k_3k_{123}}+\text{2 perms} \bigg] \propto  \FR{u_2u_3}{u_1}\mathcal{S}_1(u_1)+\text{2 perms} .
\end{align}
As a result, if we look at the squeezed limit $k_1\ll k_2\simeq k_3$, or equivalently, $u_1\ll 1$ and $u_2\simeq u_3\simeq 1$, we have:  
\bge
  \lim_{k_1\ll k_2} \mb{S}_\si\propto u_1^{-1}\mathcal{S}_1(u_1)+2u_1\mathcal{S}_1(1).
\ede
Therefore, to get the correct squeezed limit including a nonoscillatory background, we need to evaluate the degenerate star at its folded limit $\mathcal{S}_N(1)$, which is not entirely trivial. We will come back to this point later when we present the more explicit squeezed-limit results.

On the other hand, the triple-exchange diagram is more challenging. Its numerical evaluation was successfully done in the original work on QSFI \cite{Chen:2009zp} by a direct implementation of 4-layer time integrals in the in-in formalism. Since then, there have been many attempts to simplify or to generalize the computation. Notable progress includes the use of analytical mixed propagators in \cite{Chen:2017ryl}, the numerical implementation of mixed propagator by solving the coupled mode equations \cite{Chen:2015dga,An:2017hlx}, as well as numerically solving the kinematic differential equations for the correlator \cite{Werth:2023pfl,Pinol:2023oux,Werth:2024aui}.  

Despite all this progress, an analytical expression for this graph turns out to be notoriously hard to get. With our results for arbitrary degenerate star graphs, it appears straightforward to get the result for the triple-exchange graph, as illustrated in Fig.\;\ref{fig_triple}. Using the notation for the $N$-pointed degenerate star in Sec.\;\ref{sec_star}, we have: 
\begin{align}
   \la\varphi_{\bm k_1}\varphi_{\bm k_2}\varphi_{\bm k_3}\ra'_{\si^3}
   =&~\FR{\mu^3 \lam_3 k_{123}^3}{512 k_1^3 k_2^3 k_3^3} \mathcal{S}_3(0;k_1,k_2,k_3)\big|_{p_0=-4,p_1=p_2=p_3=-2}\n\\
   =&~ \FR{\mu^3 \lam_3 k_{123}^3}{512 k_1^3 k_2^3 k_3^3}\sum_{\ec_1,\ec_2,\ec_3= 0,\pm 1}  {2\cos[\fr{\pi}2\Delta_{123}^\ec]}\mathcal{C}^{\ec}_1\mathcal{C}^{\ec}_2 \mathcal{C}^{\ec}_3  u_1^{\Delta_1^{\ec}}u_2^{\Delta^{\ec}_2}u_3^{\Delta^{\ec}_3}\n\\
  &\times {}^{1+2}\mathcal{F}_{0+2}\left[\bgm \Delta_{123}^\ec-3\\-\edm\middle|\bgm \Delta^{\ec}_1-1,1;\Delta^{\ec}_2-1,1; \Delta^{\ec}_3-1,1\\ \Lambda_{1+}^{\ec},\Lambda_{1-}^{\ec}; \Lambda_{2+}^{\ec},\Lambda_{2-}^{\ec}; \Lambda_{3+}^{\ec},\Lambda_{3-}^{\ec}\edm\Bigg|u_1,u_2,u_3\right],
\end{align}
where $u_i = {2k_i}/{k_{123}}$. Also, we should set $\wt\nu_1=\wt\nu_2=\wt\nu_3=\wt\nu$ to be the mass of the $\si$ field, as well as $p_0=-4$ and $p_1=p_2=p_3=-2$. Consequently, we should set $\Delta_i^\pm=\fr32\pm\ii\wt\nu\equiv\Delta^\pm$ and $\Delta_i^0=2\equiv\Delta^0$ for all $i=1,2,3$, and $\Delta_0^0=0$.

Although we are able to express the three-point function in terms of hypergeometric functions, it is not trivial to evaluate them since the arguments are constrained by $u_{123}=2$. We can try to proceed like the case of $\mathcal{S}_2$, expanding the result in small $u_1$ (namely, small $k_1$), which is physically interesting because the cosmological collider signal is dominant in this squeezed limit:
\begin{align}
    \label{eq_triExAna}
    \mb{S}_{\si^3}
 =&-\frac{\mu^3 \lam_3/H^4}{128\pi P_\zeta^{1/2}}\FR{1}{u_1u_2u_3}\mathcal{S}_3(0;k_1,k_2,k_3)\big|_{p_0=-4,p_1=p_2=p_3=-2}\n\\
 =&-\frac{\mu^3 \lam_3/H^4}{128\pi P_\zeta^{1/2}}\FR{1}{u_1u_2u_3}\sum_{\ec=-1}^{+1}\sum_{m=0}^\infty \ii^{m}u_1^{m+\Delta^{\ec}_1}\mathcal{C}^{\ec}_1\Gamma\bgb m+\Delta^{\ec}_1-1 \\ m+\Lambda_1^+,m+\Lambda_1^-\edb \mathcal{B}^{\ec}_{m}(u_2,u_3)+\text{c.c.},
\end{align}
where we introduced the function $\mathcal{B}^{\ec}_{m}(u,v)$, defined by the following series when it is convergent and by analytical continuation outside the convergence domain:
\begin{align}
\label{eq_Bfunction}
    \mathcal{B}^{\ec}_{m}(u,v) \equiv &~\sum_{\ec_2,\ec_3= -1}^{+1}e^{-\ii\pi(m+\Delta^{\ec}_{123})/2}\mathcal{C}^{\ec}_2 \mathcal{C}^{\ec}_3 u_2^{\Delta^{\ec}_2}u_3^{\Delta^{\ec}_3}\n\\
  &~\times {}^{1+2}\mathcal{F}_{0+2}\Bigg[\bgm m+\Delta_{123}^\ec-3\\-\edm\Bigg|\bgm \Delta^{\ec}_2-1,1; \Delta^{\ec}_3-1,1\\ \Lambda_{2+}^{\ec},\Lambda_{2-}^{\ec}; \Lambda_{3+}^{\ec},\Lambda_{3-}^{\ec} \edm\Bigg|u,v\Bigg],
\end{align}
where ${}^{1+2}\mathcal{F}_{0+2}$ is a dressed KdF function of two variables, whose definition can be found in \eqref{eq_KdFfun2}. Clearly, the physically interesting region $u_2\simeq u_3\simeq 1$ is outside the convergence domain of the above series and therefore the series is useless for numerical computation of the $\mathcal{B}^{\ec}_{m}$ function. Fortunately, the function $\mathcal{B}^{\ec}_{m}$ can be numerically evaluated with high efficiency by performing a Wick-rotated time integral \cite{Chen:2017ryl}, which is motivated by the fact that it is related to the 2-pointed degenerate star via $\mathcal{B}^{\ec}_{m}(u,v) = \mathcal{S}^+_2[K_1;K_2,K_3] |_{p_0\to -4+m+\Delta^{\ec}}$. As a result, we can numerically compute the following integral:
\begin{align}
    \mathcal{B}^{\ec}_{m}(u,v)
    =&~64 \int_0^{\infty} \mathrm{d}z ~(-2\ii z)^{-4+m+\Delta^{\ec}}e^{-(2-u-v)z} \mathscr{D}_+(\ii uz)\mathscr{D}_+(\ii vz).
\end{align}
Here $\mathscr{D}_+(z)$ is a dimensionless mixed propagator as introduced in \cite{Chen:2017ryl}:
\begin{align}
    \mathscr{D}_+(z) = \frac{-1}{2} \sum_{\aa=\pm}\int_{-\infty}^0 \frac{\di z'}{(-z')^2} \ii \aa e^{\ii \aa z'}\wt D_{+\aa}(z,z').
\end{align} 
The mixed propagator has an analytic expression \cite{Chen:2017ryl}, as follows:
\begin{align}
    \mathscr{D}_+(z) =&~ \frac{\pi}{8}z^{3/2} e^{-\pi\wt\nu}\Big\{\big[\sqrt{2\pi}e^{\ii\pi(1/4-\ii\wt\nu/2)}\text{sech}(\pi\wt\nu)-\ii(\coth(\pi\wt\nu)+1){F}_{\wt\nu}(z)+\ii\text{csch}(\pi\wt\nu){F}_{-\wt\nu}(z)\big]\mathrm{H}^{(2)}_{-\ii\wt\nu}(z)\n\\
    &+\big[\sqrt{2\pi}e^{-\ii\pi(1/4+\ii\wt\nu/2)}\text{sech}(\pi\wt\nu)+\ii(\coth(\pi\wt\nu)+1){F}_{-\wt\nu}(z)-\ii\text{csch}(\pi\wt\nu){F}_{\wt\nu}(z)\big]\mathrm{H}^{(1)}_{\ii\wt\nu}(z)\Big\},
\end{align}
where ${F}_{\wt\nu}(z)$ is defined by:
\begin{align}
    {F}_{\wt\nu}(z) = \frac{z^{\ii\wt\nu+1/2}}{2^{\ii\nu}(\ii\wt\nu+1/2)\Gamma(\ii\wt\nu+1)}\,{}_2\text{F}_2\Bigg[\bgm \fr12+\ii\wt\nu,\fr12+\ii\wt\nu\\\fr32+\ii\wt\nu,1+2\ii\wt\nu \edm\Bigg|-2\ii z\Bigg].
\end{align}

\subsection{Squeezed limit}
In CC physics, the oscillatory signal is most prominent in the squeezed limit $k_1\ll k_2\simeq k_3$, and thus it would be desirable to have a simpler characterization of the inflaton bispectrum in this limit, for both the oscillatory signal and the smooth background. Below, we provide the leading order results in the squeezed limit for the triple-exchange bispectrum, together with the single-exchange and double-exchange bispectra. All three cases feature similar behavior in the squeezed limit, including a smooth background and an oscillatory signal. The scaling behaviors of the background and the signal as functions of $k_1/k_2$ are the same for all three cases. However, one point we would like to highlight is that the relative sizes of the signal and the background differ for the three cases due to the different structure of channel permutations. Therefore, barring other contributions to the background which could be subdominant as we are assuming now, a comparison between the size of the signal and the background can tell the difference among the three processes even in the squeezed limit.\footnote{These are the information we can extract from the squeezed limit. When moving to not-too-squeezed limit, we can try to tell the difference by utilizing the phase information \cite{Qin:2022lva}. Also, in a given model such as (\ref{eq_QSFIcouplings}),   more than one of the three processes could contribute at the same time, but the couplings of each process are related to each other as they are all derived from (\ref{eq_SQSFI}). In this case, the relative sizes of the signal and the background are also useful observables to test the model. We leave more discussions on these interesting phenomenological questions for future work. }

\begin{figure}[t]
\centering
\includegraphics[width=0.95\textwidth]{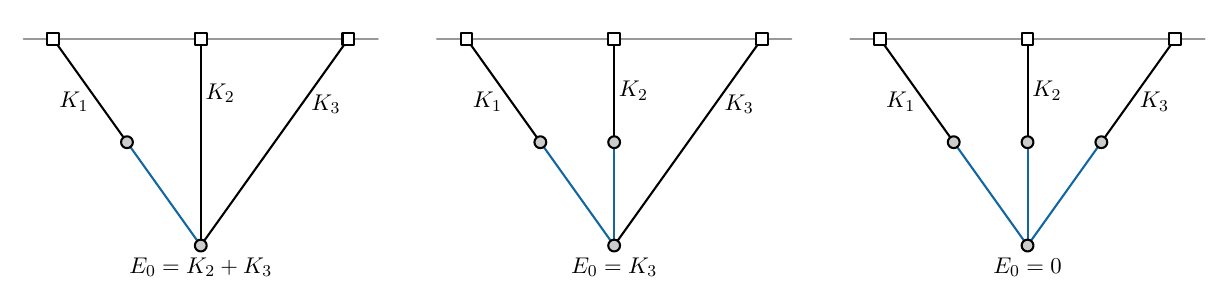} 
\caption{The inflaton bispectrum with single, double, and triple exchanges of a massive scalar $\si$. }
\label{fig_bispectrum}
\end{figure}

For definiteness, we consider the three processes of exchanging a single species of massive scalar $\si$ with interactions in (\ref{eq_QSFIcouplings}), as shown in Fig.\;\ref{fig_bispectrum}. Then, denoting the shape functions of the single and double exchanges by $\mb{S}_{\si^1}$ and $\mb{S}_{\si^2}$ respectively, we have:
\begin{align}
    \mb{S}_{\si^1}=&-\frac{\mu \lambda_1/H^2 }{128\pi P^{1/2}_{\zeta}} \frac{u_2 u_3}{u_1} \mathcal{S}_1(k_{23};k_1)\big|_{p_0=0,p_1=-2} + \text{2 perms},\\
    \mb{S}_{\si^2}=&-\frac{\mu^2 \lambda_2/H^3 }{128\pi P^{1/2}_{\zeta}} \frac{ u_3}{u_1 u_2} \mathcal{S}_2(k_{3};k_1,k_2)\big|_{p_0=p_1=p_2=-2} + \text{2 perms},
\end{align}
while the shape function for the triple exchange $\mb{S}_{\si^3}$ has been given in (\ref{eq_triExAna}). 

In the squeezed limit, all three shapes $\mb{S}_{\si^i}$ $(i=1,2,3)$ can be put into a unified form:
\begin{align}
    \lim_{k_1 \ll k_2\simeq k_3}\mb{S}_{\si^i} = &~\frac{\mu^i \lambda_i/H^{i+1} }{32\pi P^{1/2}_{\zeta}} \bigg[ C^{(i)}_{\text{B}}(\wt\nu)\FR{k_1}{k_2}+ C^{(i)}_{\text{S}+}(\wt\nu)\Big(\FR{k_1}{4k_2}\Big)^{1/2+\ii\wt\nu}+C^{(i)}_{\text{S}+}(-\wt\nu)\Big(\FR{k_1}{4k_2}\Big)^{1/2-\ii\wt\nu} \bigg]\n\\
 =&~\frac{\mu^i \lambda_i/H^{i+1} }{32\pi P^{1/2}_{\zeta}} \bigg\{ C^{(i)}_\text{B}(\wt\nu)\FR{k_1}{k_2}+C^{(i)}_\text{S}(\wt\nu)\Big(\FR{k_1}{k_2}\Big)^{1/2}\cos\Big[\wt\nu\log\FR{k_1}{4k_2}+\vartheta^{(i)}(\wt\nu)\Big] \bigg\},
\end{align}
Here the real coefficient $C^{(i)}_\text{B}(\wt\nu)$ gives the size of the smooth background as a function of the mass $\wt\nu$, and the complex coefficients $C_{\text{S}+}^{(i)}(\wt\nu)$ characterize the oscillatory signal which can be further decomposed into the amplitudes $C_\text{S}^{(i)}(\wt\nu)\equiv|C_{\text{S}+}^{(i)}(\wt\nu)|$ and the phases $\vartheta^{(i)}(\wt\nu)\equiv\,\text{Arg}\,C_{\text{S}+}^{(i)}(\wt\nu)$. For the three cases considered here, their explicit expressions are:
\begin{align}
    &C_\text{B}^{(1)}(\wt{\nu})\equiv \frac{1}{\fr14+\wt\nu^2}-\fr12\text{Re}[\mathcal{A}_0(\wt\nu)],&&C_{\text{S+}}^{(1)}(\wt{\nu}) \equiv 2f(\wt\nu)\cos\big[\fr\pi4(3+2\ii\wt\nu)\big]\Gamma\big(\fr52+\ii\wt\nu\big) ;\\
    &C_\text{B}^{(2)}(\wt{\nu})\equiv -\frac{\fr12\mathcal{A}_{0}(\wt\nu) }{\fr14+\wt\nu^2}-\fr12\Re[\mathcal{B}^0_0(1,1)],&&C_{\text{S+}}^{(2)}(\wt{\nu}) \equiv 2f(\wt\nu)\mathcal{A}_{\ii\wt\nu-1/2}(\wt\nu);\\
    &C_\text{B}^{(3)}(\wt\nu)\equiv- \FR{\fr12\Re[\mathcal{B}^0_0(1,1)]}{\fr14+\wt\nu^2},
 &&C_\text{S+}^{(3)}(\wt\nu)\equiv f(\wt\nu)\big[\mathcal{B}^+_{0}(1,1)+\mathcal{B}^{-*}_{0}(1,1)\big].
\end{align}
Here the functions $f(\wt\nu)$ and $\mathcal{A}_p$ are defined by:
\begin{align}
    f(\wt\nu)\equiv&~ \frac{\pi^{3/2}\cos[\frac\pi 2(\ii \wt\nu-\frac12)]}{\sin(2\ii\pi\wt\nu)\Gamma[1+\ii\wt\nu]},\\
 \label{eq_doubleFoldedLimitSingle}\mathcal{A}_{p}(\wt\nu)=&~ \frac{2\ii \cos\big[\fr{\pi}{2}\big(p\!+\!\ii\wt\nu\!+\!\fr32\big)\big]\cos\big[\fr{\pi}{2}\big(\ii\wt\nu\!-\!\fr12\big)\big]}{\sinh(\pi\wt\nu)} \Gamma(p+3)\n\\
    &\times{}_3\mathcal{F}_2\Bigg[\bgm \fr12+\ii\wt\nu,\fr12+\ii\wt\nu,p+\fr52+\ii\wt\nu\\ 1+2\ii\wt\nu,p+\fr72+\ii\wt\nu\edm \Bigg| 1 \Bigg]+(\wt\nu\to-\wt\nu), 
\end{align}
and $\mathcal{B}_0^\ec(u,v)$ is defined in (\ref{eq_Bfunction}). 

\begin{figure}[t]
\centering
\includegraphics[width=0.45\textwidth]{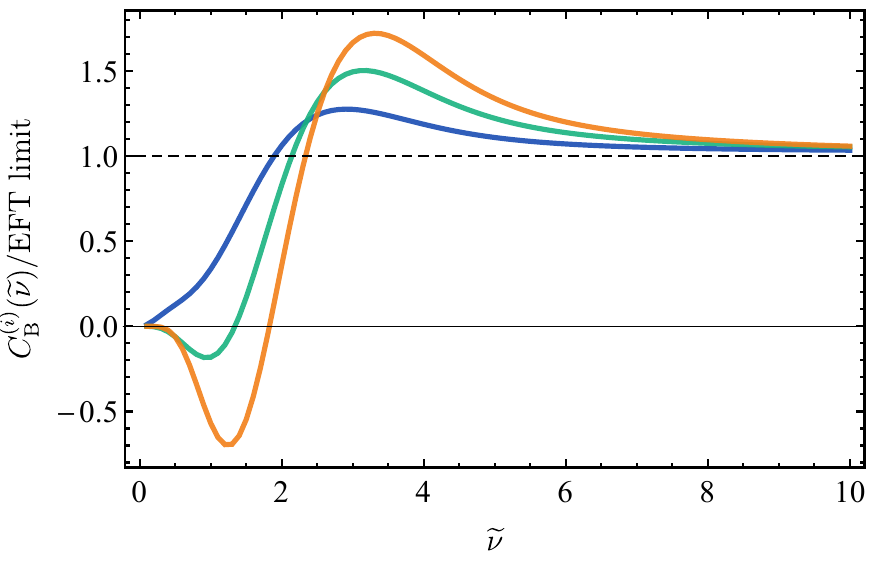}~~~~~
\includegraphics[width=0.45\textwidth]{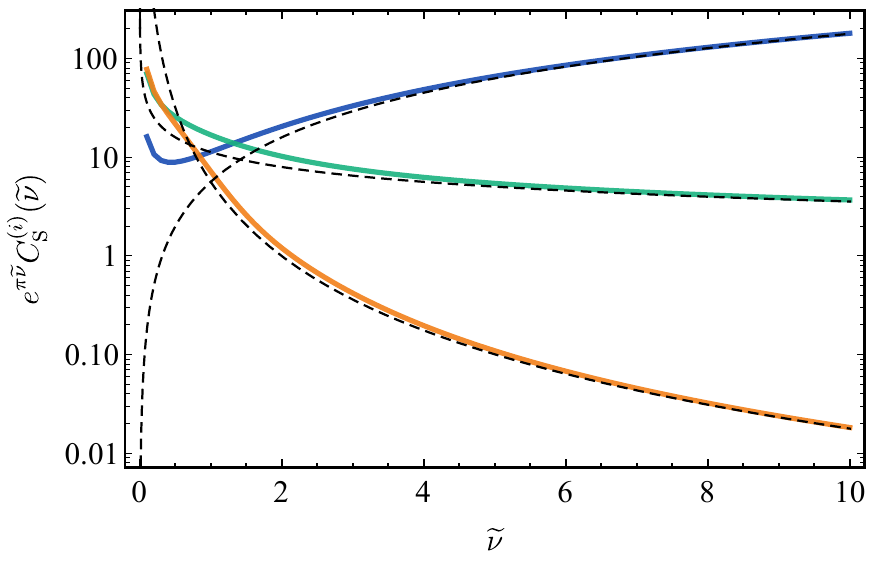} 
\includegraphics[width=0.45\textwidth]{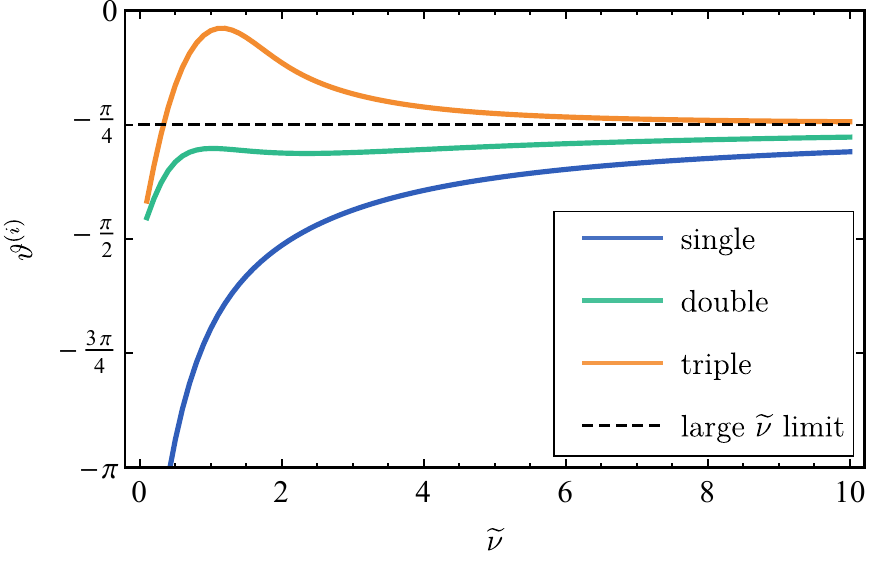} 
\caption{The size of the nonoscillatory backgrounds $C_\text{B}^{(i)}$, the size $C_\text{S}^{(i)}$ and the phase $\vartheta^{(i)}$ of the oscillatory signals as functions of the mass $\wt\nu$ of the intermediate states in the squeezed limit of the shape functions of inflaton bispectra $\mb S_{\si^i}$ with single, double, and triple massive exchanges. The black dashed curves show the results in the large-mass limit in (\ref{eq_CLargeMass1})-(\ref{eq_CLargeMass3}). The backgrounds $C_\text{B}^{(i)}$ are normalized to their large-mass values, with the three curves showing $C_\text{B}^{(1)}/(3\wt\nu^{-2})$, $C_\text{B}^{(2)}/(3\wt\nu^{-4})$, and $C_\text{B}^{(3)}/\wt\nu^{-6}$, respectively. Also, the common Boltzmann factor $e^{-\pi\wt\nu}$ is removed from the there signal sizes $C_\text{S}^{(i)}$.}
\label{fig_SignalPar}
\end{figure}

It is also useful to look at the large mass limits of these functions. For $\wt\nu\gg 1$, they become:
\begin{align}
\label{eq_CLargeMass1}
    &C_\text{B}^{(1)}(\wt{\nu})\sim 3\wt\nu^{-2},&&C_{\text{S}}^{(1)}(\wt{\nu}) \sim  \pi^{3/2}\wt\nu^{3/2}e^{-\pi\wt\nu},\\
\label{eq_CLargeMass2}
    &C_\text{B}^{(2)}(\wt{\nu})\sim 3\wt\nu^{-4},&&C_{\text{S}}^{(2)}(\wt{\nu}) \sim 2\pi^{3/2}\wt\nu^{-1/2}e^{-\pi\wt\nu},\\
\label{eq_CLargeMass3}
    &C_\text{B}^{(3)}(\wt\nu)\sim \wt\nu^{-6},
 &&C_\text{S}^{(3)}(\wt\nu)\sim  \pi^{3/2}\wt\nu^{-5/2}e^{-\pi\wt\nu},
\end{align}
and $\vartheta^{(i)}\sim -\pi/4$ for all $i=1,2,3$. We show the background coefficients $C_\text{B}^{(i)}$, the signal amplitudes $C_\text{S}^{(i)}$, and the signal phases $\vartheta^{(i)}$ together with their large $\wt\nu$ limits in Fig.\;\ref{fig_SignalPar}. From the figure we can observe an interesting feature of the background coefficients:   $C_\text{B}^{(2)}(\wt\nu)$ changes sign at $\wt\nu\simeq 1.3$ and so does $C_\text{B}^{(3)}(\wt\nu)$ at $\wt\nu\simeq 1.8$, while $C_\text{B}^{(1)}$ remains positive for all $\wt\nu>0$. So there are values of $\wt\nu$ for double and triple exchange processes where the background is accidentally suppressed relative to the signal in the squeezed limit. Of course, this does mean that the background vanishes altogether since $C_\text{B}^{(i)}$ corresponds only to the leading-order behavior in the squeezed limit. Still, there could be interesting implications for phenomenology that merit further studies. 

Finally, from the above results, we see that the three shape functions in the squeezed and large mass limits have the following simple analytical expressions:
\begin{align}
  \lim_{\wt\nu\gg 1}\lim_{k_1\ll k_2}\mb{S}_{\si^1}
  \simeq&~ \FR{\mu\lam_1/H^2}{32\pi P_\zeta^{1/2}\wt\nu^2}\bigg\{3\FR{k_1}{k_2}+\pi^{3/2} e^{-\pi\wt\nu}\wt\nu^{7/2}\Big(\FR{k_1}{k_2}\Big)^{1/2}\cos\Big[\wt\nu\log\FR{k_1}{4k_2}-\FR{\pi}4\Big]\bigg\},\\
  \lim_{\wt\nu\gg 1}\lim_{k_1\ll k_2}\mb{S}_{\si^2}
  \simeq&~ \FR{\mu^2\lam_2/H^3}{32\pi P_\zeta^{1/2}\wt\nu^4}\bigg\{3\FR{k_1}{k_2}+2\pi^{3/2} e^{-\pi\wt\nu}\wt\nu^{7/2}\Big(\FR{k_1}{k_2}\Big)^{1/2}\cos\Big[\wt\nu\log\FR{k_1}{4k_2}-\FR{\pi}4\Big]\bigg\},\\
  \lim_{\wt\nu\gg 1}\lim_{k_1\ll k_2}\mb{S}_{\si^3}
  \simeq&~ \FR{\mu^3\lam_3/H^4}{32\pi P_\zeta^{1/2}\wt\nu^6}\bigg\{\FR{k_1}{k_2}+\pi^{3/2} e^{-\pi\wt\nu}\wt\nu^{7/2}\Big(\FR{k_1}{k_2}\Big)^{1/2}\cos\Big[\wt\nu\log\FR{k_1}{4k_2}-\FR{\pi}4\Big]\bigg\}.
\end{align}
As expected, the background parts all scale like an equilateral shape $\sim k_1/k_2$ with the coefficients $\sim \wt\nu^{-2i}$ $(i=1,2,3)$ in agreement with the EFT results. (The coefficients of 3 in single and double exchanges are from the channel permutations.) On the other hand, all signals scale as $\sqrt{k_1/k_2}$ in the squeezed limit and have a uniform phase $-\pi/4$ in the large mass limit. Their coefficients also feature a uniform dependence on the mass $\sim e^{-\pi\wt\nu}\wt\nu^{7/2}$ in which we have a familiar Boltzmann suppression factor and also a power-law factor whose exponent $7/2$ is determined in all cases. We note that even in the squeezed limit, the relative sizes between backgrounds and the signals differ in three cases, being $3:1$, $3:2$, and $1:1$ for single, double, and triple exchanges. The difference arises purely from channel permutations and may provide useful information in phenomenological studies, as discussed earlier.

\section{Conclusion and Outlook}
\label{sec_conclusion}

The analytical studies of massive inflation correlators have revealed their rich and interesting underlying structures, which have certainly enriched our understanding of QFT amplitudes in cosmological spacetimes. Meanwhile, there is an intriguing interplay between this analytical program of correlators and high-energy particle phenomenology, which has received great boosts through the program of cosmological collider physics. The interplay is nicely illustrated by the degenerate kinematics from the bilinear mixing between the massless inflaton mode and a massive field, which is the central topic of this work. 

To study degenerate kinematics, we proposed to use a new class of kinematic variables. After adopting this change of variables, the result gives us two small but nice surprises: First, when we use a series ansatz to solve the PDE system, we get first-order recursion relations, making it much easier to find a general term formula in terms of hypergeometric functions. Second, when we take the folded limit to reach bilinear mixings, the result features a reduction of transcendental weight. These are surprises to us since we are not mathematically sophisticated enough to elucidate the underlying reasons. Nevertheless, we can exploit these facts and find significantly simpler expressions for graphs with bilinear mixings, including the classic examples of degenerate stars that are of great importance for CC phenomenology. 

There are obvious applications of our new variables, including graphs with spins and chemical potentials. They certainly deserve future explorations, but we do not have much to say about them here. Instead, we want to finish this work with a discussion on a ``tension'' between the transcendental weight of a graph and its degenerate kinematics. 

It may be useful to note that the reduction of transcendental weight is not automatic in two senses: A trivial one is that it does not show up with the ``old variables'' of $r_{(\al i)}=K_\al/E_i$; Less trivial is that, even with the new variables, the weight reduction does not happen for \emph{every} degenerate limit. We have examples for the latter: Consider the two-site chain $\G'(E_1,E_2,K)$, which is a weight-2 object, namely, with two-layer summations. When taking a single folded limit $E_2\to K$, we get a weight-1 result as elaborated in this work. However, when we take one more folded limit $E_{1,2}\to K$, the result is a number independent of any kinematic variable but is still a weight-1 series evaluated at a particular kinematic point. Another example is the 3-point function with triple massive exchange. It can be thought of as a particular limit of the 4-site star graph $\G'(E_0,E_1,E_2,E_3,K_1,K_2,K_3)$ which is a weight-6 object. In this work, we have shown that taking three folded limits $E_i\to K_i$ ($i=1,2,3$) all reduce the weight, leading to the 3-pointed degenerate star $\mathcal{S}_3$ which is weight-3. However, it seems that the weight does not reduce further when we take one more limit $E_0\to 0$. Currently, we are not aware of any new variables that could achieve this reduction, and we suspect that the $E_0\to 0$ limit does not reduce weight whatsoever. In our opinion, this may be the main reason for the difficulty of computing the triple-exchange bispectrum that people have considered for many years.

At a more conjectural level, the above observation also applies to loop graphs: From known examples, loop graphs are also represented by hypergeometric series. In this regard, increasing the number of loops would increase the weight but not the number of independent variables. Thus, a loop graph can also be thought of as a graph with degenerate kinematics. The absence of weight reduction for these ``degenerate kinematics'' may be a reason for the difficulty of loops as well.

Certainly, it remains a very interesting problem to find a series expression for the triple-exchange bispectrum that converges in at least part of the physical region. On the other hand, for tree graphs with degenerate kinematics, we have good numerical strategies to find their values quite efficiently. Thus, finding such a series expression seems to be more of a theoretical curiosity than a practical necessity. On the other hand, loops remain formidable even numerically. Therefore, we expect that our methods here may be of some use in tackling those more challenging problems. 

\paragraph{Acknowledgments} We thank Bingchu Fan, Soubhik Kumar, Haoyuan Liu, Qianshu Lu, Zhehan Qin, Jiayi Wu, Hongyu Zhang, and Yisong Zhang for useful discussions. Partial results of this work have been presented at the workshop ``Cosmology Beyond the Analytic Lamppost (CoBALt)'' at the Institut Pascal at Université Paris-Saclay, and we thank the organizers and participants of the workshop for useful discussions and feedback. This work is supported by NSFC under Grants No.\ 12275146 and No.\ 12247103, the National Key R\&D Program of China (2021YFC2203100), and the Dushi Program of Tsinghua University.

\newpage
\begin{appendix}

\section{Generic Massive Trees: A Summary} 
\label{app_masstree}

In this appendix, we summarize the key results of \cite{Liu:2024str}, which form the basis of this work. 

\paragraph{Dimensionless tree graphs}
As in \cite{Liu:2024str}, we consider the most general massive tree-graph contribution to the  $N$-point correlation function of a conformal scalar $\varphi$ with mass $m^2=2$, in which all external lines (bulk-to-boundary propagators) propagate the conformal scalar $\varphi$, and internal lines (bulk propagators) exchange principal scalars with mass parameter $\wt\nu_\al^2=m^2_\al-9/4>0$. We also call $\wt\nu$ the mass when no confusion arises. Correlators involving external massless scalar and/or derivative couplings can be trivially reduced to graphs considered here, by appropriately adjusting parameters such as $p_i$ (introduced below) and $\wt\nu_\al$.

For a tree graph $\wh\G$ contribution to an inflation correlator with $N$ external lines,  $I$ internal lines, and $V=I+1$ vertices, it proves convenient to rescale $\wh\G$ in the following way:
\bge
\label{eq_Grescale}
  \wh{\mathcal{G}}(\bm k_1,\cdots,\bm k_N)=\bigg(\prod_{n=1}^N\FR{-\tau_f}{2k_n}\bigg)\bigg(\prod_{i=1}^V\FR{1}{E_{i}^{1+p_i}}\bigg)\bigg(\prod_{\al=1}^I \FR{1}{K_\al^3}\bigg) \mathcal{G}(\{E\},\{K\}),
\ede
where $\bm k_1,\cdots, \bm k_N$ are 3-momenta for $N$ external lines, $k_i\equiv|\bm k_i|$ $(i=1,\cdots,N)$. On the right hand side, $\tau_f\to 0$ is a late-time cutoff for the external lines; $E_i$ is the magnitude sum of the momenta of all external lines attached to Vertex $i$ and is called the \emph{vertex energy}; $K_\al$ is the magnitude of the 3-momentum flowing in (internal) Line $\al$, and is called the \emph{line energy}; Finally, $p_i$ is called the \emph{twist} of Vertex $i$, which captures the time dependence of the coupling (see below). The convenience of the above rescaling is that the resulting dimensionless graph $\G(\{E\},\{K\})$ is a function of vertex and line energies only. Using the diagrammatic rule in Schwinger-Keldysh formalism \cite{Chen:2017ryl}, we can write down an integral representation for $\G$:
\begin{align}
\label{eq_dimlessG}
  \G (\{E\},\{K\})=&\sum_{\aa_1,\cdots,\aa_V=\pm}\int_{-\infty}^0\prod_{i=1}^V\Big[\di z_i\, \ii\aa_i (-z_i)^{p_i}e^{\ii \aa_i z_i}\Big]\prod_{\al=1}^I \wt{D}_{\aa_{i}\aa_{j}}^{(\wt\nu_\al)}(r_{(\al i)}z_{i},r_{(\al j)}z_{j}),
\end{align}
where $z_j$ is the integral variable of Vertex $j$ and is related to the conformal time $\tau_j$ via $z_j=-E_j\tau_j$. Importantly, there are $2I$ independent energy ratios $r_{(\al i)}\equiv K_\al/E_i$ which fully specify the kinematics of a tree graph. So, in summary, a dimensionless tree graph is specified by assigning a twist $p_i$ to each vertex, a mass $\wt\nu_\al$ to each internal line, and $2I$ energy ratios $r_{(\al i)}$. As in \cite{Liu:2024str}, we assume that all integrals are infrared finite, namely, they are convergent in the upper limit. (The convergence in the lower limit is guaranteed in the physical region by $\ii\ep$-prescription.) 
 
\paragraph{Differential equations} In \cite{Liu:2024str}, a set of differential equations was found for the dimensionless graph $\G$ in (\ref{eq_dimlessG}). For each energy ratio $r_{(\al i)}$, we have an equation: 
\begin{align} 
\label{eq_generalDE}
  &\bigg[ \Big(\vartheta_{(\al i)}-\FR{3}{2}\Big)^2+\wt\nu_\al^2-r_{(\al i)}^2\big(\vartheta_{\{i\}}+p_i+2\big)\big(\vartheta_{\{i\}}+p_i+1\big)\bigg]\G=\FR{r_{(\al i)}^{p_j+4}r_{(\al j)}^{p_i+4}}{\big[r_{(\al i)}+r_{(\al j)}\big]^{p_{ij}+5}}\mathsf{C}_{\al}[\G].
\end{align} 
Here, $\vt_{(\al i)}\equiv r_{(\al i)}\pd_{r_{(\al i)}}$ is called the Euler operator of $r_{(\al i)}$, and $\vt_{\{i\}}$ is the sum of Euler operators of energy ratios $r_{(\al i)}$ with $E_i$ in the denominator. For instance, if a vertex with vertex energy $E_i$ is attached to four internal lines with energies $K_1,\cdots,K_4$, then $\vt_{\{i\}}=\vt_{(1i)}+\vt_{(2i)}+\vt_{(3i)}+\vt_{(4i)}$. On the right hand side, we have the \emph{contraction} of $\G$ over Line $\al$, denoted by $\mathsf{C}_{\al}[\G]$. The contraction is defined by removing Line $\al$ in $\G$, pinching its two endpoints, say Vertex $i$ and Vertex $j$, into a new vertex, and assigning the new vertex an energy $E_i+E_j$ and a twist $p_i+p_j+4$. Altogether, we have $2I$ coupled second-order PDEs for $\G$, which is a function of $2I$ independent variables. We can uniquely fix $\G$ by solving these equations with appropriate boundary conditions. See \cite{Liu:2024str} for details. 

\paragraph{Solutions} The full solution for $\G$ can be expressed as a completely inhomogeneous solution (CIS) of (\ref{eq_generalDE}), denoted as $\text{CIS}\,[\G]$, plus all of its cuts: 
\bge
\label{eq_fullsolutionG}
  \G=\text{CIS}\,\big[\G\big]+\sum_\al \mathop\text{Cut}_{K_\al} \big[\G\big]+\sum_{\al\neq\be} \mathop\text{Cut}_{K_\al,K_\be} \big[\G\big]+\cdots+\mathop\text{Cut}_{\text{all~}K} \big[\G\big].
\ede
It is important that the separation of $\G$ into a CIS and cuts depends crucially on the relative sizes (orderings) of energies $E_i$ and $K_\al$. The following expressions work for all $K_\al$ smaller than all $E_i$ (This condition can often be loosened.) 

Also, we call the largest vertex energy $E_1$. Once the vertex of the largest vertex energy is decided, a tree graph is naturally endowed with a family-tree structure, in which $E_1$ is the ancestor site, and any line acquires a direction flowing from the mother site to the daughter site. 

Then, the CIS is the part of the correlator that is not annihilated by any single differential operators on the left-hand side of (\ref{eq_generalDE}) and is fully analytic $E_1$ as $E_1\to \infty$ except for an overall factor $E_1^{-p_{1\cdots V}}$. At the integral level (see (\ref{eq_dimlessG})), it represents the most nested part of the time integral, and can be expressed as a hypergeometric series of $2I$ variables:
\begin{align}
\label{eq_MFT}
  \text{CIS}\,\big[\G\big]=&~\sum_{\{\ell,m\}} 2^{V}\cos(\pi p_{1\cdots V}/2)\Gamma(q_1+p_1+1)  \n\\
  &\times\prod_{i=2}^V \FR{(-1)^{\ell_i}}{\ell_i!\big(\fr{\ell_i+q_i+p_i}{2}+\fr54\pm\fr{\ii\wt\nu_i}2\big)_{m_i+1} }
  \Big(\FR{K_i}{2E_1}\Big)^{2m_i+3}\Big(\FR{E_i}{E_1}\Big)^{\ell_i+p_i+1}.
\end{align} 
Here $\{\ell,m\}$ denotes the collection of $2I$ summation variables $\ell_2,\cdots,\ell_V$, and $m_2,\cdots,m_V$. The line variables $K_i$ and $m_i$ share the same index $i$ with the vertex to which the line flows. The parameter $q_i\equiv \wh{\ell}_i+2\wh{m}_i+\wh{p}_i+4N_i$ encodes the family-tree structure of the CIS. Here $\wh\ell_i$ denotes the sum of $\ell$'s of all descendant sites of $i$, and other hatted quantities are defined likewise. Due to the natural family-tree structure of this series, we also call $\text{CIS}\,[\G]$ a \emph{massive family tree}. The term ``massive'' is added to distinguish it from the previously defined family trees for conformal scalars \cite{Fan:2024iek}. 
 
We adopt the notations of \cite{Fan:2024iek,Liu:2024str} to express (massive) family trees: 1) Number all vertices (and thus lines) as above. 2) Within a pair of square brackets ($\ft{\cdots}$ for conformal family trees and $\mft{\cdots}$ for massive family trees), write ancestor-descendant lines from left to right. 3) Whenever a mother has more than one daughter, enclose each daughter and her whole subfamily in parentheses. Several examples suffice to illustrate the rules:%
{\allowdisplaybreaks
\begin{eqnarray} 
\parbox{35mm}{ 
\includegraphics[width=35mm]{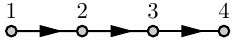}}&=&\mft{1234};\\
\parbox{35mm}{ 
\includegraphics[width=35mm]{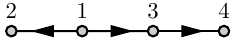}}&=&\mft{1(2)(34)};\\
\parbox{35mm}{ 
\includegraphics[width=35mm]{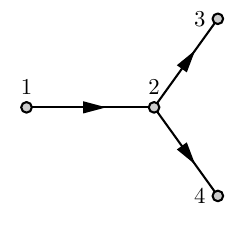}}&=&\mft{12(3)(4)};\\
\parbox{35mm}{ 
\includegraphics[width=35mm]{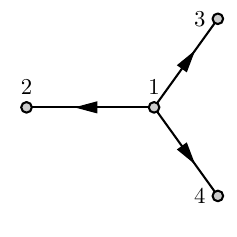}}&=&\mft{1(2)(3)(4)};
\end{eqnarray}
}%
With the family-tree notation, we can state what we mean by a cut. A cut of $\G$ with respect to Line $\al$ that connects Vertices $i$ and $j$ means the following:
\begin{align}
\label{eq_SingleCut}
  \mathop\text{Cut}_{K_\al} \big[\G\big] = \mft{\wh{1}\cdots i^\sharp\cdots }\Big\{\mft{ \cdots j^{\sharp}\cdots } + \mft{\cdots j^{\flat}\cdots }\Big\} + (\wt\nu_\al\to-\wt\nu_\al).
\end{align} 
That is, we separate $\G$ into two pieces by removing Line $\al$, \emph{augment} ($\sharp$) the site (Site $i$) that shares the same subgraph with the maximal energy site (Site 1), but including both augmentation ($\sharp$) and flattening ($\flat$) of the other site (Site $j$). The augmentation and flattening are defined by the following dressings of an otherwise ordinary CIS of a subgraph:
\begin{align}
\label{eq_augm}
  \mft{\cdots i^\sharp \cdots}\equiv &\sum_{m=0}^\infty\sqrt{\FR{2}{\pi}}\FR{\Gamma(-m-\ii\wt\nu_\al)}{m!}\Big(\FR{K_\al}{2E_i}\Big)^{2m+\ii\wt\nu_\al+3/2}\mft{\cdots i \cdots}_{p_i\to p_i+2m+\ii\wt\nu_\al+3/2}~,\\ 
\label{eq_flat}
  \mft{\cdots i^\flat \cdots}\equiv &\sum_{m=0}^\infty\sqrt{\FR{2}{\pi}}\FR{\Gamma(-m+\ii\wt\nu_\al)}{m!} \Big(\FR{K_\al}{2E_i}\Big)^{2m-\ii\wt\nu_\al+3/2}\n\\
  &\times \bigg\{\FR{\cos\big[\fr{\pi(p_\text{tot}+2\ii\wt\nu_\al)}2\big]}{\cos\big(\fr{\pi p_\text{tot}}2\big)}\mft{\cdots i  \cdots}\bigg\}_{p_i\to p_i+2m-\ii\wt\nu_\al+3/2}~,
\end{align}
where $p_\text{tot}$ is the sum of all twists in the MFT. The cut with more than one line can be defined similarly, and we refer readers to \cite{Liu:2024str} for details and explicit examples.

From the viewpoint of SK time integrals, an $N$-site MFT is computed by an $N$-fold nested time integral. The nest appears only for equal-sign propagators $\wt D_{\pm\pm}^{(\wt\nu_\al)}$, implying that the SK indices at all sites must take the same value, either $+$ or $-$, and the MFT is the sum of these two choices:
\bge
  \mft{\mathscr{P}(\wh 1\dots N)}=\mft{\mathscr{P}(\wh 1\dots N)}_++\mft{\mathscr{P}(\wh 1\dots N)}_-.
\ede
Here $\mathscr{P}(\wh 1\cdots N)$ means an arbitrary partial order of $1\cdots N$ with Site 1 being the root, and $\mft{\cdots}_\pm$ means to single out the $\pm$ branch of the MFT. It was shown in \cite{Liu:2024str} that $\mft{\cdots}_\pm$ is related to the full MFT by:
\begin{align}
\label{eq_branchRelation}
    \FR{\mft{\mathscr{P}(\wh 1\dots N)}_\pm}{\mft{\mathscr{P}(\wh 1\dots N)}} = \frac{e^{\mp \ii \pi p_\text{tot}/2}}{2 \cos(\pi p_\text{tot}/2)}.
\end{align} 

\section{Proof of the Compact Formula} 
\label{app_proof}

In this appendix, we provide an inductive proof of the compact formula \eqref{eq_DegStarSol} for degenerate stars $\mathcal{S}_N$. The key point is to show that multiple cuts of $\mathcal{S}_N$ can generally be written as
\begin{align}
\label{eq_starMultipleCuts}
    \mathop{\text{Cut}}_{\{K\}}\,\big[\mathcal{S}_N^+\big] =&~e^{-\ii{\pi}\wt\Delta_0^\ec/2}\sum_{\{m\}}\Gamma(\wh{m}_0+\wt\Delta_0^\ec-3)
    \prod_{i=1}^N\mathcal{C}^{\ec}_i\Gamma\bgb m_i+\Delta^{\ec}_i-1 \\ m_i+\Delta^{\ec}_{i}+\Delta^{\pm}_i-2\edb u_\be^{m_\be+P^{\cc_\be}_\be}+\text{shadows,}
\end{align}
where $\{K\}$ denotes the set of all cut lines, $\ec_i=0$ corresponds to the uncut lines, and $\ec_i=1$ to the cut lines. The $+$ superscript in $\mathcal{S}_N^+$ means that we consider specifically the branch where the SK index of the vertex $E_0$ is fixed to be $+$. If \eqref{eq_starMultipleCuts} is correct, then the compact expression follows naturally as a total sum of the CIS and all possible cuts.

To proceed, we first need to adapt the cutting rule \eqref{eq_cuttingRule} to the $+$ branch of a degenerate star. It can be achieved by picking up the $+$ branch of the left subgraph in equation \eqref{eq_cutG}. Therefore, we can rewrite \eqref{eq_cutG} as:
\begin{align}
    \mathop{\text{Cut}}_{K_j}\,[\mathcal{G'}^+_{N}] = \sum_{\{\cc,m\}}\Big[ A^{+\ss_i}_{m_1,\cc_1}(u_i)A^{-\ss_j *}_{m_2,-\cc_2}(u_j)\mathcal{L}^{\cc_1}_+ \mathcal{R}^{\cc_2}_+ + A^{-\ss_i *}_{m_1,\cc_1}(u_i)A^{-\ss_j}_{m_2,-\cc_2}(u_j)\mathcal{L}^{-\cc_1}_+ \mathcal{R}^{-\cc_2}_-\Big],
\end{align}
where $\mathcal{G'}^+_{N}$ is a non-degenerate star diagram without taking any folded limit. Instead of applying the relation \eqref{eq_branchRelation} to both subdiagrams, we in this case only use it to rewrite $\mathcal{R}_\pm$ into an SK-branch-independent form, but keep $\mathcal{L}_+$ unchanged. After some algebra, we get the single cut of $+$ branches:
\begin{align}
    \mathop{\text{Cut}}_{K_j}\,[\mathcal{G'}^+_N] = \mft{\wh{0}^\sharp\cdots (\sla{j}) \cdots}'_+\Big\{ \mft{ j^\sharp }' +\mft{j^\flat}'\Big\}+(\wt\nu_j \to -\wt\nu_j)
\end{align}
where the shifted massive family trees adapted for $+$ branches are defined as follows:
\begin{align}
   \mft{\wh{0}^\sharp\cdots (\sla{j}) \cdots}'_+  \equiv &~ \FR{-e^{\ii \pi m/2}}{2\cos(\pi\Delta_j^+)}\sum_{m=0}^{\infty} \Gamma\bgb \Delta_j^+ +m-1\\ 2\Delta_j^+ +m- 2\edb\FR{1}{m!}\Big(\FR{2 K_j}{\wh E_0}\Big)^{m+\Delta_j^+} \n\\ &\times\big\{\mft{\wh{0}\cdots (\sla{j}) \cdots}'\big\}_{p_0\to p_0+m+\Delta_j^+,E_i\to E_0+K_j},
\end{align}
On the other hand, we can get the folded limit of a cut leaf from \eqref{eq_foldedCut}, again keeping only the $+$ branch of the left subgraph:
\begin{align}
    \label{eq_foldedCut+}
    \mathop{\text{Fold}}_{E_j}\Big\{\mathop{\text{Cut}}_{K_j}\,[\mathcal{G'}_N^+]\Big\} = 2\cos\big[\fr{\pi}{2}\big(\Delta^0_j+\Delta^\pm_j\big)\big]\Gamma\Bigg[ \bgm \Delta^0_j+\Delta^\pm_i-3\\ \Delta^0_j-1\edm \Bigg] \mft{\wh{0}\cdots i^+ \cdots}'_+ +(\wt\nu_\alpha \to -\wt\nu_\alpha).
\end{align}
After taking folded limit at all leaves, we get the single cut for the $+$ branch of a degenerate star $\mathcal{S'}_N^+$, as:
\begin{align}
    \label{eq_starCut+}
    \mathop{\text{Cut}}_{K_j}\,\big[\mathcal{S'}_N^+(E_0;\cdots)\big]
    =~& \sum_{m_j=0}^{\infty} e^{\ii\pi m_j/2}\mathcal{S'}_{N-1}^+(E_0+K_j;\cdots\sla{K}_j\cdots)_{p_0\to p_0+m_j+\Delta^+_j}\n\\ &\times \mathcal{C}^1_j \Gamma\Bigg[ \bgm m_j+\Delta^+_j-1\\ m_j+2\Delta^+_j-2\edm \Bigg]\FR{u_i^{m_j+\Delta^+_j}}{m_j!}  +(\wt\nu_j\to -\wt\nu_j).
\end{align}
Since the above derivations do not rely on any properties that are exclusive to CISs, we immediately see that the above expression  can be directly generalized to multiple cuts as follows:
\begin{align}
    \label{eq_starCuttingRule}
    \mathop{\text{Cut}}_{K_j}\,\big[\mathop{\text{Cut}}_{\{K\}}\,[\mathcal{S'}^+]\big]
    =~& \sum_{m_j=0}^{\infty} e^{\ii\pi m_j/2}\mathop{\text{Cut}}_{\{K\}}\,[\mathcal{S'}^+]_{p_0\to p_0+m_j+\Delta^+_j}\n\\ &\times \mathcal{C}^1_j \Gamma\Bigg[ \bgm m_j+\Delta^+_j-1\\ m_j+2\Delta^+_j-2\edm \Bigg]\FR{u_j^{m_j+\Delta^+_j}}{m_j!}  +(\wt\nu_j\to -\wt\nu_j).
\end{align}
Now, we have all the ingredients for the proof.

To prove the equation \eqref{eq_starMultipleCuts}, we proceed by mathematical induction. First, the $+$ branch of an uncut CIS can be easily computed by substituting the relation \eqref{eq_branchRelation} into \eqref{eq_CISStar}, and we can easily check that the result is identical to the equation \eqref{eq_starMultipleCuts} with the set of cut lines $\{K\}$ being empty. This is our base case. Next, assuming that the formula \eqref{eq_starMultipleCuts} holds for some arbitrary set $\{K\}$. Using the cutting rule \eqref{eq_starCuttingRule}, we can easily prove that the formula continues to hold for the set $\{K_j\}\cup \{K\}$ under this hypothesis. Thus, the equation \eqref{eq_starMultipleCuts} holds for arbitrary sets of cut lines. This completes our proof of the compact formula for degenerate stars. 

\section{Partial-Energy Limit at a Single Site}
\label{app_PElimit}
In this appendix, we derive the exact series representation of a general tree graph in the small partial-energy region at any given site. As mentioned in Sec.\;\ref{subsec_partialenergy}, if we naively apply the tuned MFT formula \eqref{eq_multituning} to a single-site graph, we would get:
\begin{align}
\label{eq_partialEnergySingularitySeries}
    \sum_{T_1,\cdots,T_M=\sharp,\flat} \mft{i^{T_1\cdots T_M}}' = & \sum_{\ec=\pm}\sum_{\{m\}}  2\cos\Big[\fr{\pi(p_i+\Delta^+_{1\cdots M})}{2}\Big] \Gamma[p_i+m_{1\cdots M}+\Delta^\ec_{1\cdots M}+1]\n\\
    &\times\prod_{\al=1}^M \frac{-1}{2\cos(\pi\Delta_\alpha^\ec)} \Gamma\bgb \Delta_\al^\ec+m_\al-1 \\ 2\Delta_\al^\ec+m_\al-2\edb\FR{1}{m_\al!}\Big(\FR{2 K_\al}{\E_i}\Big)^{m_\al+\Delta_\al^\ec}.\n\\
\end{align}
However, this series is clearly divergent in the partial-energy limit $\E_i$, which calls for analytical continuation. This can be achieved by rewriting \eqref{eq_partialEnergySingularitySeries} in the form of an MB integral, as follows:
\begin{align}
    &\sum_{T_1,\cdots,T_M=\sharp,\flat} \mft{i^{T_1\cdots T_M}}'\n\\ = ~& 2\cos\Big[\fr{\pi(p_i+\Delta^+_{1\cdots M})}{2}\Big]\int_{s_1,\cdots,s_M} \Gamma[p_i-s_{1\cdots M}+1] \prod_{\alpha=1}^M \frac{\Gamma[-s_\alpha-1,s_\alpha+\Delta^{\pm}_\alpha]}{\text{sech}(\pi\wt\nu_\alpha)} \Big(\FR{2 K_\al}{\E_i}\Big)^{-s_\alpha}
\end{align}
We can compute this MB integral by summing up the residues of relevant poles. Suppose we are working in the kinematic region $K_1>K_2>\cdots>K_M>\E_i$. The general rule for computing MB integrals is that we collect all the possible sets of poles which result in convergent series in the desired parametric region \cite{Fan:2025scu}. For our kinematic region, we need to sum up $M+1$ sets of poles, which we label by $j=1,2,\cdots,M+1$. For the $j^{\text{th}}$ set of poles, we collect the poles from $\Gamma(-s_\alpha-1)$ for $1\le\alpha <j$, $\Gamma(p_i-s_{1\cdots M}+1)$ and $\Gamma[s_\beta+\Delta^\pm_\alpha]$ for $j<\beta\le M$. Taken together, for a given $j\in\{1,\cdots,M+1\}$, the poles are located at:
\bge
\begin{cases}
    s_\alpha = m_\alpha-1, &(1\le~\alpha<j)\\
    s_\beta=-m_\beta-\Delta^\ec_\beta, & (j<\beta\le M)\\
    s_j=p_i-m_{1\cdots(j-1)}+m_{(j+1)\cdots M}+\Delta^{\ec}_{(j+1)\cdots M}+j.
\end{cases} 
\ede
By the residue theorem, we sum up all the residues from these poles and get:
{\small
\begin{align}
\label{eq_partialEnergyLimit}
    &\frac{1}{\E_i^{p_i+1}}\sum_{T_1,\cdots,T_M=\sharp,\flat} \mft{i^{T_1\cdots T_M}}'= 2\cos\Big[\fr{\pi(p_i+\Delta^+_{1\cdots M})}{2}\Big]\n\\
    &\times \sum_{\{m\}} \Bigg\{\frac{1}{\E_i^{p_i+1}}\Gamma[p_i-m_{1\cdots M}+M+1]\prod_{\alpha=1}^M \Gamma[m_\al-1+\Delta^\pm_\al]\frac{(-1)^{m_\alpha+1}}{m_\al!\text{sech}(\pi \wt\nu_\al)}\Big(\frac{\E_i}{2 K_\alpha}\Big)^{m_\alpha-1}\n\\
    &+ \sum_{\ec,j=1}^M \Big(\frac{1}{2 K_j}\Big)^{p_i+1}\Big(\FR{\E_i}{2 K_j}\Big)^{m_j}\prod_{\alpha=1}^{j-1}\frac{(-1)^{m_\al+1}\Gamma[m_\al-1+\Delta^\pm_\al]}{m_\al!\text{sech}(\pi \wt\nu_\al)}\Big(\frac{K_j}{ K_\alpha}\Big)^{m_\alpha-1}\n\\
    & \times\prod_{\beta=j+1}^{M}\frac{(-1)^{m_\beta}\Gamma[m_\beta+\Delta^\ec_\be-1,-m_\beta-2\Delta^{\ec}_\beta+3]}{m_\beta!\text{sech}(\pi \wt\nu_\be)}\Big(\frac{K_\beta}{ K_j}\Big)^{m_\beta+\Delta_\beta^{\ec}}\n\\
    & \times \frac{(-1)^{m_j+1}\Gamma[-p_i\!+\!m_{1\cdots (j-1)}\!-\!m_{j\cdots M}\!-\!\Delta_{(j+1)\cdots M}^\ec\!-\!j\!-\!1, p_i-m_{1\cdots (j-1)}\!+\!m_{j\cdots M}\!+\!\Delta_{(j+1)\cdots M}^\ec\!+\!\Delta^\pm_j+j]}{m_j!\text{sech}(\pi \wt\nu_j)} \Bigg\}.
\end{align}
}%
In the partial-energy limit $\E_i\to 0$, the term in the 3rd line of \eqref{eq_partialEnergyLimit} becomes singular. So we have the singular part of this expression, denoted by $\text{Sg}[\cdots]$, as follows:
\begin{align}
     & \text{Sg}\,\Bigg\{ \FR{1}{\E_i^{p_i+1}} \sum_{T_1,\cdots,T_M=\sharp,\flat} \mft{i^{T_1\cdots T_M}}' \Bigg\} =  \FR{2\cos\big[\fr\pi2(p_i+\Delta^+_{1\cdots M})\big]}{\E_i^{p_i+1}}  \n\\
    &\times \sum_{\{m\}}\Gamma[p_i-m_{1\cdots M}+M+1]\prod_{\alpha=1}^M \FR{\Gamma[m_\al-1+\Delta^\pm_\al]}{m_\al!\text{sech}(\pi\wt\nu_\al)} \Big(\frac{-\E_i}{2 K_\alpha}\Big)^{m_\alpha-1}.
\end{align}

When the twist $p_i$ is an integer, the above result should be treated with care. The factor $\Gamma[p_i-m_{1\cdots M}+M+1]$ can be divergent with respect to integer $p_i$, but the divergence cancels out in the full result of the single-site tree $\mft{i^{T_1\cdots T_M}}'$ and we are left with a logarithmic divergence in $\E_i$ . To show this, we consider the case when $p_i=-2$ and $M=1$ as an example:
\begin{align}
     \FR{1}{\E_i^{p_i+1}} \sum_{T=\sharp,\flat} \mft{i^{T} }'=& -\frac{2\cos\big[\fr{\pi(p_i+\Delta^+)}{2}\big]}{\text{sech}(\pi\wt\nu_\al)} \sum_{m} \Bigg\{\frac{2K}{\E_i^{p_i+2}}\Gamma[p_i-m+2,m+\Delta^\pm-1]\n\\
    &+ \frac{1}{(2 K)^{p_i+1}} \Gamma[-p_i-m-2, p_i+m+\Delta^\pm+1] \Bigg\}\frac{1}{m!}\Big(\FR{-\E_i}{2 K}\Big)^{m}.
\end{align}
In the limit $p_i\to -2$, we have:
\begin{align}
    \frac{2K}{\E_i^{p_i+2}}\Gamma[p_i+2,\Delta^\pm-1] &= 2K~\Gamma[\Delta^\pm-1]\Big(\frac{1}{p+2}-\gamma-\log \E_i\Big)+\mathcal{O}(p+2),\n\\
    \frac{1}{(2 K)^{p_i+1}} \Gamma[-p_i-2, p_i+\Delta^\pm+1]&=-2K~\Gamma[\Delta^\pm-1]\Big(\frac{1}{p+2}+\gamma-\log 2K\Big)+\mathcal{O}(p+2).
\end{align}
It is obvious that the divergent term $\frac{1}{p_i+1}$ cancels out. Thus, we get the singular term that is logarithmic:
\begin{align}
    \text{Sg} \, \Bigg\{\E_i \sum_{T=\sharp,\flat} \mft{i^{T} }'\Bigg\} = \frac{4 \cos\big[\fr{\pi(p_i+\Delta^+)}{2}\big]}{\text{sech}(\pi\wt\nu_\al)}\Gamma[\Delta^\pm-1] K \log \frac{\E_i}{2K}.
\end{align}

\section{Two-Pointed Degenerate Star}
\label{app_S2}
In this appendix, we unfold (\ref{eq_S2}) in both the KdF form and the partially resummed form. For this purpose, it is useful to rewrite it as:
\bge
\label{eq_S2_cuts}
  \mathcal{S}_2 =  \underbrace{\text{CIS}\big[\mathcal{S}_2\big]}_{\ec_1=\ec_2=0}+\underbrace{\mathop\text{Cut}_{K_1}\big[\mathcal{S}_2\big]}_{\ec_1=\pm,\ec_2=0}+\underbrace{\mathop\text{Cut}_{K_2}\big[\mathcal{S}_2\big]}_{\ec_1=0,\ec_2=\pm}+\underbrace{\mathop\text{Cut}_{K_1,K_2}\big[\mathcal{S}_2\big] }_{\ec_1=\pm,\ec_2=\pm}.
\ede
Then, we can present the four terms in turn. First, the CIS is:
\begin{align}
    \text{CIS}\big[\mathcal{S}_2\big] =& ~{2\cos(\pi p_{012}/2)} \mathcal{C}^0_1\mathcal{C}^0_2~u_1^{p_1+4}u_2^{p_2+4}\times{}^{1+2}\mathcal{F}_{0+2}\Bigg[\bgm p_{012}+9\\-\edm\Bigg|\bgm p_1+3,p_2+3 ;1,1\\ p_1+\frac{7}{2}\pm\ii \wt{\nu}_1, p_2+\frac{7}{2}\pm\ii \wt{\nu}_2 \edm\Bigg|u_1,u_2\Bigg]\n\\
    =&~{2\cos(\pi p_{012}/2)\mathcal{C}_1^0\mathcal{C}_2^0}u_1^{p_1+4}u_2^{p_2+4} \n\\
    &~\times \sum_{m_1}u_1^{m_1} \Gamma\bgb m_1+p_1+3 \\ m_1+p_1+\fr72 \pm\ii \wt\nu_1\edb {}_3 \mathcal{F}_2 \Bigg[\bgm 1,p_2+3,m_1+p_{012}+9 \\ p_2+\frac72+\ii\wt\nu_2, p_2+\frac72-\ii\wt\nu_2\edm\Bigg|u_2\Bigg];
\end{align} 
The second term is:
\begin{align}
    &\mathop\text{Cut}_{K_1}\big[\mathcal{S}_2\big] ={2\cos\big[\fr{\pi}2(p_{02}+\ii\wt\nu_1+\fr32)\big]}\mathcal{C}^{1}_1\mathcal{C}^{0}_2u_1^{\frac32+\ii\wt\nu_1}u_2^{p_2+4}\n\\
    &~~~\times {}^{1+2}\mathcal{F}_{0+2}\Bigg[\bgm p_{02}+\frac{13}{2}+\ii\wt\nu_1\\-\edm\Bigg|\bgm \ii\wt\nu_1+\frac12,p_2+3 ;1,1\\ 1+\ii \wt\nu_1\pm\ii\wt\nu_1, p_2+\frac{7}{2}\pm\ii \wt{\nu}_2 \edm\Bigg|u_1,u_2\Bigg] +(\wt\nu_1\to -\wt\nu_1)\n\\
   & = {2\cos\big[\fr{\pi}2(p_{02}+\ii\wt\nu_1+\fr32)\big]}\mathcal{C}^{1}_1\mathcal{C}^{0}_2 u_1^{\frac32+\ii\wt\nu_1}u_2^{p_2+4}\n\\
    &~~~\times \sum_{m_1=0}^\infty u_1^{m_1}\Gamma\bgb m_1+\ii\wt\nu_1+\frac12 \\m_1+1+2\ii\wt\nu_1,m_1+1\edb {}_3\mathcal{F}_2\Bigg[\bgm 1,p_2+3,m_{1}+p_{02}+\frac{13}{2}+\ii\wt\nu_1\\ p_2+\frac72+\ii \wt\nu_2,p_2+\frac72-\ii\wt\nu_2\edm \Bigg| u_2 \Bigg]\n\\
    &~~~+(\wt\nu_1\to -\wt\nu_1);
\end{align}
The third term, the single cut over $K_2$, is related to the second term via the replacement $(p_1,K_1,\wt\nu_1)\leftrightarrow (p_2,K_2,\wt\nu_2)$:
\begin{align}
    ~&\mathop\text{Cut}_{K_2}\big[\mathcal{S}_2\big] = \mathop\text{Cut}_{K_1}\big[\mathcal{S}_2\big]\big|_{(p_1,K_1,\wt\nu_1)\leftrightarrow (p_2,K_2,\wt\nu_2)}\n\\
    &={2\cos\big[\fr{\pi}2(p_{01}+\ii\wt\nu_2+\fr32)\big]}\mathcal{C}^{0}_1\mathcal{C}^{1}_2 u_1^{p_1+4}u_2^{\frac32+\ii\wt\nu_2}\n\\
    &~~~\times \sum_{m_1=0}^\infty u_1^{m_1}\Gamma\bgb m_1+p_1+3 \\m_1+p_1+\frac72+\ii \wt\nu_1,m_1+p_1+\frac72-\ii\wt\nu_1\edb {}_2\mathcal{F}_1\Bigg[\bgm \frac12 +\ii\wt\nu_2,m_{1}+p_{01}+\frac{13}{2}+\ii\wt\nu_2\\ 1+2\ii\wt\nu_2\edm \Bigg| u_2 \Bigg]\n\\
    &~~~+(\wt\nu_2\to -\wt\nu_2).
\end{align}
Finally, the fourth term is:
 \begin{align}
   &\mathop{\operatorname{Cut}}_{K_1,K_2}\big[\mathcal{S}_2\big]
  = {2\cos[\fr{\pi}2(p_0+\ii\wt\nu_{12}+3)]}  \mathcal{C}^1_1\mathcal{C}^1_2 u_1^{\frac32+\ii\wt\nu_1}u_2^{\frac32+\ii\wt\nu_2}\n\\
  &~~~ \times \mathcal{F}_2\Bigg[p_0+4+\ii\wt\nu_{12}\Bigg|\bgm \frac12+\ii\wt\nu_{1},\frac12+\ii\wt\nu_2 \\ 1+2\ii \wt\nu_1,1+2\ii \wt\nu_2 \edm \Bigg| u_1,u_2\Bigg] + \text{shadows}\n\\
 & =2\cos\big[\fr{\pi}2(p_0+\ii\wt\nu_{12}+3)\big] \mathcal{C}^1_1\mathcal{C}^1_2 u_1^{\frac32+\ii\wt\nu_1}u_2^{\frac32+\ii\wt\nu_2}\n\\
  &~~~\times \sum_{m_1=0}^\infty u_1^{m_1} \Gamma\bgb m_1+\frac12+\ii\wt\nu_1\\m_1+1+2\ii\wt\nu_1,m_1+1\edb {}_2\mathcal{F}_1\Bigg[\bgm \frac12 +\ii\wt\nu_2,m_{1}+p_{0}+4+\ii\wt\nu_{12}\\ 1+2\ii\wt\nu_2\edm \Bigg| u_2 \Bigg] 
 +\text{shadows}.
\end{align}
Here $(\cdots)+\text{shadows}$ means $\big[(\cdots)+(\wt\nu_1\to-\wt\nu_1)\big]+(\wt\nu_2\to-\wt\nu_2)$.

\section{Special functions}
\label{app_spcfun}

In this appendix, we collect special functions used in the main text. First, following our previous works on similar topics, we use shorthand notations for the products and fractions of Euler $\Gamma$ functions:
\begin{align}
  \Gamma\left[ z_1,\cdots,z_m \right]
  \equiv&~ \Gamma(z_1)\cdots \Gamma(z_m) ,\\
  \Gamma\left[\bgm z_1,\cdots,z_m \\w_1,\cdots, w_n\edm\right]
  \equiv&~\FR{\Gamma(z_1)\cdots \Gamma(z_m)}{\Gamma(w_1)\cdots \Gamma(w_n)}.
\end{align}
In the main text, we also frequently use the Pochhammer symbol $(a)_n\equiv \Gamma(a+n)/\Gamma(a)$.

Next, we list useful hypergeometric functions. In the literature, a collection of hypergeometric series have been well studied and designated with special names and we mention a few used functions here. More details about these functions can be found in \cite{Slater:1966,srivastava1985multiple}. First, consider the hypergeometric functions with a single variable. The (generalized) hypergeometric function ${}_p\text{F}_q$ is defined by the following way when the series converges: 
\begin{align}
\label{eq_pFq}
  {}_p\text{F}_q\left[\bgm a_1,\cdots,a_p \\ b_1,\cdots,b_q \edm  \middle| z \right]
  =\sum_{n=0}^\infty
  \FR{(a_1)_n\cdots (a_p)_n}{(b_1)_n\cdots(b_q)_n}\FR{z^n}{n!}.
\end{align}
In the main text, we use the following dressed version of the hypergeometric function: 
\begin{align}
\label{eq_dressedF}
  {}_p\mathcal{F}_q\left[\bgm a_1,\cdots,a_p \\ b_1,\cdots,b_q \edm  \middle| z \right]
  =&~\Gamma\bgb a_1,\cdots,a_p \\ b_1,\cdots,b_q\edb
  {}_pF_q\left[\bgm a_1,\cdots,a_p \\ b_1,\cdots,b_q \edm  \middle| z \right]\n\\
  =&~\sum_{n=0}^\infty\Gamma\left[\begin{matrix}
        a_1+n, \cdots, a_p+n \\
        b_1+n, \cdots, b_q+n
    \end{matrix}\right]\FR{z^n}{n!}.
\end{align}

Next, we come to hypergeometric functions with two variables. First, there are four Appell functions $\text{F}_1,\cdots,\text{F}_4$. One of them, $\text{F}_2$, is used in the main text, whose dressed version is defined as:
\begin{align}
\label{eq_dressedF2}
  \mathcal{F}_2\left[a \middle| \bgm b_1,b_2\\ c_1,c_2 \edm\middle| x,y\right]= \sum_{m,n=0}^\infty\Gamma\bgb a+m+n,b_1+m,b_2+n \\ c_1+m, c_2+n \edb\FR{x^my^n}{m!n!}.
\end{align}
Next, a more general class of bivariate hypergeometric functions is called the Kampé de Fériet (KdF) function in the literature, whose definition is: 
\begin{align}
  &{}^{p+q}\text{F}_{r+s}\left[\bgm a_1,\cdots,a_p\\ c_1,\cdots,c_r\edm \middle| \bgm b_1,\cdots,b_q;b_1',\cdots,b_q'\\ d_1,\cdots,d_s;d_1',\cdots,d_s'\edm\middle|x,y \right]\n\\
  =&\sum_{m,n=0}^\infty\FR{(a_1)_{m+n}\cdots(a_p)_{m+n}}{(c_1)_{m+n}\cdots(c_r)_{m+n}}\FR{(b_1)_{m}(b_1')_{n}\cdots(b_q)_{m}(b_q')_{n}}{(d_1)_{m}(d_1')_{n}\cdots(d_s)_{m}(d_s')_{n}}\FR{x^my^n}{m!n!}.
\end{align}
Again, we use the dressed version in the main text:
\begin{align}
\label{eq_KdFfun2}
  &{}^{p+q}\mathcal{F}_{r+s}\left[\bgm a_1,\cdots,a_p\\ c_1,\cdots,c_r\edm \middle| \bgm b_1,\cdots,b_q;b_1',\cdots,b_q'\\ d_1,\cdots,d_s;d_1',\cdots,d_s'\edm\middle|x,y \right]\n\\
  =&\sum_{m,n=0}^\infty 
  \Gamma\bgb a_1+m+n,\cdots,a_p+m+n\\ c_1+m+n,\cdots,c_r+m+n\edb
  \Gamma\bgb b_1+m,\cdots,b_q+m\\ d_1+m,\cdots,d_s+m\edb
  \Gamma\bgb b_1'+n,\cdots,b_q'+n\\ d_1'+n,\cdots,d_s'+n\edb
   \FR{x^my^n}{m!n!}.
\end{align}
Finally, the  functions can be generalized into multivariate hypergeometric functions \cite{srivastava1985multiple}. Here, we present the trivariate generalization, which is used in the main text:
\begin{align}
\label{eq_KdFfun3}
    &{}^{p+q}\mathcal{F}_{r+s}\left[\bgm a_1,\cdots,a_p\\ c_1,\cdots,c_r\edm \middle| \bgm b_1,\cdots,b_q;b_1',\cdots,b_q';b_1'',\cdots,b_q''\\ d_1,\cdots,d_s;d_1',\cdots,d_s';d_1'',\cdots,d_s''\edm\middle|x,y ,z\right]\n\\
  =&\sum_{m,n,k=0}^\infty 
  \Gamma\bgb a_1+m+n+k,\cdots,a_p+m+n+k\\ c_1+m+n+k,\cdots,c_r+m+n+k\edb
  \n\\
  &\times\Gamma\bgb b_1+m,\cdots,b_q+m\\ d_1+m,\cdots,d_s+m\edb
  \Gamma\bgb b_1'+n,\cdots,b_q'+n\\ d_1'+n,\cdots,d_s'+n\edb\Gamma\bgb b_1''+k,\cdots,b_q''+k\\ d_1''+k,\cdots,d_s''+k\edb
   \FR{x^my^nz^k}{m!n!k!}.
\end{align}

\end{appendix}

\newpage
\bibliography{CosmoCollider} 
\bibliographystyle{utphys}

\end{document}